\input harvmac.tex

\newdimen\tableauside\tableauside=1.0ex
\newdimen\tableaurule\tableaurule=0.4pt
\newdimen\tableaustep
\def\phantomhrule#1{\hbox{\vbox to0pt{\hrule height\tableaurule width#1\vss}}}
\def\phantomvrule#1{\vbox{\hbox to0pt{\vrule width\tableaurule height#1\hss}}}
\def\sqr{\vbox{%
  \phantomhrule\tableaustep
  \hbox{\phantomvrule\tableaustep\kern\tableaustep\phantomvrule\tableaustep}%
  \hbox{\vbox{\phantomhrule\tableauside}\kern-\tableaurule}}}
\def\squares#1{\hbox{\count0=#1\noindent\loop\sqr
  \advance\count0 by-1 \ifnum\count0>0\repeat}}
\def\tableau#1{\vcenter{\offinterlineskip
  \tableaustep=\tableauside\advance\tableaustep by-\tableaurule
  \kern\normallineskip\hbox
    {\kern\normallineskip\vbox
      {\gettableau#1 0 }%
     \kern\normallineskip\kern\tableaurule}%
  \kern\normallineskip\kern\tableaurule}}
\def\gettableau#1 {\ifnum#1=0\let\next=\null\else
  \squares{#1}\let\next=\gettableau\fi\next}

\tableauside=1.0ex
\tableaurule=0.4pt
\input epsf
\noblackbox

\def\ra{{\rightarrow}}

\def\wphi{{\widetilde \phi}}

\def\l{{\lambda}}
\def\bbbone{{\mathchoice {\rm 1\mskip-4mu l} {\rm 1\mskip-4mu l}
          {\rm 1\mskip-4.5mu l} {\rm 1\mskip-5mu l}}}
\def\bbbt{{\mathchoice {\setbox0=\hbox{$\displaystyle\rm
T$}\hbox{\hbox to0pt{\kern0.3\wd0\vrule height0.9\ht0\hss}\box0}}
{\setbox0=\hbox{$\textstyle\rm T$}\hbox{\hbox
to0pt{\kern0.3\wd0\vrule height0.9\ht0\hss}\box0}}
{\setbox0=\hbox{$\scriptstyle\rm T$}\hbox{\hbox
to0pt{\kern0.3\wd0\vrule height0.9\ht0\hss}\box0}}
{\setbox0=\hbox{$\scriptscriptstyle\rm T$}\hbox{\hbox
to0pt{\kern0.3\wd0\vrule height0.9\ht0\hss}\box0}}}}
\def\bbbti{{\mathchoice {\setbox0=\hbox{$\displaystyle\rm
T$}\hbox{\hbox to0pt{\kern0.3\wd0\vrule height0.9\ht0\hss}\box0}}
{\setbox0=\hbox{$\textstyle\rm T$}\hbox{\hbox
to0pt{\kern0.3\wd0\vrule height0.9\ht0\hss}\box0}}
{\setbox0=\hbox{$\scriptstyle\rm T$}\hbox{\hbox
to0pt{\kern0.3\wd0\vrule height0.9\ht0\hss}\box0}}
{\setbox0=\hbox{$\scriptscriptstyle\rm T$}\hbox{\hbox
to0pt{\kern0.3\wd0\vrule height0.9\ht0\hss}\box0}}}}


\def\unlockat{\catcode`\@=11}
\def\lockat{\catcode`\@=12}

\unlockat

\def\newsec#1{\global\advance\secno by1\message{(\the\secno. #1)}
\global\subsecno=0\global\subsubsecno=0\eqnres@t\noindent
{\bf\the\secno. #1}
\writetoca{{\secsym} {#1}}\par\nobreak\medskip\nobreak}
\global\newcount\subsecno \global\subsecno=0
\def\subsec#1{\global\advance\subsecno
by1\message{(\secsym\the\subsecno. #1)}
\ifnum\lastpenalty>9000\else\bigbreak\fi\global\subsubsecno=0
\noindent{\it\secsym\the\subsecno. #1}
\writetoca{\string\quad {\secsym\the\subsecno.} {#1}}
\par\nobreak\medskip\nobreak}
\global\newcount\subsubsecno \global\subsubsecno=0
\def\subsubsec#1{\global\advance\subsubsecno
\message{(\secsym\the\subsecno.\the\subsubsecno. #1)}
\ifnum\lastpenalty>9000\else\bigbreak\fi
\noindent\quad{\secsym\the\subsecno.\the\subsubsecno.}{#1}
\writetoca{\string\qquad{\secsym\the\subsecno.\the\subsubsecno.}{#1}}
\par\nobreak\medskip\nobreak}

\def\subsubseclab#1{\DefWarn#1\xdef
#1{\noexpand\hyperref{}{subsubsection}%
{\secsym\the\subsecno.\the\subsubsecno}%
{\secsym\the\subsecno.\the\subsubsecno}}%
\writedef{#1\leftbracket#1}\wrlabeL{#1=#1}}
\lockat

\def\IL{{\relax{\rm I\kern-.18em L}}}
\def\IH{{\relax{\rm I\kern-.18em H}}}
\def\IR{{\relax{\rm I\kern-.18em R}}}
\def\IE{{\relax{\rm I\kern-.18em E}}}
\def\IC{{\relax\hbox{$\inbar\kern-.3em{\rm C}$}}}
\def\IZ{{\relax\ifmmode\mathchoice
{\hbox{\cmss Z\kern-.4em Z}}{\hbox{\cmss Z\kern-.4em Z}}
{\lower.9pt\hbox{\cmsss Z\kern-.4em Z}}
{\lower1.2pt\hbox{\cmsss Z\kern-.4em Z}}\else{\cmss Z\kern-.4em
Z}\fi}}

\def\CR {{\cal R}}

\def\CF {{\cal F}}
\def\CJ {{\cal J}}
\def\CP {{\cal P }}
\def\CL {{\cal L}}
\def\CV {{\cal V}}
\def\CO {{\cal O}}
\def\CZ {{\cal Z}}
\def\CE {{\cal E}}

\def\CC {{\cal C}}

\def\CS {{\cal S}}


\def\CO {{\cal O}}

\def\CP {{\cal P }}

\def\CE{{\cal E }}
\def\CV{{\cal V }}
\def\CZ {{\cal Z }}
\def\CS {{\cal S }}
\def\CT{{\cal T}}

\def\wy{{\widetilde Y}}

\font\manual=manfnt \def\dbend{\lower3.5pt\hbox{\manual\char127}}

\def\IZ{{\relax\ifmmode\mathchoice
{\hbox{\cmss Z\kern-.4em Z}}{\hbox{\cmss Z\kern-.4em Z}}
{\lower.9pt\hbox{\cmsss Z\kern-.4em Z}}
{\lower1.2pt\hbox{\cmsss Z\kern-.4em Z}}\else{\cmss Z\kern-.4em
Z}\fi}}
\def\half {{1\over 2}}

\def\CW{{\cal W}}
\def\CJ {{\cal J}}

\def\CO {{\cal O}}

\def\CP {{\cal P }}

\def\CE{{\cal E }}
\def\CV{{\cal V }}
\def\CZ {{\cal Z }}
\def\CS {{\cal S }}

\def\om{{\overline M}}

\def\oY{{\overline Y}}
\def\w\phi{{\widetilde \phi}}


\def\IZ{{\relax\ifmmode\mathchoice
{\hbox{\cmss Z\kern-.4em Z}}{\hbox{\cmss Z\kern-.4em Z}}
{\lower.9pt\hbox{\cmsss Z\kern-.4em Z}}
{\lower1.2pt\hbox{\cmsss Z\kern-.4em Z}}\else{\cmss Z\kern-.4em
Z}\fi}}
\def\IB{{\relax{\rm I\kern-.18em B}}}
\def\IC{{\relax\hbox{$\inbar\kern-.3em{\rm C}$}}}
\def\ID{{\relax{\rm I\kern-.18em D}}}
\def\IE{{\relax{\rm I\kern-.18em E}}}
\def\IF{{\relax{\rm I\kern-.18em F}}}
\def\IG{{\relax\hbox{$\inbar\kern-.3em{\rm G}$}}}
\def\IGa{{\relax\hbox{${\rm I}\kern-.18em\Gamma$}}}
\def\IH{{\relax{\rm I\kern-.18em H}}}
\def\II{{\relax{\rm I\kern-.18em I}}}
\def\IK{{\relax{\rm I\kern-.18em K}}}
\def\IP{{\relax{\rm I\kern-.18em P}}}

\def\l{{\lambda}}
\def\wXi{{\widetilde \Xi}}
\def\wG{{\widetilde G}}
\def\wcv{{\widetilde {\cal V}}}

\def\inbar{\,\vrule height1.5ex width.4pt depth0pt}

\font\cmss=cmss10 \font\cmsss=cmss10 at 7pt
\def\IR{\relax{\rm I\kern-.18em R}}
\def\IT{\relax{\rm I\kern-.58em T}}

\def\Tr{{\rm Tr}}

\def\wt{{\widetilde T}}
\def\wz{{\widetilde Z}}

\def\wJ{{\widetilde J}}
\def\wbeta{{\widetilde \beta}}

\def\wcz{{\widetilde{\cal Z}}}
\def\wC{{\widetilde {\cal C}}}
\def\wU{{\widetilde U}}
\def\wV{{\widetilde V}}
\def\wW{{\widetilde W}}
\def\wT{{\widetilde T}}
\def\ybox{{\tableau{1}}}

\def\boxit#1{\vbox{\hrule\hbox{\vrule\kern8pt
\vbox{\hbox{\kern8pt}\hbox{\vbox{#1}}\hbox{\kern8pt}}
\kern8pt\vrule}\hrule}}
\def\mathboxit#1{\vbox{\hrule\hbox{\vrule\kern8pt\vbox{\kern8pt
\hbox{$\displaystyle #1$}\kern8pt}\kern8pt\vrule}\hrule}}


\def\inbar{\,\vrule height1.5ex width.4pt depth0pt}

\font\cmss=cmss10 \font\cmsss=cmss10 at 7pt
\def\IR{\relax{\rm I\kern-.18em R}}

\def\Tr{{\rm Tr}}


\def\half{{1\over 2}}
\def\ra{{\longrightarrow}}

\def\sin{{\rm sin}}

\def\wbeta{{\widetilde \beta}}
\let\includefigures=\iftrue
\newfam\black
\includefigures

\input epsf
\def\plb#1 #2 {Phys. Lett. {\bf B#1} #2 }
\long\def\del#1\enddel{}
\long\def\new#1\endnew{{\bf #1}}
\let\<\langle \let\>\rangle

\def\figin{\epsfcheck\figin}\def\figins{\epsfcheck\figins}
\def\epsfcheck{\ifx\epsfbox\UnDeFiNeD
\message{(NO epsf.tex, FIGURES WILL BE IGNORED)}
\gdef\figin##1{\vskip2in}\gdef\figins##1{\hskip.5in} blank space instead
\else\message{(FIGURES WILL BE INCLUDED)}
\gdef\figin##1{##1}\gdef\figins##1{##1}\fi}
\def\DefWarn#1{}
\def\figinsert{\goodbreak\midinsert}
\def\ifig#1#2#3{\DefWarn#1\xdef#1{fig.~\the\figno}
\writedef{#1\leftbracket fig.\noexpand~\the\figno}
\figinsert\figin{\centerline{#3}}\medskip
\centerline{\vbox{\baselineskip12pt
\advance\hsize by -1truein\noindent
\footnotefont{\bf Fig.~\the\figno:} #2}}
\bigskip\endinsert\global\advance\figno by1}
\else
\def\ifig#1#2#3{\xdef#1{fig.~\the\figno}
\writedef{#1\leftbracket fig.\noexpand~\the\figno}
\figinsert\figin{\centerline{#3}}\medskip
\centerline{\vbox{\baselineskip12pt
\advance\hsize by -1truein\noindent
\footnotefont{\bf Fig.~\the\figno:} #2}}
\bigskip\endinsert
\global\advance\figno by1}
\fi

\input xy
\xyoption{all}
\font\cmss=cmss10 \font\cmsss=cmss10 at 7pt
\def\inbar{\,\vrule height1.5ex width.4pt depth0pt}
\def\IC{{\relax\hbox{$\inbar\kern-.3em{\rm C}$}}}
\def\IP{{\relax{\rm I\kern-.18em P}}}
\def\IF{{\relax{\rm I\kern-.18em F}}}
\def\IZ{\relax\ifmmode\mathchoice
{\hbox{\cmss Z\kern-.4em Z}}{\hbox{\cmss Z\kern-.4em Z}}
{\lower.9pt\hbox{\cmsss Z\kern-.4em Z}}
{\lower1.2pt\hbox{\cmsss Z\kern-.4em Z}}\else{\cmss Z\kern-.4em
Z}\fi}
\def\IR{{\relax{\rm I\kern-.18em R}}}
\def\IQ{\relax\hbox{\kern.25em$\inbar\kern-.3em{\rm Q}$}}

\def\pmb#1{\setbox0=\hbox{#1}%
 \kern-.025em\copy0\kern-\wd0
 \kern.05em\copy0\kern-\wd0
 \kern-.025em\raise.0433em\box0 }
\font\cmss=cmss10
\font\cmsss=cmss10 at 7pt
\def\rlx{\relax\leavevmode}
\def\Cop{\relax\,\hbox{$\inbar\kern-.3em{\rm C}$}}
\def\Rop{\relax{\rm I\kern-.18em R}}
\def\Nop{\relax{\rm I\kern-.18em N}}
\def\Pop{\relax{\rm I\kern-.18em P}}
\def\Zop{\rlx\leavevmode\ifmmode\mathchoice{\hbox{\cmss Z\kern-.4em Z}}
 {\hbox{\cmss Z\kern-.4em Z}}{\lower.9pt\hbox{\cmsss Z\kern-.36em Z}}
 {\lower1.2pt\hbox{\cmsss Z\kern-.36em Z}}\else{\cmss Z\kern-.4em
 Z}\fi}

\def\inbar{\,\vrule height1.5ex width.4pt depth0pt}
\def\IC{{\relax\hbox{$\inbar\kern-.3em{\rm C}$}}}
\def\IP{{\relax{\rm I\kern-.18em P}}}
\def\IF{{\relax{\rm I\kern-.18em F}}}
\def\IZ{\relax\ifmmode\mathchoice
{\hbox{\cmss Z\kern-.4em Z}}{\hbox{\cmss Z\kern-.4em Z}}
{\lower.9pt\hbox{\cmsss Z\kern-.4em Z}}
{\lower1.2pt\hbox{\cmsss Z\kern-.4em Z}}\else{\cmss Z\kern-.4em
Z}\fi}
\def\IR{{\relax{\rm I\kern-.18em R}}}

\def\half{{1\over 2}}

\def\by{{\overline y}}

\def\by{{\overline Y}}

\def\wY{{\widetilde Y}}
\def\wX{{\widetilde X}}
\def\wC{{\widetilde C}}
\def\wF{{\widetilde F}}

\def\wt{{\widetilde t}}
\def\om{{\overline M}}

\def\oY{{\overline Y}}
\def\oL{{\overline L}}
\def\oX{{\overline X}}
\def\wZ{{\widetilde Z}}
\def\wh{{\widetilde h}}

\def\wy{{\widetilde Y}}
\def\wcz{{\widetilde{\cal{Z}}}}
\def\wTheta{{\widetilde \Theta}}

\def\tq{{\tilde q}}
\def\tz{{\tilde z}}
\def\ttt{{\tilde t}}
\def\tzeta{{\tilde \zeta}}
%
%
%
\nref\AVi{M. Aganagic and C. Vafa, ``Mirror
Symmetry, D-Branes and Counting Holomorphic Discs'',
hep-th/0012041.}
\nref\AKV{M. Aganagic, A. Klemm and C. Vafa,
``Disk Instantons, Mirror Symmetry and the Duality Web'', Z.
Naturforsch. {\bf A 57} (2002) 1, hep-th/0105045.}
\nref\AKMV{M. Aganagic, A. Klemm, M. Mari\~no and C. Vafa,
``Matrix Model as a Mirror of Chern-Simons Theory'', hep-th/0211098.}
\nref\AViv{M. Aganagic and C. Vafa, ``$G_2$
Manifolds, Mirror Symmetry and Geometric Engineering'',
hep-th/0110171.}
\nref\AMV{M. Aganagic, M. Mari{\~n}o and C. Vafa,
``All Loop Topological String Amplitudes from Chern-Simons
Theory'', hep-th/0206164.}
\nref\ACJM{A.C. Avram, P. Candelas, D. Jan$\check {\rm{c}}$i{\'c} and 
M. Mandelberg, ``On the Connectedness of the Moduli Space 
of Calabi-Yau Manifolds'', Nucl. Phys. {\bf B465} (1996) 458,
hep-th/9511230.}
\nref\Ba{V. Batyrev, ``Dual Polyhedra and Mirror Symmetry for 
Calabi-Yau Hypersurfaces in Toric Varieties'', J. Algebraic Geometry, 
{\bf 3} (1994) 493.} 
\nref\BKK{P. Berglund, S. Katz and A. Klemm, 
``Mirror Symmetry and the Moduli Space for Generic Hypersurfaces 
in Toric Varieties'', Nucl. Phys. {\bf B456} (1995) 153, hep-th/9506091.}
\nref\B{K. Behrend, ``Gromov-Witten Invariants in Algebraic Geometry'', 
Invent. Math. {\bf 127} (1997) 601.}
\nref\BF{K. Behrend and B. Fantechi, ``The
Intrinsic Normal Cone'', Invent. Math. {\bf 128} (1997) 45.}
\nref\CdFKM{P. Candelas, X. de la Ossa, A. Font, S. Katz and
 D.R. Morrison, ``Mirror Symmetry for Two Parameter Models -- I'',
 Nucl. Phys. {\bf B416} (1994) 481, hep-th/9308083.}
\nref\CK{D.A. Cox and S. Katz, 
{\it ``Mirror Symmetry and Algebraic Geometry'',} Mathematical Surveys and 
Monographs, Vol. 68, Providence 1999.} 
 \nref\CKYZ{T.-M.
Chiang, A. Klemm, S.-T. Yau and E. Zaslow, ``Local Mirror
Symmetry: Calculations and Interpretations'', ATMP {\bf 3} (1999)
495, hep-th/9903053.}
 \nref\HC{H. Clemens, ``Double Solids'', Adv.
Math. {\bf 47} (1983) 107.}
 \nref\DFGi{D.-E. Diaconescu, B. Florea
and A. Grassi, ``Geometric Transitions and Open String
Instantons'', ATMP {\bf 6} (2002), hep-th/0205234.}
\nref\DFGii{D.-E. Diaconescu, B. Florea and A. Grassi, 
``Geometric Transitions, del Pezzo Surfaces and Open String Instantons'', 
ATMP {\bf 6} (2002), hep-th/0206163.} 
\nref\DVi{R. Dijkgraaf and C. Vafa, 
``Matrix Models, Topological Strings, and Supersymmetric Gauge Theories'', 
Nucl. Phys. {\bf B644} (2002) 3, hep-th/0206255.}
\nref\DVii{R. Dijkgraaf and C. Vafa, ``On Geometry and Matrix Models'', 
Nucl. Phys. {\bf B644} (2002) 21, hep-th/0207106.}
\nref\FP{C. Faber and R. Pandharipande, ``Hodge Integrals and
Gromov--Witten Theory'', Invent. Math. {\bf 139} (2000) 139,
math.AG/9810173.}
\nref\GP{T. Graber and R. Pandharipande,
``Localization of Virtual Classes'', Invent. Math. {\bf 135}
(1999) 487, math.AG/9708001.}
\nref\GVi{R. Gopakumar and C. Vafa, ``Topological Gravity as Large
$N$ Topological Gauge Theory'', ATMP {\bf 2} (1998) 413,
hep-th/9802016.}
\nref\GVii{R. Gopakumar and C. Vafa, `` M-Theory and Topological
Strings -- I'', hep-th/9809187;  `` M-Theory and Topological Strings
-- II'', hep-th/9812127.}
\nref\GViii{R. Gopakumar and C. Vafa, ``On the Gauge Theory/Geometry
Correspondence'', ATMP {\bf 3} (1999) 1415, hep-th/9811131.}
\nref\GJT{S. Govindarajan, T. Jayaraman and T. Sarkar, ``Disc
Instantons in Linear Sigma Models'', hep-th/0108234.}
\nref\GZ{T. Graber and E. Zaslow, ``Open String Gromov-Witten
Invariants: Calculations and a Mirror 'Theorem' '', hep-th/0109075.}
\nref\GMS{B.R. Greene, D.R. Morrison and A. Strominger, 
``Black Hole Condensation and the Unification of String Vacua'', 
Nucl. Phys. {\bf B451} (1995) 109, hep-th/9504145.}
\nref\HKTY{S. Hosono, A. Klemm, S. Theisen and S.-T. Yau, ``Mirror
Symmetry, Mirror Map and Applications to Calabi-Yau Hypersurfaces'',
Commun. Math. Phys. {\bf 167} (1995) 301, hep-th/9308122.}
\nref\HLY{S. Hosono, B.H. Lian and S.-T. Yau, ``GKZ-Generalized
Hypergeometric Systems in Mirror Symmetry of Calabi-Yau
Hypersurfaces'', Commun. Math. Phys. {\bf 182} (1996) 535, hep-th/9511001.}
\nref\Iq{A. Iqbal, ``All Genus Topological String Amplitudes and 
5-Brane Webs as Feynman Diagrams'', hep-th/0207114.}
\nref\IK{A. Iqbal and A.-K. Kashani-Poor, 
``Instanton Counting and Chern-Simons Theory'', hep-th/0212279.}
\nref\KKLMi{S.
Kachru, S. Katz, A. Lawrence and J. McGreevy, ``Open String
Instantons and Superpotentials'', Phys. Rev. {\bf D62} (2000)
026001, hep-th/9912151.}
\nref\KKLMii{S. Kachru, S. Katz, A.
Lawrence and J. McGreevy, ``Mirror Symmetry for Open Strings'',
hep-th/0006047.}
\nref\KL{S. Katz and C.-C. M. Liu, ``Enumerative Geometry of Stable
Maps with Lagrangian Boundary Conditions and Multiple Covers of
the Disc'', ATMP {\bf 5} (2001) 1, math.AG/0103074.}
\nref\KSAS{S. Katz and S.A. Str\o mme, ``{\tt SCHUBERT}, a Maple
Package for Intersection Theory and Enumerative Geometry'', {\tt 
http://www.mi.uib.no/schubert.}} 
\nref\kkv{A. Klemm, S. Katz and C. Vafa, ``M-Theory,
Topological Strings and Spinning Black Holes'', ATMP {\bf 3}
(1999) 1445, hep-th/9910181.}
\nref\MK{M. Kontsevich, ``Enumeration of Rational Curves via Torus Actions'',
{\it The Moduli Space of Curves}, 335-368, Progr. Math. {\bf 129},
Birkh\"auser Boston, MA, 1995.}
\nref\KS{M. Kreuzer and H. Skarke, ``{\tt PALP}: A Package for Analyzing
Lattice Polytopes with Applications to Toric Geometry'', {\tt
http://hep.itp.tuwien.ac.at/$\,^\sim$kreuzer/CY.html}, math.SC/0204356.}
\nref\LMi{J.M.F. Labastida and M. Mari\~no, ``Polynomial Invariants for
Torus Knots and Topological Strings'', Commun. Math. Phys. {\bf 217} (2001)
423, hep-th/0004196.}
\nref\LMV{J.M.F. Labastida, M. Mari\~no and C. Vafa, ``Knots,
Links and Branes at Large $N$'', JHEP {\bf 11} (2000) 007, hep-th/0010102.}
\nref\LMii{J.M.F. Labastida and M. Mari\~no, ``A New Point of View in the
Theory of Knot and Link Invariants'', math.QA/0104180.}
\nref\LM{W. Lerche and P. Mayr, ``On ${\cal N}=1$ Mirror Symmetry
for Open Type II Strings'', hep-th/0111113.}
\nref\LMWi{W. Lerche, P. Mayr and N. Warner, 
``Holomorphic ${\cal N}=1$ Special Geometry of Open--Closed Type II Strings'', 
hep-th/0207259.}
\nref\LMWii{W. Lerche, P. Mayr and N. Warner, 
``${\cal N}=1$ Special Geometry, Mixed Hodge Variations and Toric Geometry'', 
hep-th/0208039.}
\nref\LT{J. Li~and G. Tian,
``Virtual Moduli Cycles and Gromov-Witten Invariants of Algebraic Varieties'',
J. Amer. Math. Soc. {\bf 11} (1998) 119.}
\nref\LS{J. Li and Y.S. Song, ``Open String Instantons and Relative Stable
Morphisms'', hep-th/0103100.}
\nref\YM{Y.I. Manin, {\it Frobenius Manifolds, Quantum Cohomology, and Moduli 
Spaces,} American Mathematical Society Colloquium Publications, Vol. 47, 
1999.} 
\nref\MV{M. Mari\~no and C. Vafa, ``Framed Knots at Large $N$'',
hep-th/0108064.}
\nref\Mi{P. Mayr, ``${\cal N}=1$ Mirror Symmetry and Open/Closed String
Duality'', hep-th/0108229.}
\nref\Mii{P. Mayr, ``Summing up Open String Instantons and ${\cal
N}=1$ String Amplitudes'', hep-th/0203237.}
\nref\DRM{D.R. Morrison, ``Through the Looking Glass'', 
{\it Mirror Symmetry III}, D.H. Phong, L. Vinet, and S.-T. Yau, eds., 
International Press, 1999, 263.}
\nref\OP{T. Oda and H.S. Park, ``Linear Gale Transforms and
Gel'fand-Kapranov-Zelevinskij Decompositions'', Tohoku Math. J. {\bf
43} (1991) 375.}
\nref\OV{H. Ooguri and C. Vafa, ``Knot Invariants and Topological
Strings'', Nucl. Phys. {\bf B 577} (2000) 419, hep-th/9912123.}
\nref\OVii{H. Ooguri and C. Vafa,
``Worldsheet Derivation of a Large $N$ Duality'', hep-th/0205297.}
\nref\RS{P. Ramadevi and T. Sarkar,
``On Link Invariants and Topological String Amplitudes'',
Nucl. Phys. {\bf B600} (2001) 487, hep-th/0009188.}
\nref\Str{A. Strominger, 
``Black Hole Condensation and Duality in String Theory'', 
Nucl. Phys. Proc. Suppl. {\bf 46} (1996) 204, hep-th/9510207.}
\nref\CVi{C. Vafa, ``Extending Mirror Conjecture
to Calabi-Yau with Bundles'', hep-th/9804131.}
\nref\sts{C. Vafa, ``Superstrings and Topological
Strings at Large $N$'', J. Math. Phys. {\bf 42} (2001) 2798,
hep-th/0008142.}
\nref\EWii{E.
Witten, ``Chern-Simons Gauge Theory as a String Theory'',
``The Floer Memorial Volume'', H. Hofer et al, eds, Birkh\"auser 1995, 637,
hep-th/9207094.}
\Title{
\vbox{
\baselineskip12pt
\hbox{hep-th/0302076}
\hbox{RUNHETC-2003-05}}}
{\vbox{\vskip 39pt
\vbox{\centerline{Large $N$ Duality for Compact Calabi-Yau Threefolds}}
}}
\vskip 15pt
\centerline{Duiliu-Emanuel Diaconescu and Bogdan Florea} 
\bigskip
\medskip
\centerline{{\it Department of Physics and Astronomy,
Rutgers University,}}
\centerline{\it Piscataway, NJ 08855-0849, USA}
\bigskip
\bigskip
\bigskip
\bigskip
\smallskip
\noindent 
We study geometric transitions for topological strings on compact Calabi-Yau 
hypersurfaces in toric varieties. Large $N$ duality
predicts an equivalence between topological open and closed string
theories connected by an extremal transition.
We develop new open string enumerative techniques and perform 
a high precision genus zero 
test of this conjecture for a certain class of toric extremal transitions. 
Our approach is based on $a)$ 
an open string version of Gromov-Witten theory with convex 
obstruction bundle and $b)$ an extension of Chern-Simons theory 
treating the framing as a formal variable.  

\vfill
\Date{February 2003}

\newsec{Introduction}

Large $N$ duality \refs{\GVi,\GViii,\OVii} predicts a highly nontrivial 
relation between open and closed topological strings on 
Calabi-Yau threefolds connected by an extremal transition. 
Originally formulated for local conifold transitions, the duality 
has been extended in various directions in 
\refs{\AKV-\AMV, \DFGi-\DVii,\Iq,\IK,\LMi-\LMii,\MV,
\OV,\RS}. 
The generalizations considered in 
\refs{\AViv,\AMV,\DFGi,\DFGii} are especially 
interesting since they shed a new light on the structure of Gromov-Witten 
invariants of toric Calabi-Yau manifolds. These developments are based 
on an earlier idea of Witten \EWii,\ who outlines a beautiful approach 
to {\bf A}-model topological open string amplitudes. 
The main idea of this approach is based on a subtle 
combination of Chern-Simons theory and open string enumerative geometry, 
which will be discussed in detail below. 

At the present stage open string enumerative geometry is not a fully 
developed mathematical formalism. The basic principles understood so far 
have been developed in \refs{\GZ,\KL,\LS} in the context of 
toric Calabi-Yau threefolds. Additional work on the subject exhibiting 
various points of view can be found in 
\refs{\AVi,\AKV,\DFGi,\DFGii,\GJT,\KKLMi,
\KKLMii,\LM-\LMWii,\Mi,\Mii}. In particular, the work of 
\refs{\DFGi,\DFGii} was successfully applied to geometric transitions for 
noncompact Calabi-Yau hypersurfaces in toric varieties. 

In this paper we consider large $N$ duality 
for extremal transitions between compact Calabi-Yau hypersurfaces 
in toric varieties. Geometric transitions for compact Calabi-Yau 
manifolds have been proposed in \sts,\ but no concrete results 
are known up to date. There are many conceptual as well as technical 
issues that have to be addressed in this context. Perhaps the most 
challenging obstacle is the absence of a rigorous mathematical 
formalism for open string amplitudes on compact Calabi-Yau threefolds. 

Keeping the details to a minimum, let us summarize our results. 
We develop a computational approach to genus zero 
open string {\bf A}-model 
amplitudes on a certain class of 
compact Calabi-Yau target manifolds. This approach 
follows the basic principles outlined in \EWii\ supplemented by a 
heavy use of equivariant enumerative techniques. 
There are two key aspects in the whole process. First, we 
develop an open string version of the 
convex obstruction bundle approach usually encountered in 
Gromov-Witten theory. Second, we propose a formal extension of 
standard Chern-Simons theory in which the framing is regarded 
as a formal variable. In particular it could take fractional values. 
Using these two elements, plus a great deal of patience, we 
run a successful genus zero test of large $N$ duality for compact 
Calabi-Yau hypersurfaces in toric varieties. Along the way, 
we gain new insights into the structure of Gromov-Witten 
invariants, and we also clarify some aspects of local 
geometric transitions \refs{\AMV,\DFGi,\DFGii}. 
A more detailed overview of the general set-up and the 
results is included in the next section. 

The paper is structured as follows. In section two we present the stage, 
the cast of characters, and outline the main plot. Section three consists 
of a conceptual discussion of open string enumerative geometry for 
compact Calabi-Yau manifolds. In section four we describe the main 
examples to be used for concrete computations. We revisit local 
geometric transitions in section five, focusing on the interplay between 
equivariant enumerative geometry and Chern-Simons theory. The main 
outcome of this discussion is a notion of formal Chern-Simons 
expansion which will prove crucial for compact threefolds. 
Sections six and seven are devoted to an implementation 
of our program in a concrete model. 

{\it Acknowledgments.} This work is a spin-off of a broader research program 
initiated by Ron Donagi, Antonella Grassi, Tony Pantev and us. It is a 
pleasure to thank them for collaboration and sharing their ideas with 
us during the completion of this project. D.-E.D. would like to thank 
Mina Aganagic, Marcos Mari\~no and Cumrun Vafa for collaboration 
on a related project and valuable discussions. B.F. would like to thank 
Philip Candelas and Harald Skarke for helpful conversations. The work of D.-E.D. 
and B.F. has been partially supported by DOE grant DOE-DE-FG02-96ER40959.

\newsec{Large $N$ Duality for Compact Threefolds -- Preliminary Remarks}

Although the problem can be formulated in a more general context, 
in this paper we will consider only toric extremal transitions 
between Calabi-Yau hypersurfaces \refs{\ACJM,\BKK,\DRM}. 
Recall \Ba\ that a family of Calabi-Yau hypersurfaces $Y$ in a toric variety $\CZ$ 
is described by a reflexive Newton polyhedron $\IP_{\Delta_{Y}}$. 
The class of extremal transitions we are 
interested in is described by an embedding of reflexive 
polyhedra $\IP_{\Delta_\wY}\subset \IP_{\Delta_{Y}}$. It has been 
shown in \refs{\ACJM,\BKK,\DRM} that such an embedding gives rise 
to a commutative diagram of the form 
\eqn\extransA{
\xymatrix{ \wY\ar[r]\ar[d] & \wcz\ar[d] \\
Y_0 \ar[r] & \CZ.\\}}
The vertical arrows are extremal contractions, and 
$Y_0$ is a singular hypersurface in the family $Y$ corresponding 
to special values of complex structure parameters. 
The toric extremal contraction 
$\wcz\ra \CZ$ admits a simple toric description in terms 
of a similar embedding of dual reflexive polyhedra $\IP_{\nabla_{Y}}
\subset \IP_{\nabla_\wY}$. The vertices of $\IP_{\nabla_\wY}$ not 
belonging to $\IP_{\nabla_{Y}}$ correspond to toric divisors 
on $\wcz$ which are contracted in the process. 
In the following we will only consider singular hypersurfaces  
$Y_0$ with isolated ordinary double points. Moreover, the 
contraction $\wY\ra Y$ is a simultaneous crepant resolution 
of the nodes. Also, we will assume $\CZ, \wcz$ to be smooth 
toric fourfolds. This can always be achieved by triangulating the toric 
fans. 

To be more concrete, let us consider a one parameter family 
$Y_\mu$ of hypersurfaces in $\CZ$ which degenerates to $Y_0$ 
for $\mu=0$. We denote by $v$ the number of nodal singularities 
of $Y_0$ and by 
$[L_1],\ldots, [L_v]$ the corresponding homology classes of 
vanishing cycles on $Y_\mu$.
The $[L_i]$, $i=1,\ldots, v$ generate a 
rank $(v-r)$
sublattice $H_v\subset H_3(Y_\mu, \IZ)$, where $r$ is the number 
of relations among vanishing cycles. 
Without loss of generality, we can assume that $[L_{r+1}], \ldots, [L_v]$
is an integral system of generators for $H_v$. 
Therefore we have the following 
linear relations with integral coefficients
\eqn\vanrelA{
[L_i] = \sum_{m=r+1}^{v} a_{i,m}[L_m]}
for $i=1,\ldots, r$. 

The exceptional locus of the crepant resolution $\wY\ra Y$ consists of 
$v$ isolated $(-1,-1)$ curves $C_1,\ldots,C_v$ on $\wY$. 
One can show \CK\ that the curve classes $[C_1],\ldots,[C_v]\in H_2(\wY,\IZ)$ 
are subject to $(v-r)$ relations, so that we are left with $r$ independent 
curve classes on $\wY$. 
Without loss of generality we can take them to be  
$[C_1],\ldots, [C_r]$. The numbers $(v,r)$ can be related to the toric data 
described in the previous paragraph using Batyrev's formulae \Ba.\ 
We will work under the simplifying assumption that all extra $r$ 
$(1,1)$ classes on ${\tilde Y}$ are toric, that is $r=h^{1,1}(\wcz)-h^{1,1}(\CZ)$.
Therefore the curves $C_1,\ldots,C_r$ determine $r$ independent curve 
classes on $\wcz$ as well. We will understand why this 
assumption is necessary by the end of this section.
 
In this geometric context, we consider an open string ${\bf A}$ 
model specified by wrapping $N_i$ topological branes on a collection 
of vanishing cycles 
$L_i\in [L_i]$ \EWii.\ 
In order to obtain a well defined topological open string 
${\bf A}$ model, the cycles $L_i$ should be lagrangian. 
Physical considerations 
\refs{\GMS,\Str} suggest that each class $[L_i]$ should contain a unique 
special lagrangian cycle homeomorphic to $S^3$ 
in order for the transition to make sense in string theory.
Unfortunately, this does not seem to be a well established mathematical
result for a generic deformation $Y_\mu$. 
For the remaining part of this section, we will assume this to be true. 
In later sections we will work with a special degeneration of $Y_\mu$ 
where one can explicitly construct such lagrangian representatives. 

In a topological theory, there should not be any constraint 
on the $N_i$ arising from flux conservation. 
If we regard our model as the topological sector of an 
underlying superstring theory, the total D-brane charge has to be zero 
because ${\widetilde Y}$ is compact. 
Therefore we must have 
\eqn\chargerelA{
\sum_{i=1}^vN_i [L_i] =0.} 
Although this is not strictly a necessary condition in the topological 
theory, it seems to be required in order to match the number of 
parameters in the context of large $N$ duality. This can be seen by 
substituting \vanrelA\ in \chargerelA\ and using the fact that 
$[L_{r+1}], \ldots,[L_{v}]$ form a basis. We obtain 
\eqn\chargerelB{
N_m+\sum_{i=1}^r N_i a_{i,m}=0,\quad m\geq r+1.}
This shows that there are only $r$ independent D-brane 
charges $N_1, \ldots, N_r$
in one to one correspondence with the exceptional curves 
$C_1,\ldots,C_r$. 

Since topological {\bf A}-model amplitudes depend on K\"ahler 
moduli, we choose specific parameterizations of the K\"ahler 
cones of $\CZ, \wcz$ as follows
\eqn\kahlerA{
J = \sum_{\alpha =1}^{h^{1,1}(\CZ)} t_\alpha J_\alpha,\qquad  
\wJ = \sum_{\gamma=1}^{h^{1,1}(\wcz)} \wt_\gamma \wJ_\gamma.}
Here $J_\alpha$, $\alpha=1, \ldots, h^{1,1}(\CZ)$ and respectively 
$\wJ_\gamma$, $\gamma=1,\ldots,h^{1,1}(\wcz)$ are K\"ahler 
cone generators. Note that $t_\alpha, \wt_\gamma$ are classical 
K\"ahler parameters. The topological amplitudes should be written 
in terms of flat K\"ahler parameters, which include instanton corrections. 
Abusing notation, we will use the same symbols for classical 
and flat K\"ahler parameters. The distinction should be clear from the 
context. 

After all these preliminary remarks, we are ready to discuss the 
large $N$ duality conjecture in the present context. Following the 
general philosophy of \GVii,\ large $N$ duality should predict a 
relation between topological closed strings on $\wY$ and topological 
open strings on $(Y,L)$, where $L=\cup_{i=1}^v L_i$. 
Therefore it would be tempting to conjecture 
a direct relation between the Gromov-Witten expansion of $\wY$ 
and a hypothetical open string expansion associated to
$(Y,L)$. While such a relation may exist abstractly, at the present 
stage such a direct approach is out of reach. Instead we will 
concentrate only on the genus zero part of the duality and 
follow one of the main lessons of mirror 
symmetry for Calabi-Yau hypersurfaces. 
In the linear sigma model approach, the genus zero Gromov-Witten 
expansion of a hypersurface $\wY$ is defined extrinsically 
in terms of maps to the ambient toric variety $\wcz$, 
using the convex obstruction bundle approach. 
Roughly, this means that one counts holomorphic maps 
to $\wcz$, subject to an algebraic constraint.
This produces a genus zero closed string expansion of the form 
\eqn\GWA{
\CF_{\wy;cl}^{(0)}(g_s,\wt_\gamma)=g_s^{-2} 
\sum_{\wbeta\in H_2(\wcz,\IZ)}\wC_{0,\wbeta} e^{-\langle \wJ,\wbeta\rangle}.}
where the coefficients $\wC_{0,\wbeta}$ have a standard definition in terms 
of intersection theory on the moduli space of stable maps to $\wcz$ which 
will be reviewed in the next section. It is worth noting here 
that only the data of $\wcz$ and the linear system $|\wY|$ enters 
the definition of $\wC_{0,\wbeta}$. Therefore these coefficients are 
manifestly independent of complex structure deformations of $Y$. 

Following the same idea one would like to define a similar instanton sum 
for topological open strings on $(Y,L)$ using maps to the ambient 
space $\CZ$ with boundary conditions on $L$. 
This is a very delicate construction for many reasons 
explained in detail in section three. One of the obvious problems 
is that such a construction would not be manifestly independent on 
complex structure deformations of $Y$, since the cycles $L$ 
move with $Y$. Moreover, for fixed $(Y,L)$, there is no rigorous 
mathematical formalism for counting open string maps 
in an appropriate sense. Finally, in open string theories one 
has to sum over highly degenerate 
maps which contract a bordered Riemann surface into an infinitely thin 
ribbon graph \EWii.\ 
From a physical point of view this yields a Chern-Simons sector of 
the topological string theory consisting of $v$ Chern-Simons theories 
with gauge groups $U(N_i)$ supported 
on the cycles $L_i$, $i=1,\ldots,v$ \EWii.\ 
The coupling of this sector to open string instantons is a very subtle issue 
not fully understood at the present stage. This subject will be discussed 
in detail in section five for local models and section seven for 
compact Calabi-Yau threefolds. 

We will not attempt to fill all the gaps mentioned in the 
previous paragraph in the present paper. 
Instead we will assume invariance under complex structure deformations, 
and construct a genus zero open string generating functional using a special 
degeneration $({\overline Y}, {\overline L})$ 
of the pair $(Y,L)$. The construction relies heavily 
on the existence of a toric action on $\CZ$ which induces a subtorus 
action on ${\overline Y}$. The existence of a suitable degeneration places 
important restrictions on the range of validity of this 
approach. As explained in sections three and six, such a direct construction 
can be carried out only for extremal transitions in which the nodal points 
are fixed points of the torus action on $\CZ$. Without giving more details 
here, let us note that in favorable cases we can construct a genus zero open 
string generating functional of the form\foot{Here we ignore some subtleties
related to open string quantum corrections to the flat K\"ahler parameters.
Since this is a general discussion, these corrections can be absorbed in 
the definition of $F_{\beta}^{(0)}(g_s,\lambda_i)$.} 
\eqn\openA{
\CF_{(Y,L);op}^{(0)}(g_s,t_\alpha,\lambda_{i}) = \CF_{Y;cl}^{(0)}(g_s,t_\alpha)
+\sum_{\beta\in H_2(\CZ,\oL;\IZ)} 
F_{\beta}^{(0)}(g_s,\lambda_{i}) e^{-<J,\beta>}.}
Here $\lambda_i=g_sN_i$ are the 't Hooft coupling constants 
for the $v$ Chern-Simons theories on $L_i$. 
Given the relations 
\chargerelB,\  we have only $r$ independent 't Hooft couplings 
$\lambda_{i}$, $i=1,\ldots, r$ in formula \openA.\ 
This matches the number of exceptional curve classes on $\wz$, according to 
the paragraph below equation \vanrelA.\ 
In the right hand side of \openA,\ $\CF_{Y;cl}(g_s,t_\alpha)^{(0)}$ is the 
genus zero closed string Gromov-Witten expansion of $Y$
\eqn\GWB{
\CF_{Y;cl}^{(0)}(g_s,t_\alpha)=g_s^{-2}
\sum_{\beta\in H_2(\CZ,\IZ)} C_{0,\beta} e^{-\langle J, \beta\rangle}.}
The construction of the generating functional in \openA\
is discussed from a conceptual 
point of view in section three, and carried out for a concrete compact example 
in sections six and seven. 

Granting the existence of a well defined open string partition function, 
what are the predictions of large $N$ duality? The conjecture is that 
the two generating functionals \GWA\ and \openA\ should be equal, 
subject to a certain identification of parameters which is the duality map. 
To this end, let us introduce a different system of generators 
$(\pi^*(J_\alpha), D_i)$, $\alpha=1,\ldots, h^{1,1}(\CZ)$, 
$i=1,\ldots, r$ of $H^{1,1}(\wcz)$. Here $\pi:\wcz\ra \CZ$ is the contraction 
map and $D_i$, $i=1,\ldots,r$ are divisor classes on $\wcz$ such that 
$\pi_*(D_i)=0$ and $D_i\cdot C_j=\delta_{ij}$ for $1\leq i,j\leq r$. 
Then, modulo some subtleties concerning open string K\"ahler moduli 
which will be considered later, the duality map is specified by the following 
relations 
\eqn\dualitymapA{
\wJ = \pi^*J-i \sum_{i=1}^r \l_i D_i.} 
Using this map, we will test the duality predictions for a concrete 
example in section seven. 

\newsec{Open String {\bf A}-Model Approach -- General Considerations}

In principle, large $N$ duality should hold for any conifold transition 
between Calabi-Yau threefolds. However, our current understanding of 
topological open string ${\bf A}$-model 
amplitudes is restricted to the so called local models, i.e. noncompact 
Calabi-Yau threefolds admitting a torus action. 
The case of compact Calabi-Yau threefolds is 
especially hard, no concrete results being known up to date. 
In this paper, we would like to propose an approach to this 
problem for a certain class of hypersurfaces in toric varieties. 
This section is a rather general exposition of the basic principles. 
The details 
will be worked out for concrete examples in the next sections. 

Let us start the discussion with some general considerations on 
open string {\bf A}-models, following \EWii.\ Suppose we 
have a model defined by $v$ lagrangian cycles $L_1,\ldots, L_v$ 
as in the previous 
section. The target space effective action of such a theory 
consists of $v$ Chern-Simons theories with gauge groups 
$U(N_i)$ $i=1,\ldots,v$ supported on the $v$ lagrangian cycles. 
This is a universal sector of the theory which is present for any 
target manifold. Then one has instanton corrections to 
the Chern-Simons theory obtained by summing over holomorphic maps 
$f:\Sigma_{g,h}\ra Y$ with lagrangian boundary conditions on the $L_i$.
Obviously, these corrections depend on the geometry of the target 
manifold, and the sum should be properly formulated in terms of 
intersection theory on a moduli space of stable maps 
$\om_{g,h}(Y,L)$. There is however an important subtlety in this 
approach which may be easier to understand for a concrete 
example. Suppose one can find a rigid holomorphic disc $D_i$ 
embedded in $Y$ with boundary $\Gamma_i\subset L_i$. Let $\beta 
=[D_i]\in H_2(Y,L_i;\IZ)$ and let $\gamma =[\Gamma_i] 
\in H_1(L_i,\IZ)$. 
This yields an instanton correction of the form 
\eqn\instcorrA{
e^{-\tau_i} \Tr\left( \hbox{Pexp} \int_{\Gamma_i} f^*A_i\right)}
to the $U(N_i)$ Chern-Simons theory on $L_i$. Here $\tau_i$ is 
the symplectic area of the disc and $A_i$ is the $U(N_i)$ gauge field 
on $L_i$. It is important to note that $A_i$ is not a fixed flat connection 
on $L_i$. Since the cycle is compact, $A_i$ is a dynamical variable, 
and one should integrate over all connections. 
However, this proposal immediately raises a puzzle if the disc $D_i$ 
moves in a family on $Y$. Since $A_i$ is not flat, the holonomy 
factor $\Tr \left(\hbox{Pexp} \int_{\Gamma_i} f^*A_i\right)$ 
depends on the position 
of $\Gamma$ in $L_i$. Therefore it is not clear how to write down 
the instanton corrections in this case.
Since {\bf A}-models make sense on general symplectic manifolds 
equipped with a compatible almost complex structure, one solution is 
to deform to a generic situation in which we have a finite number of 
rigid discs. Then each disc would yield a correction of the form 
\instcorrA.\ Conceptually, this is a satisfying solution, but 
it is not very effective in practice if one is interested in 
explicit numerical computations. 
The solution proposed in \refs{\DFGi,\DFGii} is based on localization with 
respect to a toric action. More precisely, if $Y$ admits a torus action which 
preserves the lagrangian cycles, the idea is to sum only over fixed 
points of the induced action on the moduli space of maps. 
To each fixed point, one can associate an instanton expansion 
consisting of terms of the form \instcorrA\ and multicovers. 
In order to obtain the topological free energy, one has to first 
sum the instanton expansions  
over all fixed points and then perform the Chern-Simons 
functional integral with all these corrections taken into account. 
Such computations have been performed for noncompact manifolds 
in \refs{\AMV,\DFGi,\DFGii}. 

At this point it may be helpful to emphasize the distinction between 
compact and noncompact cycles. If $L_i$ were a noncompact cycle 
in some noncompact Calabi-Yau manifold, one could take $A_i$ to 
be a fixed flat connection on $L_i$. For a flat 
connection the holonomy factor 
$\Tr \left(\hbox{Pexp} \int_{\Gamma_i} f^*A_i\right)$ 
depends only on the homology class
$\gamma$. 
Therefore for fixed homology classes $(\beta,\gamma)$ the instanton 
series can be thought of a formal series in two sets of variables 
\KL.\ The computation of topological amplitudes is then reduced 
to the construction of a virtual cycle of dimension zero on the 
moduli space of stable open string maps. The structure of this 
moduli space is not very well understood at the moment, hence 
there are various technical problems with such a construction. 
Nevertheless one can go a long way exploiting the presence of a 
torus action \refs{\GZ,\KL,\LS,\Mii}.
Even if the moduli space is very complicated, the fixed loci of the 
induced torus action are much simpler and can be described in 
detail. Then one can define the virtual cycle by directly summing over 
invariant maps. Since the moduli space has boundary, the resulting invariants 
will depend on a choice of boundary conditions, which can be encoded 
in a choice of the torus weights \KL.\ Therefore we obtain a family of 
invariants depending on a discrete set of choices. 

If the cycles $L_i$ are compact,
the local contribution of a fixed locus is to be interpreted as 
a series of corrections to Chern-Simons theory, as explained above. 
As a result, each term in this series will be a rational function
of the weights of the torus action. 
The final expression for the open string free energy 
is obtained by computing the free energy of the corrected 
Chern-Simons action. 
At this point we seem to obtain a puzzle 
since a priori the resulting open string amplitudes 
will depend on the weights of the 
torus action. This dependence is not physically acceptable in the context 
of large $N$ duality since 
there are no discrete ambiguities in the dual closed string expansion. 

The resolution of this puzzle is that the choice of the toric weights has to 
be correlated to the framing of the knots $\Gamma_i$ in Chern-Simons 
theory in such a way that the final result agrees with the closed 
string dual. Although this idea has been concretely applied in certain 
local models in \refs{\DFGi,\DFGii} the general rules behind 
the weights-framing correspondence are not well understood at the present 
stage. In particular, integrality of the framing imposes very strong 
constraints on the allowed weights, which may not even have solution in 
many cases. One of the main results of this work 
is a general rule for this correspondence based on a certain extension 
of the Chern-Simons expansion treating the framing as a formal variable.
This will be discussed at length for local models in section five, 
and for compact Calabi-Yau threefolds in section seven. 

For the remaining part of this section, we will focus on localization 
questions for open string maps to compact Calabi-Yau hypersurfaces. 
Our goal is to develop an algorithm for the computation of the 
local contributions associated with fixed loci in such situations. 

\subsec{Open String Morphisms -- The Convex Obstruction Bundle Approach} 

Open string {\bf A}-model localization techniques 
have been first developed in \refs{\GZ,\KL,\LS} for noncompact lagrangian 
cycles in noncompact toric threefolds. An extension of these techniques 
to compact lagrangian cycles has been discussed in \refs{\DFGi,\DFGii}. 
In the later cases, the ambient Calabi-Yau threefold was a 
noncompact hypersurface in a toric variety. The invariants computed there 
may be regarded as an open string version of closed string local 
Gromov-Witten invariants \refs{\CKYZ}. 

Briefly, the typical situation
encountered in closed string computations is the following. One has a 
toric Calabi-Yau variety $X$ isomorphic to the total space of a holomorphic 
vector bundle $N$ over a base $B$. Usually $B$ is a rational 
curve or a toric Fano surface, so that the zero section of the fibration 
$N\ra B$ is rigid in $X$. The basic data of the enumerative problem is 
given by the negative obstruction bundle $\CE= R^1\rho_*(ev^*N)$ on the 
moduli space of stable maps $\om_{g,0}(B,\beta)$, with fixed 
class $\beta \in H_2(B,\IZ)$. Here $ev:\om_{g,1}(B,\beta)\ra B$ is the 
evaluation map, and $\rho:\om_{g,1}(B,\beta)\ra \om_{g,0}(B,\beta)$ 
is the forgetful map. This is standard material. 
The local Gromov-Witten invariants are defined by integrating the 
Euler class of $\CE$ against the virtual fundamental class 
$[\om_{g,0}(B,\beta)]$. Alternatively, we can define the same invariants 
as the degree of the virtual cycle $[\om_{g,0}(X,\beta)]$, which is 
of dimension zero. One can show without too much effort that the two
definitions are equivalent (see for example section 9.2.2. of \CK).\
Since $B$ is toric, we can in fact explicitly evaluate these numbers 
using localization techniques \refs{\GP,\MK}. 

By contrast, compact Calabi-Yau threefolds, such as hypersurfaces 
in toric varieties, require a different approach. 
The obvious difficulty in applying localization techniques to this 
case is the absence of a 
torus action on a smooth generic hypersurface. For genus zero maps, 
this problem is 
elegantly solved by the convex obstruction bundle approach. 
Shifting perspective, the genus zero 
numerical invariants are formulated in terms of 
the moduli space   
$\om_{0,0}(\CZ,\beta)$ of stable maps to the ambient toric variety 
$\CZ$, which admits a toric action. 
Here $\beta\in H_2(\CZ,\IZ)$ is a fixed second homology class. 
On this moduli space one can construct a virtual fundamental class 
of positive degree. Therefore, in order to obtain numerical invariants, 
one has to intersect this class with the Euler class of the 
obstruction bundle $\CV= R^0\rho_*(ev^*\CO(-K_\CZ))$. 
Here $\rho:\om_{0,1}(\CZ,\beta)\ra \om_{0,0}(\CZ,\beta)$ is the forgetful 
map and $ev:\om_{0,1}(\CZ,\beta) \ra \CZ$ is the evaluation map as usual. 
Note that the 
line bundle $\CO(-K_\CZ)$ defines the linear system of Calabi-Yau 
hypersurfaces on $\CZ$. Therefore, the zeroes of a generic section of 
$\CV$ on $\om_{0,0}(\CZ,\beta)$ can be thought of intuitively 
as maps to a generic hypersurface $Y$. In fact one can show rigorously 
that these intersection numbers coincide with the degree of the 
virtual cycle $[\om_{0,0}(Y,\beta)]$ (see Example 7.1.5.1 in \CK).\ 
Since only the data of the linear system $|-K_\CZ|$ enters this definition,
the invariants are manifestly independent of complex structure 
deformations of $Y$. This approach is valid if the bundle $\CO(-K_\CZ)$ 
is convex i.e. $ R^1\rho_*(ev^*\CO(-K_\CZ))=0$. By a slight abuse of language, 
we will call $\CV$ a convex obstruction bundle in this case. 

Following the same general line of thought, we would like to 
propose an extension of the convex obstruction bundle approach to 
open string ${\bf A}$-model amplitudes on compact Calabi-Yau hypersurfaces. 
It should be said at the offset that the methods employed here 
are not entirely rigorous, since a rigorous mathematical formulation of 
open string Gromov-Witten invariants has not been developed
up to date. We make 
no claim of filling this gap. Moreover, since our approach  relies 
heavily on localization with respect to a toric action, it applies only 
to a special class of 
hypersurfaces and lagrangian cycles 
which will be described later. The models presented in the next
section provide a good testing ground for explicit computations which can be 
checked against large $N$ duality predictions.

The main idea is quite straightforward. Given a Calabi-Yau hypersurface
$Y\subset \CZ$ and a collection of vanishing cycles 
$L=\cup_{i}L_i$, we would like 
to consider the open string morphisms to $Z$ with certain boundary conditions 
along $L_i$, viewed as cycles in $Z$. Obviously, the $L_i$ cannot be 
lagrangian cycles in $Z$ for dimensional reasons, but this will not be a 
major obstacle. 
More precisely, one would like to consider the moduli space of stable maps 
$\om_{0,h}(\CZ, L;\beta)$ where $\beta$ denotes now a relative homology 
class $\beta\in H_2(\CZ, L;\IZ)$. Note that 
$H_2(\CZ,L;\IZ) \simeq H_2(\CZ,\IZ)$ since $L$ is a collection of 3-spheres, 
hence we can identify $\beta$ with an absolute class. 
In order to localize the open string morphisms following \refs{\GZ,\KL,\LS}, 
we need a torus action on $\om_{0,h}(\CZ, L;\beta)$ compatible with the
torus action on $\CZ$. The problem here is that the torus action on $\CZ$ does 
not preserve the vanishing cycles $L_i$ lying on a generic hypersurface $Y$. 
This is a new element compared to the closed string situation. 

In the following we would like to propose 
a solution to this problem based on the assumption that 
toplogical ${\bf A}$-model amplitudes are 
invariant under complex structure deformations.
Working under this assumption, 
it suffices to define and compute these invariants 
by specializing $(Y,L)$ to a singular hypersurface ${\overline Y}$ 
and a collection of cycles ${\overline L}$ preserved by a subtorus. 
The main problem with such an approach is that the existence of such a 
'good degeneration' of $(Y,L)$ is not by any means obvious. 
This is in fact the main restriction on 
the class of transitions which can be treated in this fashion.  
We have not been able to develop a systematic characterization 
of 'good degenerations' in arbitrary families of Calabi-Yau hypersurfaces. 
Analyzing the concrete 
examples worked out in the next sections, it seems that such a degeneration 
will exist whenever the nodal points of $Y_0$ are fixed points 
of the torus action on the ambient toric variety $\CZ$. Clearly, this point 
deserves further study, but this is the criterion that seems to emerge so far. 

Provided one can find such a degeneration ${\overline Y}$, the next step is to 
evaluate the contributions of the fixed loci in $\om_{0,h}(\CZ, \oL;\beta)$. 
This can be done by constructing the tangent-obstruction complex 
for open string morphisms to the pair $(\CZ,\oL)$. Although the target space 
is a toric fourfold, the main idea is the same as \refs{\GZ,\KL}, i.e. 
one has to consider the deformation complex of an open string map 
$f:\Sigma_{0,h}\ra \CZ$ 
subject to boundary conditions along $\oL$. 
The boundary conditions can be inferred from the concrete presentation 
of $\oL$, as we will discuss in detail in concrete examples.
In particular, in all these examples we will find a  
real subbundle $T_\IR$ of $T_{\CZ}$ restricted to $\oL$ so that 
the pair $(f^*T_\CZ, f_\partial^* T_\IR)$ forms a Riemann-Hilbert 
bundle on $\Sigma_{0,h}$. 

We need one more ingredient, which is the 
convex obstruction bundle on the moduli space.  
One would like to define the fiber 
of the obstruction bundle $\CV$ over a point $(\Sigma_{0,h}, f)$, 
as the space of holomorphic sections of $f^*\CO(-K_\CZ)$ subject 
to certain boundary conditions along $\partial \Sigma_{0,h}$. 
In order to obtain a good boundary problem for the Dolbeault 
operator, the boundary conditions 
should be specified by a real sub-bundle $\CR$ of 
$f^*\CO(-K_\CZ)|_{\partial \Sigma_{0,h}}$ \KL.\ 
In fact the pair 
$(f^*\CO(-K_\CZ), \CR)$ should define a Riemann-Hilbert bundle on 
$\Sigma_{0,h}$. Such a structure can be naturally obtained in the 
present case if we use the isomorphism $\CO(-K_\CZ)\simeq \Lambda^4(T_\CZ)$ 
where $T_\CZ$ is the holomorphic tangent bundle to $\CZ$. 
Then we can take $\CR$ to be $f_\partial^*\left(\Lambda^4 (T_\IR)\right)$. 
It is not a priori obvious that this construction is sensible, since 
the structure of the moduli space is not understood. 
However, we will show in the next sections that the results are in 
perfect agreement with large $N$ duality predictions. This is strong 
evidence in favor of this approach. 

As mentioned several times so far, evaluating the contributions 
of the fixed points is only a first step in the computation 
of open string amplitudes. In order to finish the computation, 
the contribution of each fixed point has to be interpreted as a 
series of instanton corrections in the Chern-Simons theory.
This involves another very subtle aspect of the whole approach, namely the 
correlation between the toric weights and Chern-Simons framing. 

Given the complexity of the problem, we will proceed in several stages. 
In the next section, we describe in detail the models 
considered in this paper. Although we will find explicit 
extremal transitions between compact Calabi-Yau manifolds, 
the starting point will be local del Pezzo models. 
Section five will be devoted to the weight-framing correspondence 
for local models, focusing on the idea of a formal Chern-Simons 
expansion. Finally, equipped with a better understanding of these 
points, we will address large $N$ duality for compact models 
in sections six and seven.

\newsec{The Models} 

Concrete examples in which the above program can be 
implemented\foot{These are not the only examples which can be treated this 
way. We will eventually learn along the way that all extremal transitions 
in which the nodal points are fixed by the torus action admit a similar 
approach.} 
arise by taking projective completions of the 
local Calabi-Yau models considered in 
\refs{\AMV,\DFGi,\DFGii}. Let us briefly recall 
the relevant geometric constructions. 

We will consider extremal transitions for noncompact toric 
Calabi-Yau threefolds fibered in complex lines over toric del Pezzo 
surfaces. The base $B$ of the fibration can be either $dP_2$ as in the model 
studied in \DFGii\ or $dP_3$ or $dP_5$ as in \AMV.\ In all cases 
it is easier to describe the small resolution $\wX$ first, which is 
isomorphic to the total space of the canonical bundle $\CO(K_B)$. The toric 
diagrams for the del Pezzo surfaces $dP_2, dP_3, dP_5$ are represented below. 

\ifig\delpezzos{del Pezzo surfaces.}{\epsfxsize5in\epsfbox{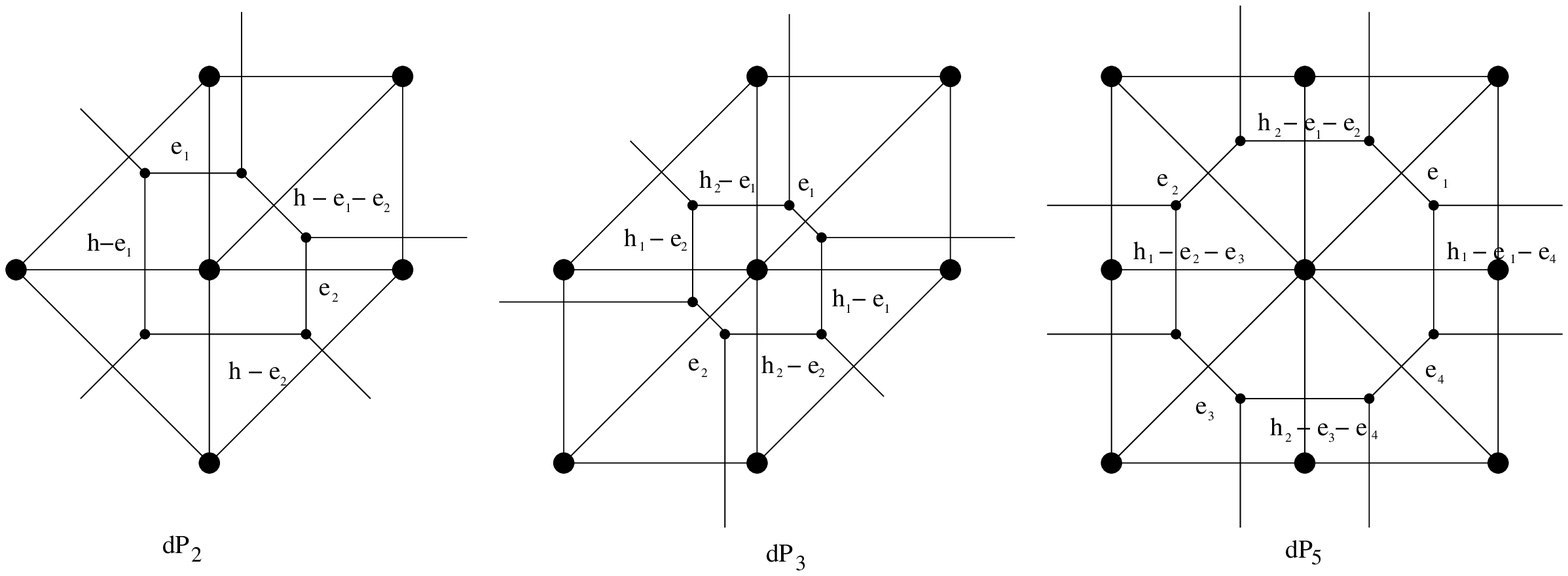}}

In each case, the toric fan of the threefold $\wX$ is a cone over the 
two-dimensional polytope represented in \delpezzos . Note that $dP_2$ is represented 
here as a two-point blow-up of $\IP^2$, and similarly $dP_3,dP_5$ are 
constructed by blowing up two and respectively four points on $\IF_0$. 
In the last case, the four points on $\IF_0$ are not in general position, 
therefore we obtain a non-generic $dP_5$ surface. This is consistent with 
the toric presentation, since the generic $dP_5$ surface is not toric. 
We denote the exceptional curves by $e_i$, $i=1,\ldots,n$, where 
$n=2,2,4$ in the three cases, as specified in \delpezzos . 
The $e_i$ can be embedded as $(-1,-1)$ curves in $\wX$ via the zero section 
$\sigma: B\ra \wX$. 

In the following we will consider extremal transitions consisting of 
a contraction of the curves $e_i$ on $\wX$ followed by a deformation
of the 
resulting nodal singularities. The singular threefolds obtained in the 
process can be represented as hypersurfaces in noncompact toric varieties 
$\CW$ \DFGii.\ 
The toric data of $\CW$ is 
\eqn\toricdataA{\eqalign{
& dP_2:\quad \matrix{ & Z_1 & Z_2 & Z_3 & U & V \cr
                     \IC^* & 1 & 1 & 1 & -1 & -2 \cr}\cr
& dP_3:\quad \matrix{ & Z_1 & Z_2 & Z_3 & Z_4 & U & V \cr
                     \IC^* & 1 & 1 & 0 & 0 & -1 & -1 \cr
                     \IC^* & 0 & 0 & 1 & 1 & -1 & -1 \cr}\cr
& dP_5:\quad \matrix{ & Z_1 & Z_2 & Z_3 & Z_4 & U & V & \cr
                     \IC^* & 1 & 1 & 0 & 0 & -2 & 0 \cr
                     \IC^* & 0 & 0 & 1 & 1 & 0 & -2. \cr}\cr}}
The equations for the nodal hypersurfaces read in the three cases
\eqn\hypereqA{\eqalign{
& dP_2:\quad UZ_1+VZ_2Z_3 =0 \cr
& dP_3:\quad UZ_1Z_4+VZ_2Z_3=0 \cr
& dP_5:\quad UZ_1Z_2+VZ_3Z_4=0.\cr}}
It is easy to check that the singular points are given by 
\eqn\singptA{\vbox{\halign{ $#$ \hfill &\qquad  $#$ \hfill &\qquad  $#$ \hfill \cr 
dP_2: & P_1=\{Z_1=Z_2=U=V=0\}, & P_2=\{Z_1=Z_3=U=V=0\};\cr
dP_3: & P_1=\{Z_1=Z_3=U=V=0\}, & P_2=\{Z_2=Z_4=U=V=0\};\cr
dP_5: & P_1=\{Z_1=Z_3=U=V=0\}, & P_2=\{Z_2=Z_4=U=V=0\},\cr
      & P_3=\{Z_1=Z_4=U=V=0\}, & P_4 = \{Z_2=Z_3=U=V=0\}.\cr}}}
The deformations are described by adding a constant term $\mu$ to the 
right hand side of the equations \hypereqA.\ The geometry of the 
resulting smooth variety can be analyzed in detail as in \DFGii.\
In particular, one can show that there are $n$ vanishing cycles on the 
deformed hypersurface $X_\mu$ (where $n=2,2,4$ as explained above)
subject to one homology relation. These cycles have the topology of a
three-sphere and they can be described 
quite explicitly in terms of local coordinates. We will give more details 
on this point at a later stage. 

Now let us describe the compact Calabi-Yau threefolds associated to the 
above local models. The idea is quite straightforward. We first take 
a projective 
completion of the ambient toric varieties $\CW$, obtaining smooth 
compact toric fourfolds $\CZ$ given by the following data 
\eqn\toricdataB{\eqalign{
&dP_2:\quad \matrix{ & Z_1 & Z_2 & Z_3 & U & V & W \cr 
\IC^* & 1 & 1 & 1 & -1 & -2 & 0 \cr
\IC^* & 0 & 0 & 0 & 1 & 1 & 1 \cr}\cr
&dP_3:\quad \matrix{ & Z_1 & Z_2 & Z_3 & Z_4 & U & V & W\cr 
\IC^* & 1 & 1 & 0 & 0 & -1 & -1 & 0 \cr
\IC^* & 0 & 0 & 1 & 1 & -1 & -1 & 0 \cr
\IC^* & 0 & 0 & 0 & 0 & 1 & 1 & 1\cr}\cr
&dP_5:\quad \matrix{ & Z_1 & Z_2 & Z_3 & Z_4 & U & V & W\cr 
\IC^* & 1 & 1 & 0 & 0 & -2 & 0 & 0 \cr
\IC^* & 0 & 0 & 1 & 1 & 0 & -2 & 0 \cr
\IC^* & 0 & 0 & 0 & 0 & 1 & 1 & 1.\cr}}}
Note that in all cases we obtain a $\IP^2$ bundle over the base $B$.
It is a simple exercise to check that the canonical 
class of $\CZ$ is given by $K_\CZ = 3\eta$, where $\eta$ is the divisor 
at infinity defined by the equation $W=0$. The generic divisor 
in the linear system $|-K_\CZ|$ is a smooth Calabi-Yau variety
defined by an equation of the form 
\eqn\hyoereqB{\eqalign{
& \sum_{a,b,c\geq 0, a+b+c=3} U^aV^bW^cf_{abc}(Z_i).\cr}}
For the $dP_2$ model $f_{abc}(Z_i)$ is a homogeneous polynomial 
of degree $a+2b$ in $(Z_1,Z_2,Z_3)$. For the other two models, $f_{abc}(Z_i)$ 
is a bihomogeneous polynomial of bidegree $(a+b,a+b)$ and respectively 
$(2a, 2b)$ in $(Z_1, Z_2)$, $(Z_3,Z_4)$. One can work out the vertices of the 
dual toric polyhedra and the Hodge numbers for all three cases\foot{The Hodge numbers of all manifolds appearing in this paper 
have been computed using the program {\tt POLYHEDRON}, written by Philip Candelas.}. 
The toric polyhedra $\nabla_Y$ are specified by the following sets of vertices 
$\CV_Y$
\eqn\toricdataC{\eqalign{
& dP_2:\quad\{{\tt (-1, 0, 0, 0),\; (-1, 1, 0, 0),\; (-1, 0, 1, 0),\;
(-1, 0, 0, 1),\; (-1, 1, 1, 1),\; (2, -1, -1, -1)}\}\cr
& dP_3:\quad\{{\tt (0, 1, 0, 0),\; (0, 0, 1, 0),\; (0, 0, 0, 1),\; (1,
-1, 0, 0),\; (1, 0, -1, 0),\; (1, 0, 0, -1),\; (-1, 0, 0, 0)}\}\cr
&dP_5:\quad\{{\tt (1, 1, 0, 0),\; (1, -1, 0, 1),\; (-1, 1, -1, 0),\; (-1, -1, 1, 0),\; (-1, -1, 0, 1)}\}.\cr}}
The Hodge numbers of the three models are respectively 
$(h_{1,1},h_{1,2})=(2,92),(3,81),(3,63)$.

The local extremal transitions can be lifted to 
transitions between compact Calabi-Yau threefolds as follows. The one 
parameter families $Y_\mu$ are given by the following equations 
\eqn\hypereqC{\eqalign{
& (UZ_1+VZ_2Z_3-\mu W)W^2+\sum_{(a,b,c)'}U^aV^bW^cf_{abc}(Z_i)=0\cr
& (UZ_1Z_4+VZ_2Z_3-\mu W)W^2 + \sum_{(a,b,c)'}U^aV^bW^cf_{abc}(Z_i)=0\cr
& (UZ_1Z_2+VZ_3Z_4-\mu W)W^2 + \sum_{(a,b,c)'}U^aV^bW^cf_{abc}(Z_i)=0\cr}}
where the $(a,b,c)'$ denote all the allowed triples $(a,b,c)$ which do not 
appear in the first three terms in each equation. 
The coefficients of the polynomials $f_{abc}(Z_i)$ are fixed at some generic 
values so that we obtain one parameter families  parameterized by $\mu$. 
These hypersurfaces are smooth for $\mu=0$ and develop isolated nodal 
singularities at $\mu =0$. The singular points are again given by \singptA,\
except that they have to be regarded as points on the compact toric fourfold 
$\CZ$. These singularities can be simultaneously resolved by performing a 
blow-up of $\CZ$ along the section $U=V=0$. Let $\wcz$ denote the 
resulting toric fourfold. The strict transform $\wY\subset \wcz$
is a crepant resolution of $Y_0$ with exceptional locus given by 
isolated $(-1,-1)$ curves $C_i$, $i=1,\ldots,n$. 

The blow-up of the ambient toric variety can be described torically by adding 
an extra vertex to the dual toric polyhedron $\nabla_Y$. One then obtains 
the following toric data for the dual polyhedron of the small resolution $\wY$
\eqn\toricdataD{\eqalign{
&dP_2:\quad {\cal V}_{\wy}={\cal V}_{Y}\cup\{\tt (-2, 1, 1, 1)\}\cr
&dP_3:\quad {\cal V}_{\wy}={\cal V}_Y\cup\{\tt (1, 0, 0, 0)\}\cr
&dP_5:\quad {\cal V}_{\wy}={\cal V}_{Y}\cup\{\tt (-1, -1, 0, 0)\}.\cr}}
The package {\tt PALP} \KS\ proved to be very useful in analyzing the above extremal
transitions between the smooth threefolds $Y$ and $\wY$. Moreover, 
the Hodge numbers of $\wY$ are respectively 
$(h_{1,1},h_{1,2})=(3,91),(4,80),(4,60)$ for the three cases. 
In the following sections, we will develop in detail a computational 
approach to open string topological amplitudes on the threefolds $Y$. 
Although we 
will work out the details only for the second model presented above, 
it is clear that the other models can be treated along the same lines. 
Moreover, these techniques seem to be valid for all extremal transitions 
in which the singular points are fixed under the generic
torus action on $\CZ$.  

\newsec{Local Transitions and Formal Chern-Simons Expansion} 

Following the program outlined in section three, here we review the 
open string localization techniques 
for local transitions developed in \refs{\DFGi,\DFGii}. The main goal of 
this review is to clarify the interplay between torus action and 
framing in Chern-Simons theory. We will find that the proper 
framework for open string amplitudes is a formal extension of the 
Chern-Simons expansion which treats the framing as a formal variable. 
This new idea will allow us to formulate general rules for the correspondence 
between toric weights and framing encompassing all cases known so far 
\refs{\AMV,\DFGi,\DFGii}. Moreover, as discussed in section six, the 
same rules play a crucial role for open string amplitudes on compact 
manifolds. 
Let us start with the first local model described in the previous
section, namely the local $dP_2$ model. The deformed hypersurface $X_\mu$ 
is given by the equation 
\eqn\hyeoreqD{
UZ_1+VZ_2Z_3=\mu} 
in the toric variety $\CW$ specified in 
\toricdataA,\ and the singular points at $\mu=0$ 
are $P_1=\{Z_1=Z_2=U=V=0\}$ and $P_2=\{Z_1=Z_3=U=V=0\}$. 
The geometry of this model has been thoroughly analyzed in \DFGii,\
so we will keep the details to a minimum. The main point is that there 
are two lagrangian spheres $L_1, L_2$ on $X_\mu$ in the same homology 
class $[L_1]=[L_2]$. Each of these cycles can be described as an intersection
of $X_\mu$ with two real hypersurfaces in $\CW$ given by the following 
equations
\eqn\hypereqE{\eqalign{
& L_1:\quad UZ_2{\overline Z}_2-W{\overline Z}_1=0,\quad 
VZ_2^2{\overline Z}_2-W{\overline Z}_3=0\cr
& L_2:\quad UZ_3{\overline Z}_3-W{\overline Z}_1=0,\quad 
VZ_3^2{\overline Z}_3-W{\overline Z}_2=0.\cr}}
Writing these equations in suitable local coordinates centered at the points 
$P_1, P_2\in \CW$, one can check that the local geometry is indeed isomoprhic 
to a local deformation of a conifold singularity. Moreover, it has been 
shown in \DFGii\ that one can choose a symplectic K\"ahler structure on 
$\CW$ so that $L_1, L_2$ are lagrangian. 

The open string topological A model considered in \DFGii\ is defined 
by wrapping $N_1,N_2$ topological branes on $L_1, L_2$. As explained there, 
and also in section three of this paper, a crucial step in the computation of 
the free energy is summing up the instanton corrections to Chern-Simons 
theory. This can be done by localization with respect to the following 
toric action 
\eqn\toractAminusone{
\matrix{ & Z_1 & Z_2 & Z_3 & U & V & W\cr 
S^1 & \lambda_1 & \lambda_2 & 0 & -\lambda_1 & -\lambda_2 & 0.\cr}}
Note that this action preserves both the hypersurface $X_\mu$ and the 
spheres $L_1, L_2$. We have to sum over open string maps which are invariant 
under the induced toric action on the moduli space. 
As usual in localization computations the image of all such maps consists of 
a collection of invariant holomorphic Riemann surfaces embedded 
in $X_\mu$. In closed string situations, these would have to be closed 
Riemann surfaces, or equivalently algebraic curves on $X_\mu$. 
Since we are doing an open string computation, we will have a 
collection of invariant bordered Riemann surfaces with boundary components 
embedded in $L_1, L_2$. According to the analysis of \DFGii,\ there are 
only three such invariant surfaces -- two discs and a cylinder -- which can be 
described as follows. 

\ifig\dpdisksancylinder{Primitive open string instantons on $X$.}{\epsfxsize3.0in\epsfbox{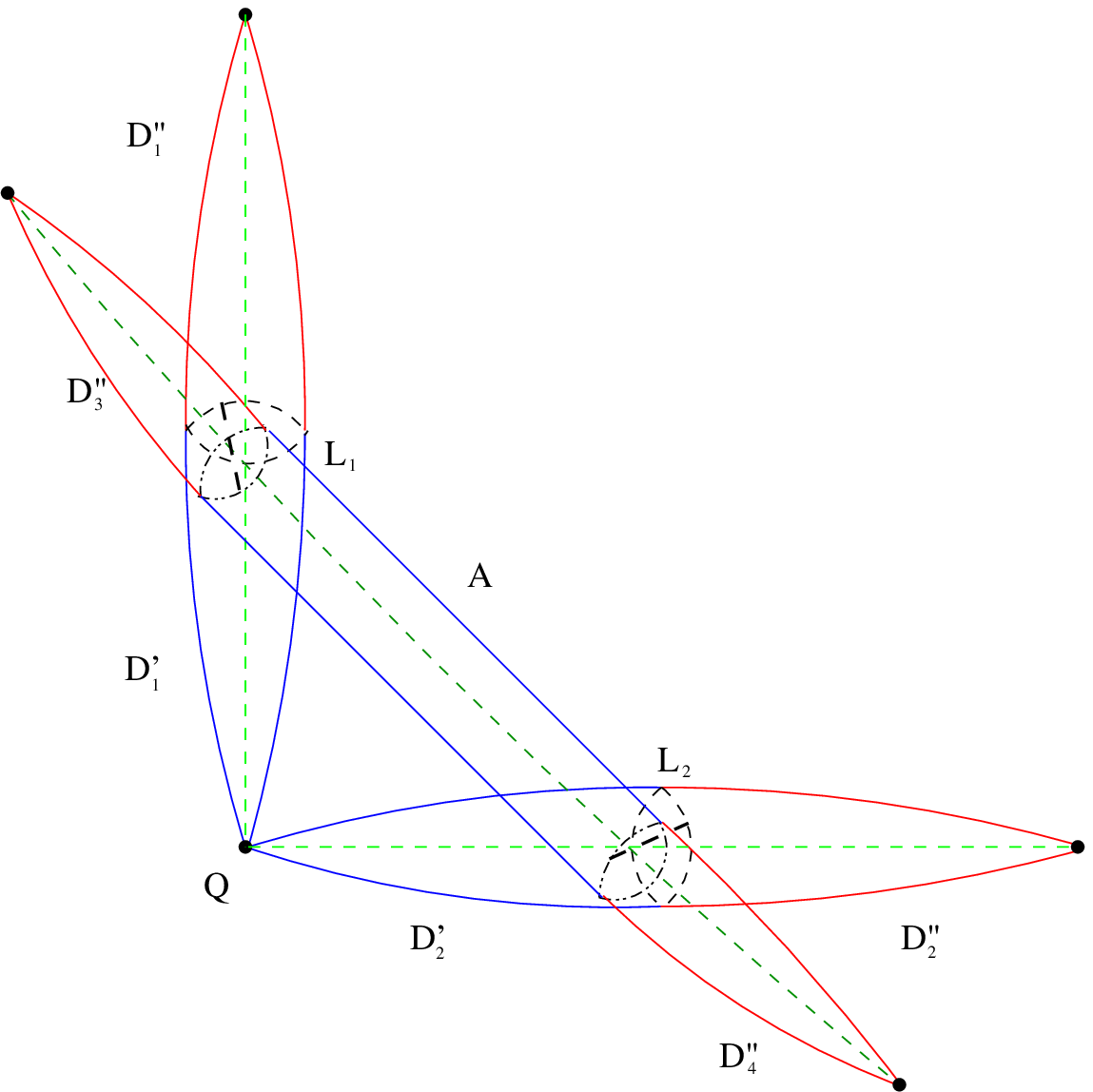}}

Let ${\overline X}_\mu$ be the (relative) projective closure of $X_\mu$ in the 
compact toric variety $\CZ$. The defining equation of ${\overline X}_\mu$ is
\eqn\hypereqF{
UZ_1+VZ_2Z_3=\mu W.}
One can check that ${\overline X}_\mu$ is a toric Fano threefold 
which contains $X_\mu$ as the complement of the divisor at infinity
$D_\infty={\overline X}_\mu \cap \{W=0\}$. The bordered Riemann surfaces 
are obtained by intersecting the spheres $L_1, L_2$ with three invariant 
curves on ${\overline X}_\mu$ given by 
\eqn\invcurvesA{\eqalign{
& C'_1:\qquad V=Z_2=0,\quad UZ_1=\mu W\cr
& C'_2:\qquad V=Z_3=0,\quad UZ_1=\mu W\cr
& C'_3:\qquad U=Z_1=0,\quad VZ_2Z_3=\mu W.\cr}}
We have the following intersections $L_1\cap C'_1=\Gamma'_1$, $L_2\cap C'_2=
\Gamma'_2$, $L_1\cap C'_3=\Xi'_1$, $L_2\cap C'_3=\Xi'_2$ where 
$\Gamma'_1,\Xi'_1$
and respectively $\Gamma'_2,\Xi'_2$ are algebraic knots in $L_1, L_2$ 
forming Hopf links with linking number $+1$. 
$\Gamma'_1$ divides $C'_1$ into two invariant discs $D'_1,D''_1$, $\Gamma'_2$ 
and divides $C'_2$ into other two invariant discs $D'_2, D''_2$. Similarly, 
$\Xi'_1, \Xi'_2$ divide $C'_3$ into two discs $D''_3, D''_4$ and a holomorphic 
annulus $A$ as in \dpdisksancylinder . 

Note that the discs $D'_1,D'_2$ have a common origin at 
$Q=\{Z_2=Z_3=V=0,~UZ_1=\mu W\}$. The origins of all other discs are points on 
the divisor at infinity $D_\infty$. 

A careful argument based on homology constraints \DFGii\ shows that only 
$D'_1, D'_2$ and $A$ contribute to the open string amplitudes in the local case. 
The other discs do not play any role for the local computation, 
but they will enter the computation of open string amplitudes 
for the compact model. 
In order to write down a general formula for the instanton corrections 
let us denote the holonomy variables associated to the four knots 
$\Gamma'_1,\Gamma'_2, \Xi'_1, \Xi'_2$ by $V'_1, V'_2, U'_1, U'_2$, respectively. The instanton 
series has the form 
\eqn\instcorrA{
F_{inst}(g_s,t'_1, t'_2, t_c, U'_1, U'_2, V'_1, V'_2) =
F^{(1)}_{inst}(g_s,t_c,U'_1,U'_2) + F^{(2)}_{inst}(g_s, t'_1, t'_2, V'_1, V'_2)}
where
\eqn\instcorrAB{\eqalign{
 F^{(1)}_{inst}(g_s,t_c, U'_1, U'_2) =
& \sum_{g=0}^\infty \sum_{h_1,h_2=0}^\infty \sum_{d=0}^\infty
\sum_{m_i\geq 0, n_j\geq 0}
i^{h_1+h_2}g_s^{2g-2+h_1+h_2} \cr
& \times C_{g,h_1, h_2}(d|m_i,n_j)
e^{-dt_c} \prod_{i=1}^{h_1} \Tr U'^{m_i}_1
\prod_{j=1}^{h_2} \Tr U'^{n_j}_2\cr}}
\eqn\instcorrAC{\eqalign{
& F^{(2)}_{inst}(g_s, t'_1, t'_2, V'_1, V'_2)=
\sum_{g=0}^\infty \sum_{h_1,h_2=0}^\infty \sum_{d_1,d_2=0}^\infty
 \sum_{m_i\geq 0, n_j\geq 0}
 i^{h_1+h_2}g_s^{2g-2+h_1+h_2} \cr
&\qquad\qquad\qquad\qquad\qquad~\times F_{g,h_1, h_2}(d_1, d_2| m_i, n_j)
e^{-d_1t'_1-d_2t'_2}  \prod_{i=1}^{h_1} \Tr V'^{m_i}_1
\prod_{j=1}^{h_2} \Tr V'^{n_j}_2.\cr}}
The first represents the contribution of multicovers of the cylinder $A$ 
and the second term represents the contribution of invariant maps with 
image contained in the union $D'_1\cup D'_2$. 
The parameters $t'_1, t'_2, t_c$ represent open string K\"ahler moduli. 
At the classical level, they should equal the symplectic areas of $D'_1, D'_2, A$
which are equal since all these surfaces are homologically equivalent. 
However, it is understood by now that these parameters can receive quantum 
corrections due to degenerate instantons, which can be in principle 
different for the three surfaces. The rest of the notation is standard: 
$d,d_1, d_2$ denote the degrees of a given map with respect to the three 
fixed surfaces $A,D'_1,D'_2$ and $(m_i, n_j)$ denote the winding numbers 
of the boundary components about the invariant knots in the target space. 

The coefficients $C_{g,h_1, h_2}(d|m_i,n_j)$ and 
$F_{g,h_1, h_2}(d_1, d_2|m_i, n_j)$ can be determined by localization 
computations for open string morphisms. This computation has been done 
in great detail in \DFGii,\ and we will not repeat it here. 
For our purposes, it suffices to recall the results for genus zero maps 
of total degree smaller or equal than three. We will rewrite all the formulae obtained in \DFGii,\ in terms of 
the ratio of toric weights $z={\lambda_2/\lambda_1}$.

Let us start with multicovers of the two discs. Since we do not make a 
particular choice of toric weights, we will have nonzero corrections 
corresponding to surfaces with one, two and three boundary components. 
We have the following expressions.

\noindent 
Degree one: 
\eqn\instcorrC{\eqalign{ 
& F_{0,1,0}(1,0|1)=F_{0,0,1}(0,1|1)=1.\cr}}
Degree two:
\eqn\instcorrD{\vbox{\halign{ $#$ \hfill &\qquad  $#$ \hfill \cr
F_{0,1,0}(2,0|2)={2z+1\over 4}, & 
F_{0,0,1}(0,2|2)={(1-z)\over 4(1+z)}, \cr
F_{0,2,0}(2,0|1,1)=-{z(1+z)\over 4},
& F_{0,0,2}(0,2|1,1)={z\over 4(1+z)^2},\cr
F_{0,1,1}(1,1|1,1) = -1. & \cr}}}
Degree three:
\eqn\instcorrE{\vbox{\halign{ $#$ \hfill &\qquad  $#$ \hfill \cr
 F_{0,1,0}(3,0|3)={(3z+1)(3z+2)\over 18}, & 
F_{0,0,1}(0,3|3)={(1-2z)(2-z)\over 18(1+z)^2},\cr
F_{0,1,1}(2,1|2,1)=-z, & F_{0,1,1}(1,2|1,2)={z\over 1+z},\cr
F_{0,2,1}(2,1|1,1,1)={z^2\over 2}, & 
F_{0,1,2}(1,2|1,1,1)={z^2\over 2(1+z)^2},\cr
F_{0,2,0}(3,0|2,1)=-{z(z+1)(2z+1)\over 3}, &
F_{0,0,2}(0,3|2,1)={z(1-z)\over 3(1+z)^3},\cr
F_{0,3,0}(3,0|1,1,1)={z^2(1+z)^2\over 6}, & 
F_{0,0,3}(0,3|1,1,1)={z^2\over 6(1+z)^4}.\cr}}}
The multicover contributions of the cylinder are independent of toric
weights. In fact one has a very simple formula of the form 
\eqn\instcorrF{
C_{0,1,1}(d|d,d)= {(-1)^d\over d}}
for multicovers of degree $d$, all other terms being zero. 
The alternating sign in this formula needs some explanation. 
A common problem with open string localization computations is that 
the moduli spaces of open string maps, and the obstruction bundles 
do not have a canonical complex structure as in the closed string case. 
Instead one has only a complex structure up to conjugation. This gives rise 
to a sign ambiguity in the open string amplitudes noticed first time in 
\refs{\GZ,\KL}. In the absence of a rigorous framework for these 
computations, this ambiguity cannot be resolved from first principles. 
Since the present open string model has a closed string dual, a practical 
solution to this problem is to fix the sign so that the final results agree 
with the closed string amplitudes. This is not a satisfactory solution, 
but it seems to be the only one available at the moment. 
The choice made here is different from the one made in \DFGii,\
where the sign was taken to be $+1$ for all $d$. It will be clear shortly 
that \instcorrF\ is a more natural choice in this context. 

The next step is to perform the Chern-Simons functional integral taking into 
account the instanton corrections \instcorrC-\instcorrF.\ At this stage, we 
need to make a choice of framing for the knots 
$\Gamma'_1, \Gamma'_2, \Xi'_1, \Xi'_2$. The idea proposed in \DFGii\ is that 
this choice is related to a choice of toric weights, namely to the 
value of $z$. Let us discuss some of the main points. 
Choose local coordinates 
$(x_1,y_1,u_1,v_1)$ centered at $Z_1=Z_2=U=V=0$ and $(x_2,y_2,u_2,v_2)$ 
centered at $Z_1=Z_3=U=V=0$. The local equations of the threefold $X_\mu$
in these patches are 
\eqn\loceqA{
x_1u_1+y_1v_1=\mu,\qquad x_2u_2+y_2v_2=\mu}
and the spheres $L_1, L_2$ are described by the real sections 
\eqn\realsectA{
L_1:\quad u_1={\overline x}_1,\ v_1={\overline y}_1,\qquad 
L_2:\quad u_2={\overline x}_2,\ v_2={\overline y}_2.}
The knots $\Gamma'_1, \Xi'_1\subset L_1$ are cut by the equations 
$y_1=v_1=0$ and respectively $x_1=u_1=0$ on $L_1$. 
In order to specify a framing of $\Gamma'_1$, $\Xi'_1$ we have to specify 
sections of the normal bundles $N_{\Gamma'_1/L_1}$ and $N_{\Xi'_1/L_1}$, respectively. 
Let us parameterize the knots by the angular variables 
$\theta_{x_1}$ and $\theta_{y_1}$ corresponding to the 
complex coordinates $x_1$ and $y_1$ respectively. Then we can write down the normal sections 
in the form 
\eqn\normsecA{\eqalign{
&\Gamma'_1:\quad (y_1,v_1) = \left(\mu^{1/2} e^{ip_1\theta_{x_1}}, \mu^{1/2} 
e^{-ip_1\theta_{x_1}}\right)\cr
&\Xi'_1:\quad (x_1,u_1) = \left(\mu^{1/2} e^{iq_1\theta_{y_1}}, \mu^{1/2} 
e^{-iq_1\theta_{y_1}}\right),\cr}}
where $(p_1, q_1)$ determine the framing of the two knots.\foot{There is a 
subtlety here pointed out in \DFGii.\ Namely, specifying a single section 
of a principal $S^1$ bundle over $S^1$ does not determine an integer number. 
One needs to specify in fact a pair of sections in order to obtain integral 
data. In the present case there is a canonical choice for a reference
section since we have explicitly constructed $(\Gamma'_1, \Xi'_1)$ as algebraic 
knots. This choice determines the framing to be $(p_1, q_1)$.} 
We can write down analogous formulae for the pair $(\Gamma'_2, \Xi'_2)$ of 
knots in $L_2$
\eqn\normsecB{\eqalign{
&\Gamma'_2:\quad (y_2,v_2) = \left(\mu^{1/2} e^{ip_2\theta_{x_2}}, \mu^{1/2} 
e^{-ip_2\theta_{x_2}}\right)\cr
&\Xi'_2:\quad (x_2,u_2) = \left(\mu^{1/2} e^{iq_2\theta_{y_2}}, \mu^{1/2} 
e^{-iq_2\theta_{y_2}}\right).\cr}}

The relation between the framings $(p_1, q_1)$ , $(p_2,q_2)$ 
and the toric weights follows by imposing 
the condition that the sections \normsecA-\normsecB\
be preserved by the torus action, as first proposed in \KL.\ 
This yields the following relations 
\eqn\framA{\eqalign{
& \lambda_{y_1} = p_1 \lambda_{x_1},\qquad 
\lambda_{x_1} = q_1 \lambda_{y_1}\cr
& \lambda_{y_2} = p_2 \lambda_{x_2},\qquad 
\lambda_{x_2} = q_2 \lambda_{y_2}\cr}}
where the $(\lambda_{x_1},\ldots,\lambda_{y_2})$ denote the weights 
of the torus action on the local coordinates, which can be easily 
expressed in terms of $(\lambda_1, \lambda_2)$.
This is a very constrained system of equations, since $(p_1,\ldots, q_2)$ 
have to be integers in order for the solution to make sense within the 
framework of standard Chern-Simons theory. In fact, it is not hard 
to show that there are no nontrivial solutions satisfying all these 
conditions. The solution proposed in \DFGii\ was to impose these 
conditions only for the knots $\Gamma'_1, \Gamma'_2$ i.e. the boundary 
components of the two discs $D'_1, D'_2$. Then, the framings $(q_1, q_2)$ 
of $(\Xi'_1, \Xi'_2)$ were determined by more indirect arguments to be $(0,0)$. 
Furthermore, imposing the equivariance conditions for the sections 
of $N_{\Gamma'_1/L_1}, N_{\Gamma'_2/L_2}$ leaves us with only two sensible 
solutions
\eqn\framB{\eqalign{
& p_1=p_2=2\ \Rightarrow \ z=0\cr
& p_1=p_2=0\ \Rightarrow \ z=1.\cr}}
The first solution was shown to be a sensible choice in \DFGii,\ although 
there were some unclear aspects of the duality map. We will shortly 
come back to this point.  

Here we would like to propose a different solution to this problem which 
involves far less obscure choices. The main idea is to relax the condition 
that the framings $(p_1, q_1, p_2, q_2)$ be integral. Instead our proposal 
is to treat them as formal variables which can in particular take fractional 
values. Although this idea seems to be at odds with standard Chern-Simons 
theory, we will show below that it makes perfect sense provided one 
treats the vacuum expectation values of Wilson loops as formal 
power series expansion in the framing variables. Following this idea, we can 
easily solve all equivariance conditions \framA\ 
\eqn\framC{
p_1=z+2,\quad q_1={1\over z+2},\quad p_2={z+2\over z+1},\quad 
q_2={z+1\over z+2}}
where $z$ is left undetermined. This fixes the framing of all knots 
in the problem with no ambiguities. At the same time, the instanton 
corrections \instcorrC-\instcorrE\ also depend on the variable $z$. 
The next step is to perform the Chern-Simons computations treating $z$ 
as a formal variable and expanding all knot and link invariants 
in powers of $z$. The truly remarkable aspect of this procedure is that 
the final result turns out to be independent of $z$, 
as it is usually the case with 
closed string localization computations! Below we will perform the 
computations for genus zero amplitudes up to degree three, finding 
very strong evidence for this conjecture. 

For concreteness, let us collect the instanton corrections 
\instcorrC-\instcorrF\ in a single formula
\eqn\instcorrG{\eqalign{
F_{inst}=& {iq'_1\over g_s}\Tr V'_1 +{iq'_2\over g_s} \Tr V'_2 - 
q_c\Tr U'_1\Tr U'_2
+{iq'^2_1\over g_s} {2z+1\over 4} \Tr V'^2_1 +{iq'^2_2\over g_s} 
{(1-z)\over 4(1+z)}
\Tr V'^2_2\cr
& + q'^2_1{z(1+z)\over 4} (\Tr V'_1)^2 -q'^2_2{z\over 4(1+z)^2} (\Tr V'_2)^2 
+q'_1q'_2 \Tr V'_1 \Tr V'_2 +\half q_c^2 \Tr U'^2_1 \Tr U'^2_2\cr
& +{iq'^3_1\over g_s}{(3z+1)(3z+2)\over 18} \Tr V'^3_1 + 
{iq'^3_2\over g_s} {(1-2z)(2-z)\over 18(1+z)^2} \Tr V'^3_2 
+ q'^2_1q'_2 z \Tr V'^2_1 \Tr V'_2 \cr
&-q'_1q'^2_2{z\over 1+z}  \Tr V'_1 \Tr V'^2_2 
+q'^3_1{z(z+1)(2z+1)\over 3} \Tr V'^2_1 \Tr V'_1 
- q'^3_2{z(1-z)\over 3(1+z)^3} \Tr V'^2_2 \Tr V'_2 \cr
&-ig_s q'^2_1 q'_2 {z^2\over 2} (\Tr V'_1)^2 \Tr V'_2 
-i g_s q'_1q'^2_2 {z^2\over 2(1+z)^2} \Tr V'_1 (\Tr V'_2)^2\cr
& -i g_s q'^3_1 {z^2(1+z)^2\over 6} (\Tr V'_1)^3 
-i g_s q'^3_2{z^2\over 6(1+z)^4} (\Tr V'_2)^3 -{1\over 3} 
q_c^3 \Tr U'^3_1 \Tr U'^3_2.\cr}}
In this formula, $q'_1=e^{-t'_1}, q'_2=e^{-t'_2}, q_c=e^{-t_c}$ 
denote the instanton factors associated to the three primitive 
open string instantons. 

The free energy of the topological open string is given by the following 
formula 
\eqn\freenA{\CF_{op}(g_s,q'_1, q'_2, q_c, \lambda_1,\lambda_2) 
= \CF_1^{CS}(\lambda_1, g_s) 
+\CF_2^{CS}(\lambda_2, g_s) + \ln \left\langle e^{F_{inst}}\right\rangle}
where $\lambda_1, \lambda_2$ are the 't Hooft couplings of the 
two Chern-Simons theories, which are related to the string coupling 
by $\lambda_{1,2} = N_{1,2} g_s$. The first two terms in the right hand 
side of this equation represent the contributions of the two 
Chern-Simons sectors while the third term represents the expectation 
value of the instanton corrections. In the following we will concentrate 
on the last term. As mentioned above, our goal is to show that the 
resulting expression for the free energy agrees with the closed string 
expansion of the local $dP_2$ model. In particular, it should 
be independent of $z$. Similar computations have been performed in 
\DFGii,\ so we will not repeat all the details here. 
The important point in the present approach is that all knot invariants 
occurring in the process have to be expanded as formal power series 
of the framing variables. 

Let us explain the algorithm for a general link $\CL$ with $c$ components 
$R_\alpha$ and framings $p_\alpha$, $\alpha=1,\dots, c$.
The framing dependence of the expectation
value $\langle W_{R_\alpha}(\CL)\rangle$ is of the form
\eqn\framdepA{
\langle W_{R_\alpha}(\CL)\rangle_{(p_1, \ldots, p_c)} =
e^{{ig_s\over 2}\sum_{\alpha=1}^c \kappa_{R_\alpha} p_\alpha}
e^{{i\lambda\over 2} \sum_{\alpha=1}^c l_\alpha p_\alpha}
\langle W_{R_\alpha}(\CL)\rangle_{(0, \ldots, 0)}}
where $l_\alpha$ is the total number of boxes in the Young tableau
of $R_\alpha$, and $\kappa_{R_\alpha}$ is a group theoretic quantity
defined as follows. Let $v=1,\ldots, r$ label the rows of the Young
tableau of a representation $R$, and $l_v$ denote the length of the
$v$-th row. Then we have \MV\
\eqn\framdepB{
\kappa_R = l + \sum_{v=1}^r (l_v^2 -2vl_v)}
where $l=\sum_{v=1}^r l_v$ is the total number of boxes.
In the process of evaluating \freenA\ we will encounter trace products 
of the form $\prod_{i} \Tr V^{n_i}$ where $V$ is any of the holonomy variables 
$V'_1,V'_2, U'_1, U'_2$. Such products have to be written as linear 
combinations 
of the form $\sum_R \Tr_R V$ in order to evaluate the Chern-Simons 
vacuum expectation values. A general rule in all such cases is that the 
number of boxes in the Young diagram of any representation $R$ in this 
sum equals the total degree $d=\sum_i n_i$ of the open string map. 
All the resulting terms will be weighted by $q^d$, where $q=q'_1, q'_2$ or 
$q_c$ is the appropriate instanton factor. This means that we will always 
obtain combinations of the form $\left(q e^{i{\lambda \over 2}p}\right)^d$ 
in the final expression for the free energy (where $p$ is the framing 
of the appropriate knot). Therefore we can simplify the computation by
absorbing all the prefactors 
$e^{{i\lambda\over 2} \sum_{\alpha=1}^c l_\alpha p_\alpha}$ in a 
redefinition of the instanton factors. Since these factors are independent 
of $g_s$ and do not play any role in the expansion, we will call them 
trivial framing factors. 
With this redefinition understood, we can use the following expression
for the link invariants in the Chern-Simons expansion 
\eqn\framdepA{
\langle W_{R_\alpha}(\CL)\rangle^\prime_{(p_1, \ldots, p_c)} =
e^{{ig_s\over 2}\sum_{\alpha=1}^c \kappa_{R_\alpha} p_\alpha}
\langle W_{R_\alpha}(\CL)\rangle_{(0, \ldots, 0).}}
As explained above, all the expectation values must be written as power 
series in the framing variables, which appear in the exponential 
prefactor of \framdepB\ multiplying the string coupling $g_s$. 
Since we want to obtain the result as a power series 
in $g_s$, the most efficient way to proceed is to expand all 
expectation values \framdepB\ in $g_s$. Given the particular 
form of the framing dependence, the resulting expressions 
will also be series expansions in $p_\alpha$. 
The expectation values in canonical framing 
$\langle W_{R_\alpha}(\CL)\rangle_{(0, \ldots, 0)}$
can be written as rational functions of the variables 
$y = e^{i\lambda/2}, x=e^{ig_s/2}$. Therefore the expansion is 
straightforward, although somewhat tedious. 
Without giving more details here, let us record the 
final answer for this computation. We will write down only the genus 
zero contribution, i.e. the coefficient of $g_s^{-2}$ in the 
final expression for the free energy, truncated to terms up to degree 
three\foot{There is a subtlety here. The instanton sum \instcorrG\ 
contains only terms of genus zero. It is not a priori clear 
that higher genus corrections do not affect the final answer 
for the genus zero free energy. In principle, this could happen since 
the Chern-Simons expansion generates various powers of $g_s$, which 
mix with the instanton corrections in a nontrivial way. 
However, in all known examples of generic transitions (present 
cases included), it is a posteriori 
clear that such effects are absent. This pattern will be confirmed 
for compact hypersurfaces in section seven.}
$$\eqalign{
&\CF^{(0)}_{op}(g_s, q'_1,q'_2,q_c, \lambda_1,\lambda_2) =
y_1^2+{1\over 8}y_1^4 +{1\over 27}y_1^6 + y_2^2+{1\over 8} y_2^4
+{1\over 27}y_2^6 \cr
& + q'_1\left(y_1-y_1^{-1}\right)+q'_2\left(y_2-y_2^{-1}\right) 
+q_c\left(y_1y_2-y_1y_2^{-1}-y_1^{-1}y_2-y_1^{-1}y_2^{-1}\right)\cr
&-{1\over 8} q'^2_1\left(7y_1^2-8+y_1^{-2}\right) -{1\over 8} 
q'^2_2\left(7y_2^2-8+y_2^{-2}\right) -
q'_1q'_2\left(y_1y_2-y_1y_2^{-1}-y_1^{-1}y_2+y_1^{-1}y_2^{-1}\right)\cr
&-{1\over 8}q_c^2\left(7y_1^2y_2^2-8y_1^2-8y_2^2+8 +y_1^2y_2^{-2}
+y_1^{-2}y_2^{2}-y_1^{-2}y_2^{-2}\right)\cr
& -q'_1q_c\left(y_1^2y_2-y_2-y_1^2y_2^{-1}+y_1^{-1}\right)
-q'_2q_c\left(y_1y_2^2-y_1-y_1^{-1}y_2^{2}+y_1^{-1}\right)\cr
& +{1\over 27}q'^3_1\left(55y_1^3-81y_1+27y_1^{-1}-y_1^{-3}\right)
+{1\over 27}q'^3_2\left(55y_2^3-81y_2+27y_2^{-1}-y_2^{-3}\right)\cr
}$$
\eqn\freenC{\eqalign{
&+q'^2_1q'_2\left(3y_1^2y_2-4y_2-3y_1^2y_2^{-1}+4y_2^{-1}+y_1^{-2}y_2
-y_1^{-2}y_2^{-1}\right)\cr
& +q'_1q'^2_2\left(3y_1y_2^2-4y_1-3y_1^{-1}y_2^{2}+4y_1^{-1}+y_1y_2^{-2}
-y_1^{-1}y_2^{-2}\right)\cr
& +q'^2_1q_c\left(3y_1^3y_2-4y_1y_2-3y_1^3y_2^{-1}+4y_1y_2^{-1}+y_1^{-1}
y_2-y_1^{-1}y_2^{-1}\right)\cr
& +q'_1q_c^2\left(3y_1^3y_2^2-4y_1^3-4y_1y_2^2+5y_1+y_1^{-1}y_2^2+y_1^3y_2^{-2}
-y_1y_2^{-2}-y_1^{-1}\right)\cr
& +q'^2_2q_c\left(3y_1y_2^3-4y_1y_2-3y_1^{-1}y_2^{3}+4y_1^{-1}y_2+y_1
y_2^{-1}-y_1^{-1}y_2^{-1}\right)\cr
& +q'_2q_c^2\left(3y_1^2y_2^3-4y_2^3-4y_1^2y_2+5y_2+y_1^{2}y_2^{-1}+y_1^{-2}
y_2^{3}-y_1^{-2}y_2-y_2^{-1}\right)\cr
& +q'_1q'_2q_c\left(3y_1^2y_2^2-4y_1^2-4y_2^2+5+y_1^2y_2^{-2}+y_1^{-2}y_2^2
-y_1^{-2}-y_2^{-2}\right)\cr
& +q_c^3\left(2y_1^3y_2^3-3y_1^3y_2-3y_1y_2^3+4y_1y_2+y_1^3y_2^{-1}
+y_1^{-1}y_2^3-y_1y_2^{-1}-y_1^{-1}y_2\right).\cr}}
A first observation is that the final answer does not depend on $z$, as 
promised above. Moreover, it is not hard to show that the above 
formula has the correct integrality properties of a genus zero 
closed string expansion. In order to compare with the Gromov-Witten 
expansion of the local $dP_2$ model, we have to rewrite \freenC\ 
in terms of closed string variables using the duality map 
\eqn\dualmapA{\eqalign{
& y_1^2 =\tq_1,\qquad y_2^2=\tq_2,\cr
& q'_1 y_1= q'_2 y_2 = q_c y_1y_2 =\tq.\cr}}
Here $\tq_1, \tq_2, \tq$ are the instanton 
factors associated to curve classes $e_1,e_2,h$ on $dP_2$. 
Then we obtain the following expression 
\eqn\freenD{\eqalign{
\CF^{(0)}_{cl}(\tq_1,\tq_2,\tq) =& \tq_1+\tq_2
+\tq(\tq_1^{-1}\tq_2^{-1}-2\tq_1^{-1}-2\tq_2^{-1}+3)
+\tq^2(-4\tq_1^{-1}\tq_2^{-1}+5\tq_1^{-1}+5\tq_2^{-1}\cr
& -6)+\tq^3(-6\tq_1^{-2}\tq_2^{-1}-6\tq_1^{-1}\tq_2^{-2}+7\tq_1^{-2}+7\tq_2^{-2}
+35\tq_1^{-1}\tq_2^{-1}-32\tq_1^{-1}\cr
& -32\tq_2^{-1}+27)+\ldots +{1\over 4}\big[\tq_1^2+\tq_2^2+\tq^2(\tq_1^{-2}\tq_2^{-2}-2\tq_1^{-2}-2\tq_2^{-2}+3)
+\ldots\big]\cr 
& +{1\over 27}\big[\tq_1^3+\tq_2^3+\tq^3(\tq_1^{-3}\tq_2^{-3}-2\tq_1^{-3}-2\tq_2^{-3}+3)+\ldots\big]+\ldots ,
}}
which is precisely the genus zero Gromov-Witten expansion of 
$dP_2$. 

This result has been obtained previously in \DFGii.\ The novelty here is 
that we have a much better understanding of the choices made in the 
process. In particular we have found and tested a general framework 
for these computations which does not require fixing the toric weights. 
This formalism is closer in spirit to closed string localization 
computations on moduli spaces of stable maps. Similar techniques can be applied at higher genus, although the 
localization computations for arbitrary $z$ become more involved.  
We will not try to pursue these computations here, but there is 
little doubt that the results will be in perfect agreement with 
the closed string expansion. 

If one is interested in computational power, a particular choice of $z$ 
may be very useful. For example we could choose $z=0$ as in \DFGii,\
in which case we would obtain a closed form for the instanton 
corrections 
\eqn\instcorrH{\eqalign{
F_{inst} = & \sum_{d=1}^\infty {iq'^d_1\over 2d\sin{dg_s\over 2}}\Tr V'^d_1 
+{iq'^d_2\over 2d\sin{dg_s\over 2}}\Tr V'^d_2 \cr
& + \sum_{d=1}^\infty {q'_1q'_2\over d} \Tr V'^d_1 \Tr V'^d_2 
+ \sum_{d=1}^\infty {(-1)^dq_c^d\over d}\Tr U'^d_1 \Tr U'^d_2.\cr}}
The framings of the four knots can be read off from \framC\
$p_1=p_2=2$, $q_1=q_2=\half$. Therefore we still have fractional 
framing for the knots $\Xi'_1, \Xi'_2$. 
Using this expression, one can perform the Chern-Simons integration 
as in \DFGii\ obtaining agreement with the closed string dual for 
all genera up to degree four. 

It is interesting to note that this system of Chern-Simons theories 
is different from the one found in \DFGii,\ although they give 
rise to identical expressions for the free energy. 
This is a very interesting phenomenon whose origin can be traced 
to a special symmetry of the Gromov-Witten expansion of the local 
$dP_2$ model. We will sketch an argument here, leaving a more thorough 
investigation of this aspect for future work.
Let us recall the expression for the instanton corrections 
found in \DFGii\ 
\eqn\instcorrI{\eqalign{
F_{inst} = & \sum_{d=1}^\infty {iq'^d_1\over 2d\sin{dg_s\over 2}}\Tr V'^d_1 
+{iq'^d_2\over 2d\sin{dg_s\over 2}}\Tr V'^d_2 \cr
& + 2\sum_{d=1}^\infty {q'_1q'_2\over d} \Tr V'^d_1 \Tr V'^d_2 
- \sum_{d=1}^\infty {q_c^d\over d}\Tr U'^d_1 \Tr U'^d_2\cr}}
with framings $p_1=p_2=2$, $q_1=q_2=0$. 
Note that the contributions of multicovers of the annulus $A$ and 
the pinched cylinder $D'_1\cap D'_2$ have different coefficients, and 
also $q_1, q_2$ have different values. 

In order to understand the meaning of these different choices, it suffices 
to consider the terms of order two in the free energy. More precisely, let us 
look at the degree two multicover contributions of the annulus $A$. 
The formula \instcorrH\ yields  
\eqn\degtwoA{
{1\over 8}q_c^2\left(y_1^2y_2^2-y_1^2y_2^{-2} -y_1^{-2}y_2^2 -y_1^{-2}
y_2^{-2}\right) -q_c^2\left(y_1^2y_2^2-y_1^2-y_2^2+1\right).}
The first term in this formula represents degree two multicovers 
of the degree one contribution. The second term represents a nontrivial 
contribution to the degree two Gromov-Witten invariants. If we perform a 
similar computation starting with \instcorrI\ we find that only 
the multicover contributions are present. Hence in this case, the 
degree two annulus corrections do not generate genuine new contributions 
to the Gromov-Witten invariants of degree two. Instead, the corrections 
corresponding to the pinched cylinder $D'_1\cap D'_2$ have to be counted 
twice in order to compensate for the missing terms. Of course, it is 
highly nontrivial that this correction yields the correct results at 
higher degree as well. 

From the closed string point of view, this phenomenon has the following 
interpretation. Let us consider for example the localization 
computations for the Gromov-Witten invariant of the local $dP_2$ 
model in curve class $2h-e_1-e_2$. Adopting a Kontsevich graph 
representation \MK\ of the fixed loci, we obtain the
graphs in fig. 3. 

\ifig\dpgraphs{Closed string Kontsevich graphs in the class $2h-e_1-e_2$ }{\epsfxsize3.0in\epsfbox{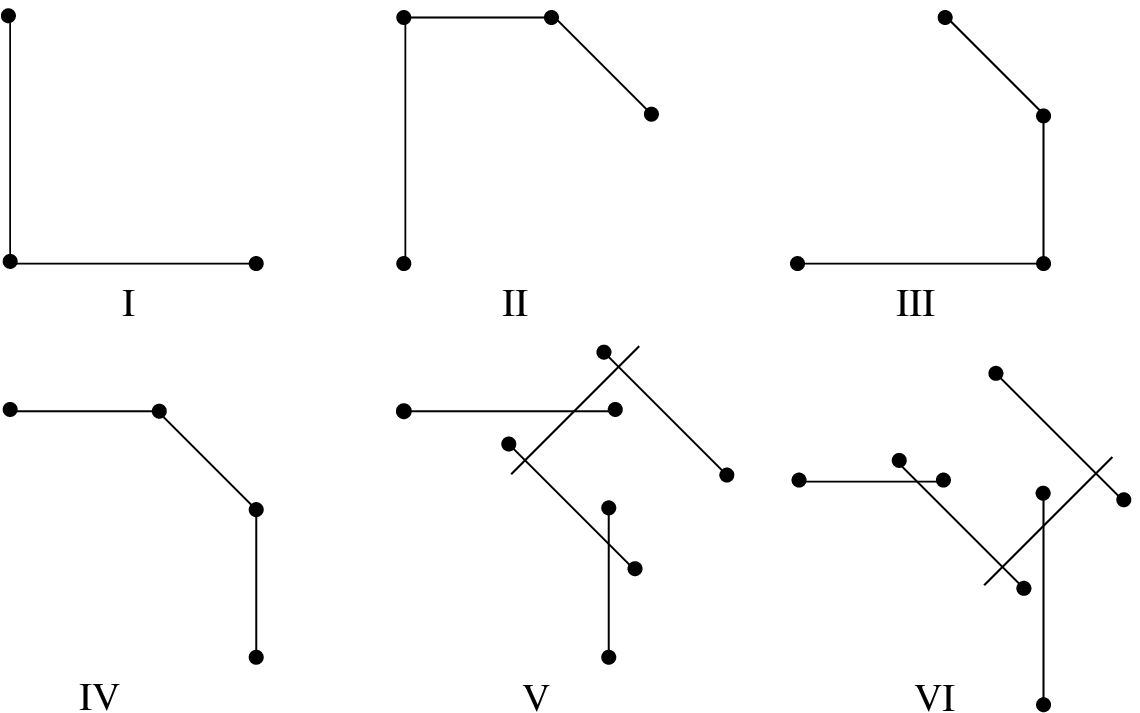}}

These graphs can be easily evaluated according to standard deformation 
theory techniques. For $z=0$, one finds that graphs ${\rm I},{\rm II}$ and ${\rm III}$ 
contribute $-1$ each while the sum of the graphs ${\rm IV},{\rm V},{\rm VI}$ is also 
$-1$. The total is $-4$  as expected. Now, we can establish a correspondence 
between these contributions and various terms of degree two in the 
open string free energy \freenC.\ The contributions of ${\rm I},{\rm II},{\rm III}$ 
correspond to the terms $-q'_1q'_2y_1^{-1}y_2^{-1}, -q'_1q_cy_1^{-1}$ 
and respectively $-q'_2q_cy_2^{-1}$. The first term arises in the 
Chern-Simons evaluation of the correction $q'_1q'_2 \Tr V_1 \Tr V_2$ 
associated to the pinched cylinder. The next two terms appear in the 
Chern-Simons evaluation of the connected expectation values
\eqn\linksA{\eqalign{
& (-i)q'_1q_c\left[\left\langle \Tr V'_1 \Tr U'_1 \right\rangle - 
\left\langle \Tr V'_1 \right \rangle \left\langle \Tr U'_1 \right \rangle\right] 
\left\langle \Tr U'_2 \right\rangle \cr
& (-i)q'_2q_c\left[\left\langle \Tr V'_2 \Tr U'_2\right\rangle - 
\left\langle \Tr V'_2 \right \rangle \left\langle \Tr U'_2 \right \rangle\right] 
\left\langle \Tr U'_1 \right\rangle \cr}}
of the links $(\Gamma'_1, \Xi'_1)$, and respectively $(\Gamma'_2, \Xi'_2)$. 
Note that there is a clear geometric interpretation of this correspondence
since each invariant primitive open string instanton corresponds to an 
invariant primitive closed string instanton. According to this rule, we would 
expect the sum over the last three graphs to be associated to a term in the 
Chern-Simons expansion of the following annulus instanton corrections 
\eqn\degtwoB{
{q_c^2\over 2} \left[ \left\langle \Tr U'^2_1\right \rangle 
\left\langle  \Tr U'^2_2 \right\rangle + \half \left \langle (\Tr U'_1)^2 
\right\rangle \left\langle (\Tr U'_2)^2\right\rangle - 
\half   \left \langle \Tr U'_1 \right\rangle^2 
\left\langle \Tr U'_2\right\rangle^2\right].}
This is indeed true if we use the instanton series \instcorrH.\ 
The relevant term is $-q_c^2$ in \degtwoA.\ 
However if we use \instcorrI\ there is no such term in the
expression of the free energy. Instead we find that 
the term $-q'_1q'_2y_1^{-1}y_2^{-1}$ 
appears with coefficient $2$, so that the final answer in closed string 
variables is the same. Moreover, similar phenomena can be noticed 
for higher degree terms in the expansion. This suggests a nontrivial 
symmetry in the graph representation of Gromov-Witten invariants of 
the local $dP_2$ model. Perhaps this symmetry can be better understood 
is we regard the $dP_2$ as the one point blow-up of a Hirzebruch $\IF_0$ 
surface. Obviously, the Kontsevich graphs of the local $\IF_0$ 
model are symmetric with respect to reflections with respect to the 
diagonal (see fig. 4). 

\ifig\Fgraphs{$F_0$ and $dP_2$ discriminant loci.}{\epsfxsize2.0in\epsfbox{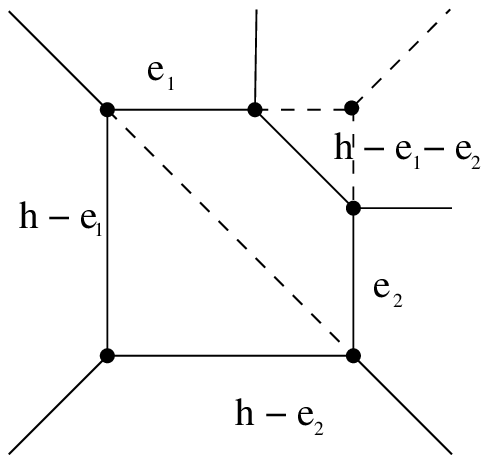}}

The above observation suggests that this symmetry is somehow present 
in a hidden form in the graph expansion of the local $dP_2$ 
model, at least for a specific choice of weights. These issues 
deserve a more detailed investigation since they are very likely related 
to localization of the integral invariants of Gopakumar and Vafa. 

To summarize this discussion, we conclude that both systems of 
Chern-Simons theories represent sensible large $N$ dual models 
for the local $dP_2$ geometry. However, the model specified by 
\instcorrH\ is preferable from a conceptual point of view since 
it was obtained as a result of a rather general set of rules. 
Essentially, the only ambiguities left in this construction 
are related to the sign of various open string corrections, which 
cannot be fixed rigorously at the present stage. 
To illustrate the general character of this approach, let us 
briefly consider other local transitions. 

\subsec{Local $dP_3$ and $dP_5$ transitions} 

The remaining local models can be analyzed along the same lines. 
Given the deformed hypersurface equations 
\eqn\hypereqG{\eqalign{
&dP_3:\quad U Z_1Z_4 + V Z_2 Z_3 = \mu, \cr
&dP_5:\quad U Z_1Z_2 + VZ_3Z_4 =\mu,\cr}}
we can find again the vanishing cycles for each model. For local 
$dP_3$ we find two spheres $L_1, L_2$ cut by the following real 
equations on $X_\mu$ 
\eqn\hypereqH{\eqalign{
& L_1: \quad UZ_2{\overline Z}_2Z_4={\overline Z}_1,\quad 
VZ_2Z_4{\overline Z}_4 = {\overline Z}_3,\cr
& L_2: \quad UZ_1{\overline Z}_1Z_3={\overline Z}_2,\quad 
VZ_1Z_3{\overline Z}_3 ={\overline Z}_4.\cr}}
In terms of local coordinates $(x_1, y_1, u_1, v_1)$, $(x_2, y_2, u_2, v_2)$ 
centered around the points $P_1, P_2$, the equations of $L_1, L_2$ take the 
canonical form 
\eqn\hypereqI{\eqalign{
& L_1: \quad x_1u_1+y_1v_1=\mu,\quad u_1={\overline x}_1,\quad 
v_1={\overline y}_1,\cr
& L_2:\quad x_2u_2+y_2v_2=\mu,\quad u_2={\overline x}_2,\quad 
v_2={\overline y}_2.\cr}}
For local $dP_5$, we obtain similarly four spheres $L_1, L_2, L_3, L_4$, 
whose local equations are identical to \hypereqI\ when written in terms 
of coordinates centered at $P_1, P_2, P_3, P_4$. The resulting geometry 
is represented in fig. 5 below. 

In order to proceed with the localization computations, we have to define 
suitable torus actions on the deformed hypersurfaces $X_\mu$, which 
preserve the vanishing cycles. It turns out that we can choose the same 
form of the 
action for both local models 
\eqn\toractB{\eqalign{
\matrix{ & Z_1 & Z_2 & Z_3 & Z_4 & U & V \cr
S^1& \lambda_1 &  0 & \lambda_3 & 0 & -\lambda_1& -\lambda_3.\cr}}}
The next step is to identify the invariant open string Riemann surfaces 
in $X_\mu$ with boundary components contained in the 3-spheres.
This can be done using the same techniques as in the previous subsection. 
Namely, we consider projective completions ${\overline X}_\mu$, and 
identify all $S^1$-invariant curves thereof which intersect the
vanishing cycles. This results in a collection of invariant discs and 
annuli embedded in ${\overline X}_\mu$. For the local transitions we have 
to keep only those surfaces which do not have points at infinity, as before. 
We will not repeat all the details of this algorithm, since it is very 
similar to the local $dP_2$ example. Moreover, we will study the compact 
version of the $dP_3$ model in great detail in the next section. 
The resulting configurations 
are represented below. 

\ifig\dpthreedisksancylinders{Primitive open string instantons on $X$.}
{\epsfxsize5.5in\epsfbox{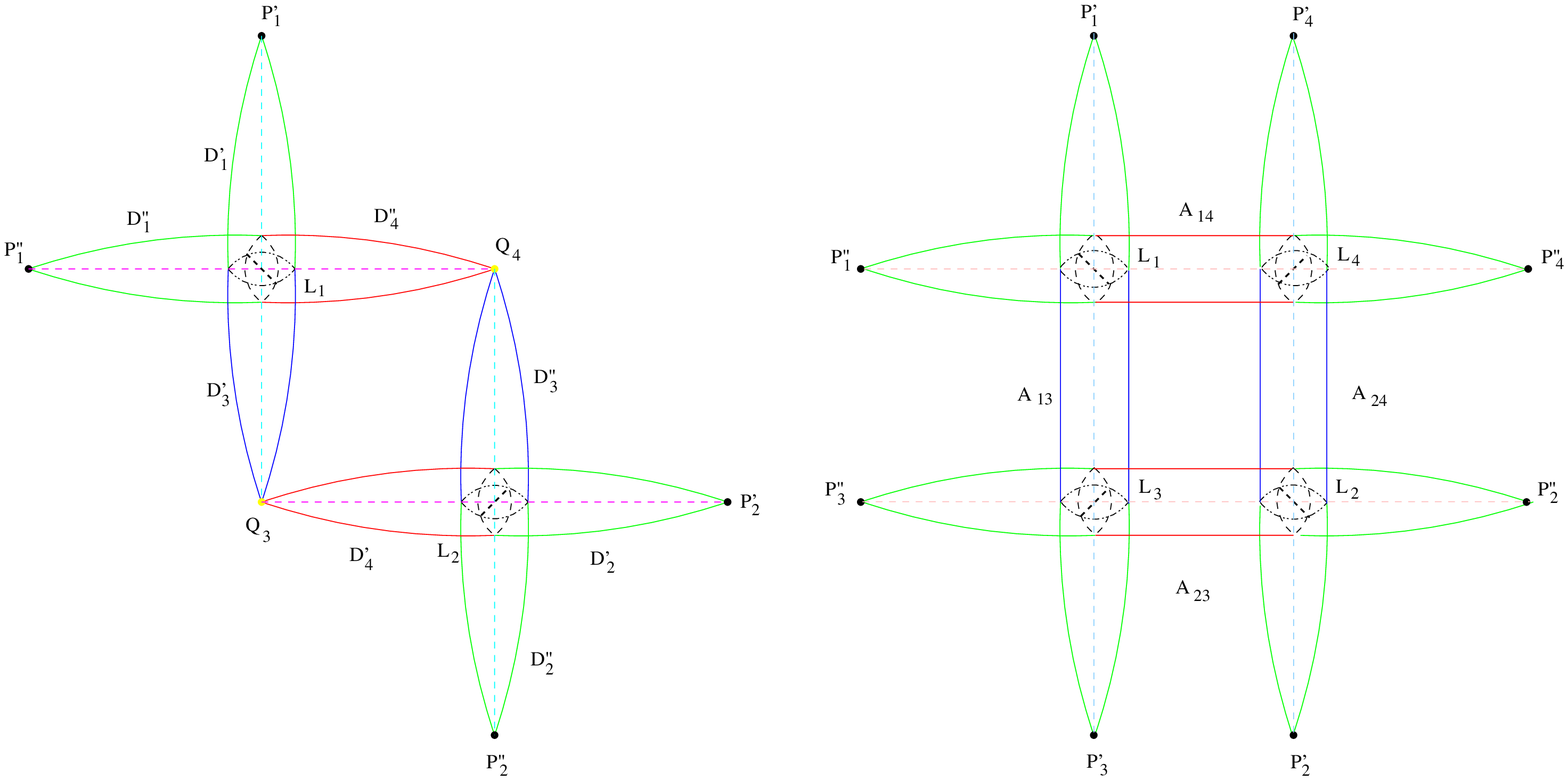}}

Note that in the $dP_3$ case the open string maps localize on a collection 
of four discs $D'_3, D'_4, D''_3, D''_4$ so that $D'_3, D''_4$ end on $L_1$, 
whereas $D''_3, D'_4$ end on $L_2$. $D'_3, D'_4$ have common origin 
$Q_3=\{Z_1=Z_4=U=0,\ VZ_2Z_3=\mu \}$ and $D''_3,D''_4$ have common origin 
$Q_4=\{Z_2=Z_3=V=0,\ UZ_1Z_4=\mu \}$. We will denote the boundary components of $D'_3, D'_4, D''_3, D''_4$ 
by $\Gamma'_3, \Gamma'_4, \Gamma''_3, \Gamma''_4 $. A short local computation shows that $(\Gamma'_3, \Gamma''_4)$ 
and respectively $(\Gamma''_3,\Gamma'_4)$ form algebraic Hopf links in $L_1, L_2$ 
with linking number $l=+1$. We will also denote the holonomy variables 
associated to the four link components by $V'_3,V'_4,V''_3,V''_4$. In the $dP_5$ example, the open string maps 
localize on a collection of 
four annuli $A_{13}, A_{14}, A_{23}, A_{24}$. The boundary components 
of each pair of annuli ending on a given sphere form again an algebraic 
Hopf link with linking number $+1$. Let us denote by $V'_1,\ldots, V''_4$
the holonomy variables associated to the eight boundary components, as shown in \dpthreedisksancylinders . 

Next, we have to compute the instanton corrections $F_{inst}$ for these 
configurations. To keep the exposition simple, for the $dP_3$ model
we will only write down genus zero corrections up to degree two. For the $dP_5$ model we can write down a closed 
form expression 
for all instanton corrections. The results of localization computations can be written as rational functions 
of $z={\lambda_3/\lambda_1}$. 
We find the following expansions. 

\noindent{Local $dP_3$:}
\eqn\dpthree{\eqalign{
F_{inst} = & -{iq'_3 \over g_s}\Tr V'_3 -{iq''_3\over g_s}  \Tr V''_3-{iq'_4\over g_s}\Tr V'_4
-{iq''_4\over g_s}\Tr V''_4 
+{iq'^2_3\over g_s}{(2-z)\over 4z}\Tr V'^2_3 
+ {i{q''^2_3}\over g_s}{(2-z)\over 4z}\Tr V''^2_3\cr
& + {iq'^2_4\over q_s}{2z-1\over 4}
\Tr V'^2_4+ {iq''^2_4\over q_s}{2z-1\over 4} \Tr V''^2_4  - q'^2_3 {z-1\over 4z^2}(\Tr V'_3)^2  -q''^2_3 {z-1\over 4z^2} 
(\Tr V''_3 )^2\cr
& -q'^2_4 {z(1-z)\over 4} (\Tr V'_4)^2-q''^2_4{z(1-z)\over 4}(\Tr V''_4)^2 
+ q'_3q'_4\Tr V'_3\Tr V'_4  +q''_3q''_4\Tr V''_3 \Tr V''_4 \cr}}
where $q'_3,q''_3,q'_4,q''_4$ are open string K\"ahler moduli 
associated to the four discs. 

\noindent{Local $dP_5$:}
\eqn\dpfive{\eqalign{
F_{inst} = & \sum_{n=1}^\infty \left({q_{13}^n\over n} \Tr V'^n_1 \Tr V'^n_3 
+ {q_{14}^n\over n} \Tr V''^n_1 \Tr V''^n_4 + {q_{23}^n \over n} \Tr V''^n_2 \Tr V''^n_3+{q_{24}^n\over n} 
\Tr V'^n_2 \Tr V'^n_4 \right)\cr}}
where $q_{13},q_{14},q_{23},q_{24}$ are open string K\"ahler moduli 
associated to the four cylinders. Note that in both expression we made particular choices of signs 
for the open string computations. For the $dP_3$ model, the correct choice
involves an alternating sign in the multicover contributions 
of a single disc. Since all open string invariant maps can be reduced to discs and closed 
string components using a normalization sequence, this choice 
fixes the sign ambiguity for all localization contributions. 
Note also that the choice made 
for the annulus corrections in the $dP_5$ example is different from 
the corresponding choice in the $dP_2$ case. We will discuss a 
potential explanation for this discrepancy after fixing the framing, 
which is the last piece of the puzzle.

In order to determine the framing for all boundary components, we 
will proceed by analogy with local $dP_2$ example. Namely, we will 
impose $S^1$-invariance for all sections to the normal bundles 
to the knots. After a straightforward computation we find the following 
values 
\eqn\framD{\eqalign{
& dP_3 :\quad p'_3= p''_3 = {1\over z}, \quad p'_4=p''_4 = z\cr
& dP_5:\quad p'_1=p'_2={1\over z},\quad p'_3=p'_4 =-{1\over z},\quad p''_1=p''_2 =z,\quad p''_3=p''_4=-z. \cr}}
As discussed before, the framings are to be thought of as formal 
variables. In order to perform the Chern-Simons functional integral 
we have to expand all expectation values of knots and links 
as formal power series in $g_s$. Following the same steps as 
in the $dP_2$ example, one can show that the resulting free 
energy agrees with the closed string expansions 
of the local 
$dP_3,~dP_5$ models.\foot{There are some subtleties 
with the closed string instanton expansion for $dP_5$ 
related to the fact that the toric model is not generic.} In particular the final answer 
is independent of $z$, without making any further choices. 

Several remarks are in order here. The $dP_5$ example has been 
considered before in \AMV.\ The dual Chern-Simons theory found there 
is formally identical to \dpfive,\ except that the framing is 
integer valued. However, it was also shown there that for 
annulus corrections, the result depends only on the effective 
framing, which is defined as the sum of the framing assigned 
to the two boundary components. In our case, we have to treat 
the framing as a formal variable, but the conclusion turns out 
to be the same. That is the final result depends only on the 
effective framing, which is an integer in all cases. 
For the annulus in the $dP_2$ model the effective framing is 
$q_1+q_2=1$, while for all annuli in the $dP_5$ model, the effective 
framing is zero. These values are in agreement with the general prescription 
given in \AMV,\ except that our algorithm is different and it applies 
equally well to invariant discs. In the later case, treating the 
framing as a formal variable is crucial since the corrections themselves 
have a similar dependence. 

The effective framing of a cylinder has a simple 
geometric dependence, which can be described as follows. 
Let us note that each invariant open 
string surface on $X_\mu$ corresponds to an invariant rational 
curve on the 
resolution ${\widetilde X}$. Using this correspondence, we can find a
correlation between the effective framing of a cylinder and the 
normal bundle of the associated curve on ${\widetilde X}$.
Suppose an annulus $A$ corresponds to a curve $C$, and 
let $S\subset X$ denote the image of the zero section. 
The normal bundle $N_{C/X}$ is an extension of $i_C^*N_{S/X}$ by 
$N_{C/S}$ which is typically split. It can be easily seen that 
the effective framing of $A$ is given by $\hbox{deg}(i_C^*N_{S/X})$. 
Presumably, this rule is consistent with the general prescription 
of \AMV,\ which is formulated in a different language.

Finally, there is one more aspect that deserves a few comments. 
We noted above that the sign ambiguity was given different resolutions 
in the $dP_2$ and respectively $dP_5$ models. Note that 
a concise rule encompassing all choices is to take the sign of 
a term of degree $n$ to be $(-1)^{np}$ 
where $p$ is the effective framing of the cylinder. This rule is very likely 
connected to a subtle sign problem found in a similar context in 
\MV.\ There it was found that the 
holonomy variable for a knot had to be redefined by a 
sign $(-1)^p$ in order to obtain the desired integrality properties. 
We believe that these problems are related, but we will not discuss 
this aspect further here.

In conclusion, the main idea developed in this section is that the 
proper framework for open string amplitudes is a formal extension 
of Chern-Simons theory which allows the framing to become fractional. 
Within this formalism we have found a natural relation between 
framing and the torus weights which can be uniformly applied to 
local transitions. In the next sections, we will extend the analysis 
of this section to transitions between compact Calabi-Yau manifolds.

\newsec{Localization of Open String Morphisms to Compact Threefolds} 

After this rather long digression we return to open string 
amplitudes on compact Calabi-Yau target spaces. We outlined the 
main principles of the approach in section three. 
In this section we will carry out this program in detail 
for one of the models introduced in section four. 
For concreteness, we focus on the compact transition 
based on the local $dP_3$ model. The other models admit 
a similar treatment.

To fix ideas let us recall the 
toric data of the ambient toric variety $\CZ$ 
\eqn\toricA{ 
\matrix{ & Z_1 & Z_2 & Z_3 & Z_4 & U & V & W \cr
\IC^* & 1 & 1 & 0 & 0 & -1 & -1 & 0 \cr
\IC^* & 0 & 0 & 1 & 1 & -1 & -1 & 0 \cr
\IC^* & 0 & 0 & 0 & 0 & 1 & 1 & 1\cr}}
and the generic Calabi-Yau hypersurface 
\eqn\defeqnA{
\sum_{a,b,c\geq 0, a+b+c=3} U^aV^bW^cf_{abc}(Z_i)=0.}
Here $f_{abc}(Z_i)$ are bihomogeneous polynomials of $(Z_1,Z_2)$ 
and respectively $(Z_3,Z_4)$ of bidegree 
$(a+b, a+b)$. 
We consider a one parameter family of hypersurfaces $Y_\mu$ defined by 
\eqn\defeqnB{
(UZ_1Z_4+VZ_2Z_3 -\mu W)W^2 + \sum_{(a,b,c)'} U^aV^bW^cf_{abc}(Z_i)=0}
where the triples $(a,b,c)'$ take all allowed values except 
$(1,0,2),(0,1,2),(0,0,3)$. The coefficients of 
$f_{abc}(Z_i)$ are fixed to some generic values. One can check that 
\defeqnB\ is smooth for generic $\mu\neq 0$ and develops two 
ordinary double points 
\eqn\singA{
P_1:\ Z_1=Z_3=U=V=0,\qquad 
P_2:\ Z_2=Z_4=U=V=0} 
at $\mu=0$. As discussed in section four, the two conifold singularities 
can be simultaneously
resolved by performing a toric blow-up of $Z$ along the section $U=V=0$. 
The strict transform $\wY$ is a crepant resolution of $Y_0$. The 
exceptional locus consists of two homologous curves $C_1, C_2$. 

The smooth fiber $Y_\mu$, $\mu \neq 0$ contains two vanishing cycles 
$L_1,L_2$ in the same homology class $[L_1] = [L_2] \in H_3(Y_\mu , \IZ)$. 
The topological open string theory considered in this section is defined 
by wrapping $N_1,N_2$ {\bf A}-branes on $L_1,L_2$. 
Then the charge constraint \chargerelB\ yields $N_1+N_2=0$. This 
means the D-brane configuration should be interpreted as a brane/anti-brane 
system. We have $N=N_1$ branes on $L_1$ and $N$ anti-branes on $L_2$, 
which can be thought of as $N$ branes wrapping the cycle $L_2$ with 
opposite orientation. This is a particular case of our general discussion 
in which $v=1$ and $r=1$, hence the lattice of vanishing 
cycles has rank one. The formula \openA\ for the genus zero 
topological open string 
free energy takes the form 
\eqn\openB{
\CF_{(Y,L);op}^{(0)}(g_s,t_\alpha,\lambda) = 
\CF_{Y;cl}^{(0)}(g_s,t_{\alpha})+\sum_{\beta \in H_2(Z,L)} 
F_{\beta }^{(0)}(g_s,\lambda) e^{-<J,\beta>}}
where $\lambda = Ng_s$ and $t_\alpha$, $\alpha=1,2,3$ are the 
K\"ahler moduli of $Y$. 
In order to compute the coefficients $F_{\beta }^{(0)}(g_s,\lambda)$ we have 
to follow the steps outlined in section three. 


Consider the following $G=(S^1)^7$ action on $\CZ$ 
\eqn\toractA{
(e^{i\phi_1},\ldots, e^{i\phi_7})\cdot (Z_1,Z_2,\ldots,W)\ra 
(e^{i\phi_1}Z_1,e^{i\phi_2}Z_2,\ldots, e^{i\phi_7} W).}
As explained in section three, one of the main problems is that the 
cycles $L_1, L_2$ are not preserved by the torus action \toractA.\  
The solution is to specialize the triple $(Y,L_1, L_2)$ 
to a degenerate hypersurface $\oY$ and two cycles 
${\overline L}_1, {\overline 
L}_2$ which are preserved by a certain subtorus. In general, 
it is not clear 
that such a limit in the complex structure moduli space always 
exists. However, 
for all the models discussed in section four, a suitable degeneration 
immediately presents itself. Recall that in the local context of the previous 
section, we had to take a projective completion ${\overline X}_\mu$ of 
$X_\mu$, which was a toric Fano threefold. For the $dP_3$ example, 
the equation of ${\overline X}_\mu$ is 
\eqn\projcompl{
UZ_1Z_4+VZ_2Z_3=\mu W.} 
Moreover, $\oX_\mu$ admits a torus action which preserves the vanishing 
cycles. The degeneration we are looking for is the reducible nonreduced 
hypersurface $\oY$ defined by 
\eqn\defeqnC{
(UZ_1Z_4+VZ_2Z_3 -\mu W)W^2=0.}
$\oY$ has two components $\oY_1=\oX_\mu$, and $\oY_2$ defined by 
$W^2=0$. The cycles $\oL_1, \oL_2$ are the vanishing cycles defined 
in equation \hypereqH,\ which are obviously embedded in $\oY$. 
By construction, the triple $(\oY,\oL_1,\oL_2)$ is preserved by a
subtorus $(S^1)^2\subset G$. 
We will give more details on the local geometry below. 

Note that similar degenerations can be found without difficulty 
for the other two models discussed in section four. Moreover, we 
believe that a suitable limit in the complex structure moduli 
space can be found for all transitions for which the singular points are 
fixed by the torus action. This is in fact the main restriction on the 
present approach to open string amplitudes on Calabi-Yau hypersurfaces.

\subsec{Local Geometry on $\CZ$}

A thorough analysis of open string maps in the present context requires 
a systematic description of the local geometry. Recall that the toric 
fourfold $\CZ$ is isomorphic to the projective bundle $\IP\left(\CO\oplus 
\CO(-1,-1)\oplus \CO(-1,-1)\right)$ over $\IF_0$,
and $U,V,W$ are relative projective coordinates. 
The torus action \toractA\ leaves 
the sections $S=\{U=V=0\}, S'=\{U=W=0\}, S''=\{V=W=0\}$ and the 
fibers $F_{1}=\{Z_1=Z_3=0\},F_2=\{Z_2=Z_4=0\}, F_3=\{Z_1=Z_4=0\}, 
F_4=\{Z_2=Z_3=0\}$ invariant. Note that $S,S',S''$ are isomorphic to $\IF_0$ 
and $F_1,\ldots, F_4$ are isomorphic to $\IP^2$. 
The fixed locus of the $G$-action on 
$\CZ$ consists of twelve isolated points $P_k,P'_k, P''_k$, $k=1,\ldots, 4$
which are intersection points of $S, S', S''$ and $F_k$. More precisely, 
$P_k= F_k\cap S, P_k'=F_k\cap S', P''_k =F_k \cap S''$, $k=1,\ldots,4$. 
For a given $k$, the points $(P_k, P_k', P''_k)$ determine three invariant 
curves ${\overline {P_kP'_k}}$, ${\overline {P_kP''_k}}$ and 
${\overline {P'_kP''_k}}$. The fixed points lying in a given section, say 
$S$, determine four invariant curves ${\overline {P_1P_3}}$, 
${\overline {P_1P_4}}$, ${\overline {P_2P_3}}$,  ${\overline {P_2P_4}}$. 
The same is true for the sections $S', S''$. Overall we obtain a 
toric skeleton consisting of twenty-four invariant curves 
intersecting at twelve fixed points as in the figure below. 

\ifig\fourfoldskeleton{
Fourfold skeleton for the generic toric action. The color coding 
is the following: green = $h_1$, blue = $h_2$, red = $h_3$, cyan = $h_1+h_2$, 
magenta = $h_1+h_3$.}
{\epsfxsize3.0in\epsfbox{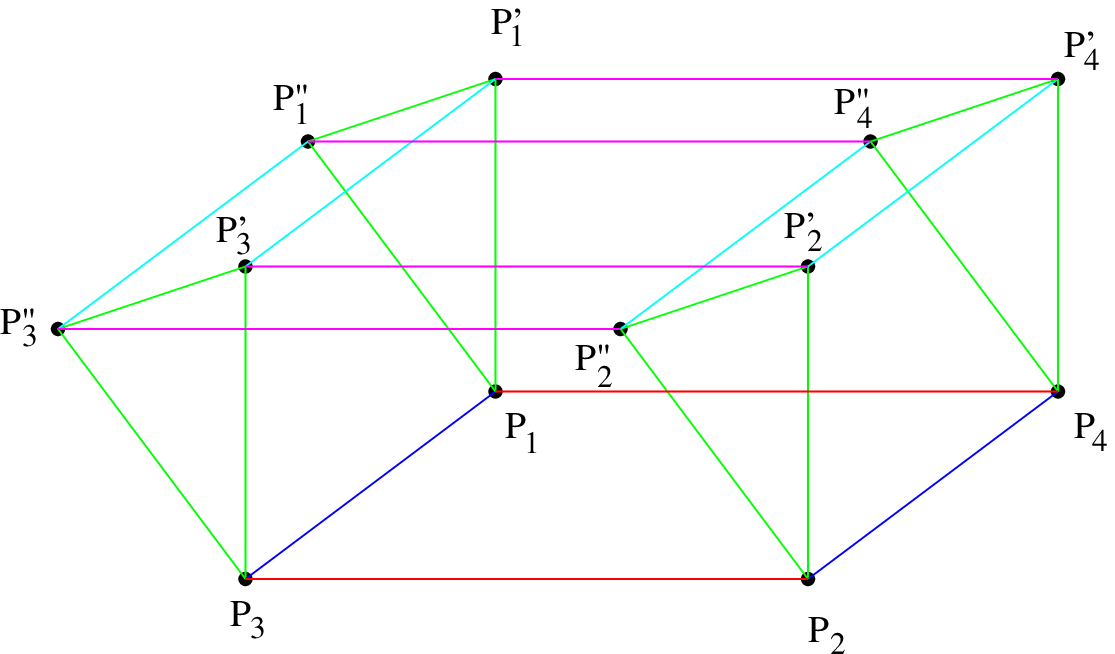}}

\noindent
For later applications, we have introduced a color coding for 
the homology classes of curves on $\CZ$ as follows.  
We choose a basis in $H_2(\CZ,\IZ)$ consisting of the Mori cone 
generators $(h_1,h_2,h_3)$. Using equation (A.1) in appendix A, 
one can check that $(h_1, h_2, h_3)$ are represented by the following 
curves on $\CZ$ 
\eqn\MorigenA{
h_1:\ Z_1=Z_3=U=0,\qquad 
h_2:\ Z_1=U=V=0,\qquad h_3:\ Z_3=U=V=0.}
By construction, $h_1$ is the hyperplane class of a $\IP^2$ fiber of 
$\CZ$ over $\IF_0$ while $h_2, h_3$ are $(1,0)$ and respectively $(0,1)$ 
classes of the zero section $U=V=0$. 
For convenience, we will call $h_1$ vertical class, 
and $h_2, h_3$ horizontal classes. 
In fig. 6, green line segments represent curves in the class $h_1$, 
blue line segments represent curves in the class 
$h_2$ and red line segments represent curves in the class $h_3$. 
We also have curves in mixed classes $h_1+h_2$ represented by cyan 
line segments and $h_1+h_3$ represented by magenta line segments. 
This color coding will also be used later for invariant discs 
in $\CZ$. 

This skeleton plays an important part in the localization computation of 
closed string Gromov-Witten invariants. In the following we will 
explain that it also plays an important role in the localization 
computation of open string instanton corrections. 

In order to analyze deformations of open string maps we will also need 
local coordinates  $(x_k,y_k,u_k,v_k)$, $(x'_k,y'_k,u'_k,v'_k)$,
$(x''_k,y''_k,u''_k,v''_k)$, $k=1,\ldots,4$ centered at each fixed point. 
We will adopt the following conventions: $(x_k,y_k), (x'_k,y'_k), 
(x''_k, y''_k)$ are horizontal coordinates along the sections $S,S',S''$ 
while $(u_k,v_k), (u'_k,v'_k), (u''_k, v''_k)$ are vertical coordinates 
along the fibers $F_k$, $k=1,\ldots,4$. 
For concreteness, let us write down the affine open subsets 
and local coordinates centered around $P_1,P_1',P_1''$ 
\eqn\loccordA{\eqalign{
& \CU_1 = \{ Z_2\neq 0, Z_4\neq 0, W\neq 0\}\cr
& \CU_1'=\{ Z_2\neq 0, Z_4 \neq 0, V\neq 0\} \cr
&\CU''_1=\{ Z_2\neq 0, Z_4 \neq 0, U\neq 0\}\cr}}
\eqn\loccoordB{\vbox{\halign{ $#$ \hfill &\qquad  $#$ \hfill &\qquad  
$#$ \hfill &\qquad  $#$ \hfill &\qquad  $#$ \hfill\cr
\CU_1: & x_1 = {Z_1\over Z_2}, & y_1={Z_3\over Z_4}, &  u_1={UZ_2Z_4\over W}, 
& v_1={VZ_2Z_4\over W}\cr
\CU_1': & x'_1={Z_1\over Z_2}, & y'_1={Z_3\over Z_4}, & u'_1={U\over V}, 
& v'_1={W\over VZ_2Z_4}\cr
\CU''_1: & x''_1={Z_1\over Z_2}, & y''_1={Z_3\over Z_4}, & u''_1={V\over U}, 
& v''_1={W\over UZ_2Z_4}.\cr}}}
The other coordinates can be obtained by permuting the indices. 
The local equations of $\oY_1$ in these coordinate patches are 
\eqn\loceqnA{\vbox{\halign{ $#$ \hfill &\qquad  $#$ \hfill\cr
\CU_1: & x_1u_1+y_1v_1=\mu\cr
\CU'_1: &  x'_1u'_1+y'_1= \mu v'_1\cr
\CU''_1: & x''_1 + y''_1v'_1 =  \mu u''_1.\cr}}}
The cycles $\oL_1, \oL_2$ on $\oY_1$ are given by the following local 
equations 
\eqn\loceqnB{\vbox{\halign{ $#$ \hfill &\qquad  $#$ \hfill &\qquad  
$#$ \hfill\cr
\oL_1: & u_1={\overline x}_1, & v_1={\overline y}_1\cr
\oL_2: & u_2={\overline x}_2, & v_2={\overline y}_2.\cr}}}
Given these local equations, 
one can deform the symplectic K\"ahler structure on $\by_1$ so that 
the cycles $\oL_1, \oL_2$ are lagrangian. The details are explained 
in appendix A of \DFGii.\ Therefore, the topological open
string A model is well defined, at least in this limit.

One can check that the triple $(\oY,\oL_1, \oL_2)$ is preserved by
a subtorus $(S^1)^2\subset G$. In fact for localization computations, 
it suffices to consider a one parameter subgroup $T\subset G$ defined by 
\eqn\subtorA{
(e^{i\phi_1},e^{i\phi_2},\ldots, e^{i\phi_7}) = 
(e^{i\lambda_1\phi},1,e^{i\lambda_3\phi},1,e^{-i\lambda_1\phi},
e^{-i\lambda_3\phi},1).}
In the next subsection we will determine the structure of $T$-invariant 
open string morphisms with boundary conditions on $\oL_1\cup \oL_2$.

\subsec{The Fixed Loci} 

At a first look, one would like to sum over open string maps to the 
singular hypersurface $\oY$. This is however a very difficult task since 
the degeneration considered here is not semistable. Therefore we do not have
a good control of intersection theory even on the moduli space of stable 
closed string maps. Instead, it is more convenient to consider open string 
maps to the ambient toric variety $\CZ$, as explained in section three. 
In order to obtain numerical invariants, we will have to develop 
an open string version of the convex obstruction bundle. 

First note that the fixed point set of the $T$-action on $\CZ$
consists of eight isolated fixed points $P_1, P_2, P'_1, P'_2, P''_1, \ldots, 
P''_4$ and two fixed curves ${\overline {P_3P'_3}}$ and 
${\overline {P_4'P''_4}}$. The presence of fixed curves in the target 
space will cause some complications at a certain point in our analysis. 

Now, let $f:\Sigma_{0,h}\ra \CZ$ be a genus zero 
open string stable morphism 
which sends the boundary $\partial \Sigma_{0,h}$ to the cycle
$\oL = \oL_1 \cup \oL_2$. We also fix the homology class  
$\beta= f_*[\Sigma_{0,h}]\in H_2(\CZ,\IZ) \simeq H_2(\CZ,\oL;\IZ)$. 
The torus action \subtorA\ induces an action on the moduli 
space of such maps $\om_{0,h}(\CZ,\oL;\beta)$. 
Our goal is to determine the structure of fixed loci for this action.
By ${T}$-invariance, the domain of such 
a map must be either $i)$ an annulus, or  
$ii)$ a nodal bordered Riemann surface  
$\Sigma_{0,h} = \Sigma_0 \cup \Delta_1 \cup\ldots \cup \Delta_h$ 
where $\Sigma_0$ is a pre-stable curve of genus $0$ and $\Delta_1, \ldots, 
\Delta_h$ are $h$ discs attached to $\Sigma_0$ by identifying the 
origins with the marked points 
$(p_1,\ldots,p_h)$. 
The data $(\Sigma_0,p_1,\ldots,p_h)$ defines a prestable marked curve 
of genus $0$. 

Obviously, the first case can be realized only if $h=2$. 
Then the image of $f$ should be an invariant holomorphic annulus 
embedded in $\CZ$ with boundary on $\oL$. 
 We will show in the following that there are no 
such annuli in the present example. In the second case, 
the morphism $f$ should map $\Sigma_0$ to a curve in $\CZ$ 
which is preserved by 
the subtorus action \subtorA.\ The disc components should be mapped to 
invariant discs embedded in $\CZ$ with boundary on $\oL$. 
One must also impose a stability constraint on $(f,\Sigma_{0,h})$ 
which makes the automorphism group finite. 

It is clear that we should start by identifying the invariant 
annuli and discs in $\CZ$ with boundary on $\oL$. 
Using an argument similar to \DFGi\ (section 5), 
one can show that any invariant embedding $f:\Sigma_{0,h}\ra \CZ$ 
can be extended to an invariant embedding 
${\overline f}:{\overline \Sigma} \ra \CZ$, 
where $\Sigma$ is a closed Riemann surface without boundary. 
Moreover, since we are only interested in discs or annuli, $\overline \Sigma$ 
will be a smooth rational curve. 
Therefore 
we have to find all ${T}$-invariant smooth rational 
curves on $\CZ$ which intersect the 
cycles $\oL_1, \oL_2$ along orbits of the torus action. Since 
$\oL_1, \oL_2$ 
are cycles in $\oY_1$, it follows that any such curve 
must be in fact contained in $\oY_1$. This reduces the analysis to 
the local case considered briefly in the previous section. 
Let us give more details here. 

Since $\oY_1$ is a toric Fano threefold, this 
is a simple question which has been addressed in similar situations in 
\refs{\DFGi,\DFGii}. 
In the present case we find the curves 
\eqn\curvesA{\vbox{\halign{ $#$ \hfill &\qquad  $#$ \hfill &\qquad  
$#$ \hfill\cr
C'_{13}: & Z_1=U=0, & VZ_2Z_3=\mu W\cr
C''_{14}: & Z_3=V=0, & UZ_1Z_4=\mu W\cr
C''_{23}: & Z_2=V=0, & UZ_1Z_4=\mu W\cr
C'_{24}: & Z_4=U=0, & VZ_2Z_3=\mu W\cr}}}
which intersect the cycles $\oL_1,\oL_2$ along invariant circles 
as follows 
\eqn\intersA{\eqalign{
&  \Gamma'_1\equiv
C'_{13}\cap \oL_1=\{x_1=u_1=0,\ |y_1|=|v_1|=\mu^{1/2}\},\qquad 
C'_{13}\cap \oL_2=\emptyset \cr
& \Gamma''_1\equiv C''_{14}
\cap \oL_1 = \{|x_1|=|u_1|=\mu^{1/2},\ y_1=v_1=0\},\qquad 
C''_{14}\cap \oL_2=\emptyset \cr
& C''_{23}\cap \oL_1=\emptyset, \qquad 
\Gamma''_2\equiv C''_{23}\cap \oL_2=\{|x_2|=|u_2|=\mu^{1/2},\ y_2=v_2=0\}\cr
& C'_{24} \cap \oL_1=\emptyset,\qquad 
\Gamma'_2\equiv 
C'_{24}\cap \oL_2=\{x_2=u_2=0,\ |y_2|=|v_2|=\mu^{1/2}\}.\cr}}
Note that all circles are algebraic 
knots with orientation induced 
by the canonical orientation of $\CZ$. Moreover $\Gamma_1,\Gamma_1'$, 
respectively  
$\Gamma'_2,\Gamma''_2$ form Hopf links in 
$\oL_1, \oL_2$ with linking number $+1$. 
For later use, let us introduce the holonomy variables 
\eqn\holvarA{
V'_i = \hbox{Pexp} \int_{\Gamma'_i} A^{(i)}, \ i=1,2 ,\qquad 
V''_i = \hbox{Pexp} \int_{\Gamma''_i} A^{(i)}, \ i=1,2}
where $A^{(1)}, A^{(2)}$ are the Chern-Simons gauge field 
fields on $\oL_1, \oL_2$.
Note that the curves $C'_{13},\ldots,C'_{24}$ are invariant under the 
$T$-action on $\CZ$, but not under the generic $G$-action. 

Each knot divides one of the curves \curvesA\ 
into two invariant discs embedded in $\CZ$ with boundary on $\oL_1$ 
or $\oL_2$ as follows 
\eqn\discsAA{\eqalign{
& C'_{13} = D'_1 \cup_{\Gamma'_1} D'_3\cr
& C''_{14} = D''_1\cup_{\Gamma''_1} D''_4 \cr
& C''_{23} = D''_2 \cup_{\Gamma''_2} D''_3\cr
& C'_{24} = D_2'\cup_{\Gamma'_2} D'_4.\cr}}
Note that $D'_1$, $D''_1$, $D'_2, D''_2$ intersect the toric skeleton 
at the fixed points $P'_1, P''_1$, $P'_2, P''_2$ lying on the 
divisor at infinity $\zeta_\infty$. The discs $D_3', D_3''$ and respectively 
$D_4', D''_4$ have common origins $Q_3, Q_4$ given by 
\eqn\discsC{\eqalign{
& D_3'\cap D' _4 = \{ Z_1=Z_4=U=0,\ VZ_2Z_3=\mu W\}=Q_3\cr
& D_3''\cap D''_4 =\{Z_2=Z_3=V=0,\ UZ_1Z_4=\mu W\}=Q_4.\cr}}
Note that $Q_3, Q_4$ lie on the $G$-invariant lines ${\overline 
{P_3P_3'}}$ and ${\overline {P_4P_4'}}$, but they are 
not fixed points of the $G$-action. The resulting configuration of 
invariant discs is represented in fig. 7.

\ifig\disksanddisksandcurves{Discs in $\CZ$ invariant under the 
restricted torus action. The points $Q_3, Q_4$ have been represented by 
yellow dots in order to distinguish them from the 
fixed points under the generic torus action. 
The color coding for discs is the same as for curve classes: 
green = $h_1$, blue = $h_2$, 
red = $h_3$.}
{\epsfxsize4.0in\epsfbox{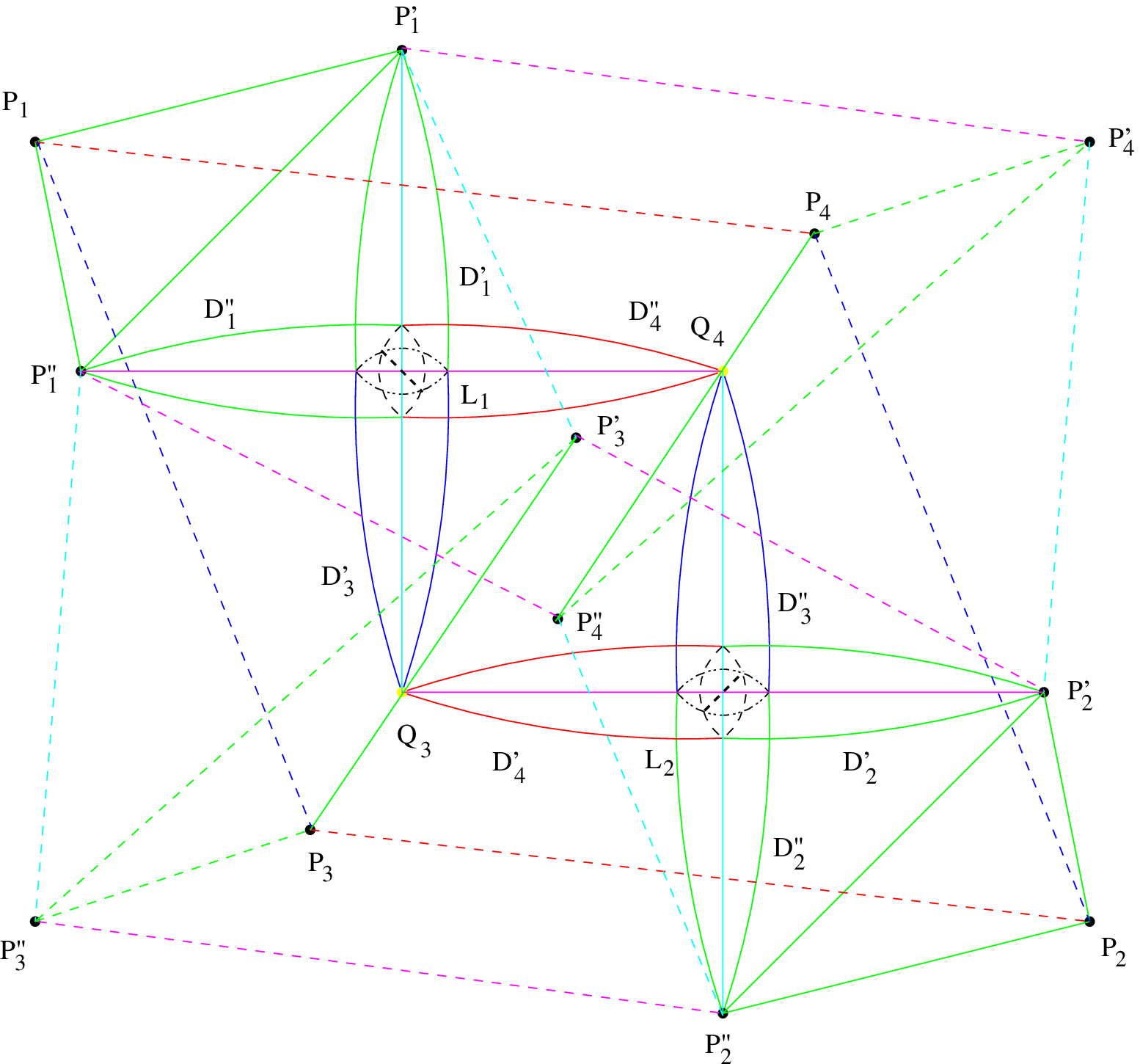}}

For future reference
let us determine the relative homology classes of the discs 
\discsAA.\ As noticed earlier, we have $H_2(\CZ,\oL;\IZ)
\simeq H_2(\CZ,\IZ)$, which is generated by the 
curve classes $h_1, h_2, h_3$ specified in \MorigenA.\ 
The homology classes of 
the discs can be determined by deforming to the singular hypersurface 
$Y_0$ as discussed in section 5 of \DFGi.\ In the present case we obtain 
\eqn\discsD{\eqalign{
& [D'_1]=[D''_1]=[D'_2]=[D''_2]=h_1,\cr
& [D'_3]=[D''_3]=h_2,\qquad [D'_4]=[D''_4]=h_3.\cr}}
Adopting the same terminology, we will call the discs $D'_1, D''_1, D'_2, 
D''_2$ vertical discs and $D'_3, D''_3, D'_4, D''_4$ horizontal discs. 
The homology classes are represented in fig. 7 using the color coding 
introduced below \MorigenA.\ 

From \discsD\ it follows that the symplectic areas of 
the discs are given at the 
classical level\foot{What is meant here is classical level
from the open string point of view. The parameters $t_1, t_2, t_3$ are 
flat coordinates on the closed string K\"ahler 
moduli space.} 
by the closed string K\"ahler parameters 
$t_1, t_2, t_3$. 
If open string quantum effects are taken into account, the 
symplectic area is shifted by a multiple of $\lambda/2$ (recall 
that $\lambda$ is the 't Hooft coupling constant) as 
discussed in \DFGi.\ Therefore we introduce the corrected 
K\"ahler parameters $\tau'_i$, $\tau''_i$, $i=1,\ldots, 4$ for 
the eight discs. The relation between $\tau'_i$, $\tau''_i$, 
and the closed string K\"ahler parameters on the small resolution $\wY$ 
is part of the prescriptions of the duality map, and 
will be discussed later. 

Having determined all invariant discs in $\CZ$, we can now understand the 
structure of a more general invariant map $f:\Sigma_{0,h}\ra \CZ$. 

As stated above the domain $\Sigma_{0,h}$ is the union of a prestable 
curve $\Sigma_0$ of genus zero with $h$ marked points $p_1,\ldots, p_h$ 
and $h$ discs $\Delta_1,\ldots, \Delta_h$ attached to 
$\Sigma_0$ at $p_1,\ldots, p_h$. 
Each disc $\Delta_a$ is mapped 
$d_a:1$ to one of the eight discs found above. 
The closed curve $\Sigma_0$ 
is mapped to a genus zero curve in $\CZ$ which is preserved by the 
restricted torus action \subtorA.\
Since we are working with the 
restricted torus action, there may be nontrivial families of invariant 
curves on $\CZ$. Therefore the connected components of the fixed loci 
may have a complicated structure, as discussed in more detail below. 

We are interested in computing all open string 
instanton corrections for a fixed relative homology class 
$\beta\in H_2(Y,L;\IZ)\simeq H_2(\CZ,\oL;\IZ)$.
Therefore we have to sum over all fixed loci 
in $\om_{0,h}(\CZ,\oL;\beta)$ which are compatible with the given homology 
class $\beta$. In order to make this sum more explicit, first we have 
to classify the fixed loci according to the images of the discs $\Delta_a$
in $\CZ$. There are eight invariant discs $D'_1,\ldots, D'_4, D''_1, \ldots, 
D''_4$ embedded in $\CZ$, and in principle any such disc can be the image of 
any component $\Delta_a$ of the domain. Let us introduce the set of 
labels $I=\{1',\ldots, 4',1'',\ldots,4''\}$ for the invariant discs in $\CZ$. 
The notation is self-explanatory. Specifying the image of each component 
$\Delta_a$ is equivalent to giving a function $\rho:\{1,\ldots,h\}\ra I$, 
and we have to sum over all such functions. The instanton series can be 
written schematically as follows 
\eqn\instcorrB{
\sum_{h=1}^\infty \sum_{(d_a,\beta)}
\sum_{\rho} C_{h,\rho}(d_a,\beta) 
e^{-\langle J, \beta-\sum_{a=1}^hd_a[D_{\rho(a)}]
\rangle } e^{-\sum_{a=1}^hd_a \tau_{\rho(a)}} \prod_{a=1}^h 
\Tr V_{\rho(a)}^{d_a}.}
Let us explain this formula in detail. The coefficients 
$C_{h,\rho}(d_a, \beta)$ 
represent the sum over all fixed loci with given $(h,\rho)$ and $(d_a,\beta)$. 
We have suppressed the genus subscript $g$ since we will exclusively consider 
genus zero maps. The exponential factors 
$e^{-\langle J, \beta-\sum_{a=1}^hd_a[D_{\rho(a)}]
\rangle } e^{-\sum_{a=1}^hd_a \tau_{\rho(a)}}$ represent the instanton factors 
associated to a given map. Naively, these factors should be written 
as $e^{-\langle J,\beta\rangle}$, where the pairing in the exponent is well 
defined since $J|_\oL=0$ by the lagrangian condition. We have modified 
this expression in order to allow for a possible shift in the open string 
K\"ahler moduli $\tau_{\rho(a)}$ with respect to the closed string parameters. 
This shift can be interpreted as an open string quantum effect and plays 
an important role in the duality map \DFGi.\ 
Therefore one has to write down separate 
expressions for the instanton factors of the open and closed string 
components, obtaining the expression in \instcorrB.\ 
The area of the closed component 
$-\langle J, \beta-\sum_{a=1}^hd_a[D_{\rho(a)}]\rangle$
must be expressed in terms of the closed string parameters 
$t_1, t_2, t_3$.
Finally, $V_{\rho(a)}$ represents the holonomy variable associated to the 
knot $\Gamma_{\rho(a)} = \partial D_{\rho(a)}$, which is defined as follows. 
If $\rho(a) \in \{1',1'',2',2''\}$, say $\rho(a)=1'$, 
$V_{\rho(a)}=V'_1$ is the holonomy of the $U(N)$ connection about the knot 
$\Gamma'_1$ defined in \holvarA.\ 
Note that the canonical orientation on
$D_1'$ induces an orientation on $\Gamma_1'$ using the inner normal 
convention. The holonomy $V_1'$ is computed with respect to this particular
orientation.
This convention is valid for all vertical discs 
$D'_1, D''_1, D'_2, D''_2$. 
If $\rho(a)\in \{3',3'',4',4''\}$, $V_{\rho(a)}= V_{i-2}^{-1}$, $i=3,4$, since
the canonical orientation on $D_3',\ldots, D_4''$ induces the opposite 
orientation on $\Gamma_1',\ldots, \Gamma_2''$ (see fig. 7).

Our main problem is to evaluate the coefficients $C_{h,\rho}(d_a,\beta)$ 
using open string localization techniques. This requires a thorough 
understanding of 
the deformation complex and the obstruction bundle 
for open string morphisms. 

\subsec{Deformation Complex and Obstruction Bundle for Open Strings 
Morphisms} 

To begin with, let us consider the definition of the instanton coefficients 
$C_{h,\rho}(d_a,\beta)$ in more detail. We stress that the following 
considerations are only heuristic arguments based on unproven assumptions. 
Essentially, we will assume that certain known results in the theory 
of closed string Gromov-Witten invariants \refs{\B,\BF,\GP,\LT}
carry over to the open string case. 
As in \refs{\GZ,\KL}, these arguments will eventually lead us to a precise 
computational definition of the enumerative invariants 
$C_{h,\rho}(d_a,\beta)$. 
The first assumption is that the 
moduli space (or, more precisely, the moduli stack) 
$\om_{0,h}(\CZ,\oL;\beta)$ exists in the appropriate category, 
and can be endowed with a perfect 
obstruction complex. Then one would have a virtual class 
$[\om_{0,h}(\CZ,\oL;\beta)]$ (of positive dimension, since $Z$ is not a 
Calabi-Yau threefold).
Following the closed string approach, we would also need an 
obstruction bundle $\CV$ on $\om_{0,h}(\CZ,\oL;\beta)$ 
of rank equal to the dimension of $[\om_{0,h}(\CZ,\oL;\beta)]$. 
Assuming these elements can be constructed, we could
define 'open string Gromov-Witten
invariants' by pairing the Euler class $e(\CV)$ with the virtual fundamental 
class. However, even in such an ideal situation the coefficients 
$C_{h,\rho}(d_a,\beta)$ could not be simply defined by this pairing
for reasons explained in section three. 
In order to write down the the couplings to Chern-Simons theory, 
one needs in fact an equivariant refinement, as discussed below. 

The holonomy $V$ about the boundary $\Gamma$ of an open string 
surface is defined with 
respect to an arbitrary (not necessarily flat) $U(N)$ connection. 
Therefore it depends continuously on the position of the boundary $\Gamma$
in the cycle $\oL$.  
In formula \instcorrB\ the $V_i$ are holonomy variables about the 
boundaries $\Gamma_i$ of the ${T}$-invariant discs in $\CZ$. 
Hence the torus action on $(\CZ,\oL)$ enters in a crucial way.
This suggests that the appropriate refinement of open string 
Gromov-Witten invariants should 
be considered in an equivariant setting. 
The torus action on $(\CZ,\oL)$ induces a torus action on 
$\om_{0,h}(\CZ,\oL;\beta)$. Under the above assumptions, we can define 
equivariant open string Gromov-Witten invariants by working with the 
equivariant virtual cycle $[\om_{0,h}(\CZ,\oL;\beta)]_{T}$
and equivariant obstruction bundle $\CV_{T}$. 

Following this line of argument, one would finally have a localization 
formula for the equivariant invariants of the form 
\eqn\locA{
\int_{[\om_{0,h}(\CZ,\oL;\beta)]_{T}} e_{T}(\CV) = 
\sum_\Xi \int_{[\Xi]_{T}} {e_{T}(\CV_\Xi) \over e_{T}({N_\Xi})}.}
In this formula we sum over all fixed loci $\Xi$ in the moduli space of open 
string maps. $e_{T}(\CV_\Xi)$ is the equivariant Euler class of the 
obstruction  bundle $\CV_T$ restricted to $\Xi$ and $e_{T}({N_\Xi})$ 
is the equivariant Euler class of the virtual normal bundle $N({\Xi})$ to 
$\Xi$ in $\om_{0,h}(\CZ,\oL;\beta)$. $[\Xi]_{T}$ is the equivariant 
fundamental class of the fixed locus $\Xi$. 
The intersection pairing in the left hand side of \locA\ takes values 
in the cohomology ring $\CR_T=H^*(B{T})\simeq H^*(BS^1)$. 
The local contributions 
in the right hand side take values in the associated 
fraction field ${\cal K}_T$. 

Now, the fixed loci $\Xi$ can be classified according to the degrees $d_a$
and the map $\rho$ as discussed in the paragraph below \instcorrB.\ 
The refinement we are looking for is defined by summing only over those 
fixed loci $\Xi$ with given $(\rho,d_a)$. We will refer to these loci as 
fixed loci of type $(\rho,d_a)$. The maps therein will be also referred to as 
open string maps of type $(\rho,d_a)$. We define 
\eqn\locB{
\CC_{h,\rho}(d_a,\beta) = \sum_{\Xi} 
\int_{[\Xi]_{T}} {e_{T}(\CV_\Xi) \over e_{T}({N_\Xi})}}
where the sum is restricted to fixed loci of type $(\rho,d_a)$. 
Clearly, the resulting invariants $\CC_{h,\rho}(d_a,\beta)$ take values in 
${\cal K}_T$. The instanton coefficients $C_{h,\rho}(d_a,\beta)$ 
are defined as the nonequivariant limit of $\CC_{h,\rho}(d_a,\beta)$ \CK.\ 
More precisely, let $i_{pt}:\{pt\}\ra B{T}$ be the embedding of a point 
in $B{T}$. Then we define $C_{h,\rho}(d_a,\beta)$ by 
\eqn\locC{
C_{h,\rho}(d_a,\beta)=i_{pt}^*\CC_{h,\rho}(d_a,\beta) \in H^*(pt).} 
If we identify $H^*(pt)$ with $\IQ$, we can regard $C_{h,\rho}(d_a,\beta)$ as 
a rational function of the toric weights $\lambda_1,\lambda_3$. 
In the following it is very convenient to regard $\lambda_1, \lambda_3$ 
as formal variables. 
This is our heuristic formula for the instanton coefficients. Besides 
lacking a rigorous formulation, this expression raises another question. 
If the $C_{h,\rho}(d_a,\beta)$ are to be thought of as rational functions 
of formal variables, what is the physical meaning of the resulting open 
string expansion? This question has been given an elegant answer 
in the previous section in the context of local transitions. 
There we found that the framing in Chern-Simons theory should also 
be treated as a formal variable, which is related to the toric weights. 
The final result was shown to be a series with numerical coefficients, 
as expected. We will show in the next section that similar considerations 
can be applied to compact models as well, although the final picture 
will be more subtle in that case. 

At this point we still have to set the formula \locB\ on firmer grounds. 
The main idea is to rewrite the open string formula \locA\ in terms of 
equivariant integrals on moduli spaces of closed string maps 
with marked points. This approach has been successfully employed 
in a local context in \GZ.\ The extension to compact threefolds 
will involve some additional (sometimes delicate) steps. 
Although our arguments are mostly heuristic, 
we will eventually obtain well defined 
integrals on well defined moduli spaces. 
Following the strategy of \refs{\GZ,\KL} we proceed with the analysis 
of the deformation-obstruction complex for open string morphisms. 

In order to simplify the exposition, let us first consider the case 
$h=1$, i.e. the domain $\Sigma_{0,1}$ has a single disc component 
$\Delta$. After a detailed treatment of this case, we will consider 
the generalization to $h>1$. 
The disc $\Delta$ can be mapped to any of the discs $D'_i$, $D''_i$, 
hence we should consider all these cases separately. 
In fact it suffices to consider only a vertical disc, say $f(\Delta)=D'_1$, 
and a horizontal one, 
$f(\Delta)=D'_3$,  since all other cases can be treated by analogy.

\bigskip
\noindent{\it Vertical Discs} 
\medskip

Let us start with $f(\Delta)=D'_1$.
We have an 
invariant map $f:\Sigma_{0,1}\ra \CZ$ where 
$\Sigma_{0,1}=\Sigma_0 \cup_p \Delta$ 
and $\Sigma_0$ is a prestable curve of genus zero with a marked point 
$p$. We denote by 
$f_0, f_\Delta$ the restrictions of $f$ to $\Sigma_0$ and respectively 
$\Delta$, and by $f_\partial$ the restriction of $f$ to the boundary 
$\partial \Sigma=\partial \Delta$. 
Moreover, from now on we will suppress the subscript $(0,1)$ on $\Sigma$. 
In local coordinates, $f_\Delta$ is given by 
\eqn\locmapA{
x'_1=u'_1=0,\qquad y'_1=v'_1=t'^{d'_1}}
where $t'$ is a local coordinate on $\Delta$. Note that there is an 
induced torus action on $\Delta$ with weight 
$\lambda_{t'}={\lambda_{v'_1}\over d'_1}$. 

The map $f$ represents a point in some component $\Xi$ 
of the fixed locus. Following our main strategy, we would like to 
identify $\Xi$ with a component $\Xi_0$ of the fixed locus 
in a certain moduli space of stable closed string maps.
Note that the triple $(\Delta, f_\Delta,p)$ defines an isolated
fixed point in the moduli space 
of $d'_1:1$ multicovers of the disc $D'_1$ with a marked point.  

If $\Sigma_0$ is empty, this is in fact an isolated fixed point 
in the moduli space of open string maps $\om_{0,1}(\CZ, \oL; d'_1h_1)$,
since $[D'_1]=h_1$, as explained in the paragraph containing 
\discsD.\
The associated coefficient $\CC_{1,1'}(d'_1, d'_1h_1)$ 
is evaluated in appendix B. We will shortly review some 
aspects of that computation. 

If $\Sigma_0$ is not empty,  
the triple $(\Sigma_0, f_0, p)$ is a stable closed string map to $\CZ$ 
with a marked point subject to the constraint $f(p)=P'_1\in \CZ$.
Moreover, the homology class of $f_0$ is determined by 
$\beta'=\beta-d'_1h_1\in 
H_2(\CZ,\IZ)$. 
Therefore $(\Sigma_0, f_0, p)$ represents a fixed point in a closed 
subspace $\om_{P'_1}$ 
of the moduli space of stable closed string maps 
$\om_{0,1}(\CZ, \beta')$ defined by the 
following commutative diagram 
\eqn\commdiagA{
\xymatrix{ \om_{P'_1} \ar[r]^{j_{P'_1}} \ar[d]_\pi & \om_{0,1}(\CZ, \beta') 
\ar[d]^{ev}\\
P'_1\ar[r]^{i_{P'_1}} & \CZ.\\}}
Note that the $T$-action on $\om_{0,1}(\CZ, \beta')$ induces a 
$T$-action on $\om_{P'_1}$ since $P'_1$ is a fixed point on $\CZ$. 
In principle, $(\Sigma_0, f_0, p)$ is not an isolated fixed point, 
but it belongs to a component $\Xi_0$ of the fixed locus of the 
induced $T$-action on $\om_{P'_1}$. 
By construction, there is an obvious $1:1$ map between $\psi:\Xi\ra \Xi_0$ 
defined by $(\Sigma,f)\ra (\Sigma_0, f_0, p)$. This map is well defined 
and $1:1$ since $(\Delta,f_\Delta,p)$ is an isolated fixed point.
Now, the fixed locus $\Xi_0$ is in fact a closed subspace of the 
moduli space $\om_{P'_1}$ (more precisely, it is a closed algebraic 
substack) and is equipped with an induced  virtual 
fundamental class $[\Xi_0]= \iota_{\Xi_0}^{!} [\om_{P'_1}]$. Here 
$\iota_{\Xi_0}:\Xi_0\ra \om_{P'_1}$ denotes the embedding map
and $\iota_{\Xi_0}^{!}:A_*(\om_{P'_1}) \ra A_*(\Xi_0)$ denotes the 
associated Gysin map. The virtual cycle $[\om_{P'_1}]$ is 
induced by the base change diagram \commdiagA,\
$[\om_{P'_1}]=i_{P'_1}^![\om_{0,1}(\CZ, \beta)]$. 
All these considerations carry over to the equivariant setting. 

As part of our assumptions, let us suppose there is a similar 
structure on $\Xi$ so that the map $\psi$ described in the above paragraph 
becomes an isomorphism. Then $\Xi$ will be also endowed with a 
virtual cycle $\psi^{!}[\Xi_0]$, where $\psi^{!}: A_*(\Xi_0) \ra 
A_*(\Xi)$ is the Gysin map. This is the first ingredient in a rigorous 
formulation of the integral in \locA.\ 
We still need to make sense of the integrand.  

To this end, we have to determine the virtual normal bundle $N_\Xi$ 
using the tangent-obstruction complex of $(\Sigma, f)$, and to construct 
the obstruction bundle $\CV$. 
First we have to introduce some notation. 
The restriction of the holomorphic tangent bundle 
$T_\CZ$ to $\oL_1$ admits a real subbundle $T_\IR$
defined as the fixed locus of the 
 local antiholomorphic involution $(x_1,y_1,u_1,v_1)\ra 
({\overline u}_1, {\overline v}_1, {\overline x}_1, {\overline y}_1)$. 
We denote by $\CT$ the sheaf of sections of Riemann-Hilbert 
bundle defined by the pair $(f^*T_\CZ, f_{\partial}^*T_\IR)$
on $\Sigma$. Note that $\CT|_{\Sigma_0}=f_0^*T_\CZ$. We will also denote 
by $\CT_\Delta$ the restriction of $\CT$ to $\Delta$.
The deformation complex for the map $f$ is of the form 
\eqn\defA{
0\ra Aut(\Sigma) \ra H^0(\Sigma, \CT)\ra {\bbbt}^1 \ra 
Def(\Sigma) \ra H^1(\Sigma, \CT) \ra {\bbbt}^2 \ra 0} 
where ${\bbbt}^1, {\bbbt}^2$ are 
the deformation and respectively obstruction spaces of the map $(\Sigma, f)$. 
Using a normalization exact sequence with respect to 
the decomposition $\Sigma = \Sigma_0 \cup_p \Delta$, one can split 
the terms in \defA\ into open string and closed string parts, 
plus corrections due to the node $p$. We will show below 
that the closed string part reduces to the standard deformation complex for 
$\Xi_0$. The open string part and the node corrections become part of the 
data of an equivariant integral on $\Xi_0$. 
In order to completely specify the integrand, we also have to construct 
the obstruction bundle $\CV$. 
As outlined in section three, the fiber of the obstruction bundle over 
a point $(\Sigma, f)$ 
is given by the space of global holomorphic sections of a Riemann-Hilbert 
bundle $\CL = (f^*\Lambda^4(T_\CZ), \CR)$ on $\Sigma$. 
The real subbundle $\CR\subset f_\partial^*(\Lambda^4(T_\CZ))|_{\oL_1}$ is defined 
as the fixed set of
the local antiholomorphic involution $(x_1,y_1,u_1,v_1)\ra 
({\overline u}_1, {\overline v}_1, {\overline x}_1, {\overline y}_1)$. 
$\CV_\Xi$ can also be decomposed in closed and respectively open string 
parts using a normalization exact sequence. 
The terms in \defA\ and in the various normalization sequences encountered 
in the process are not in general 
vector bundles over the fixed loci since the fiber dimension may jump. 
Nevertheless we will formally manipulate such objects as if they 
were locally free following for example the approach of \MK.\ 
This approach yields correct results since it is only the equivariant 
K-theory class of these objects which 
enters the computations. Equivariant K-theory classes will be denoted 
by $[\ ]$. We have to distinguish two cases. 

$i)$ If $\Sigma_0$ is empty, there is no closed string part and $\Xi$ 
is an isolated fixed point. In this case we will denote $\Xi$ by 
$\Xi_\Delta$ in order to avoid any confusion with the 
general case. The deformation complex becomes 
\eqn\defB{0\ra Aut(\Delta) \ra H^0(\Delta, \CT)\ra {\bbbt}^1 \ra 
0 \ra H^1(\Delta, \CT) \ra {\bbbt}^2 \ra 0} 
since the domain $\Delta$ has no deformations. The automorphism group 
is generated by holomorphic vector fields of the form 
$a\partial_{t'}+bt'\partial_{t'}$, where $a\in \IC$ and $b\in \IR$. 
Therefore, in terms of $T=S^1$ representations we have 
\eqn\defC{
Aut(\Delta) \simeq (0)_{\IR}\oplus (-\lambda_{t'}).}
The cohomology groups $H^0(\Delta, \CT)$ and $H^1(\Delta, \CT)$ are computed in appendix B. 
We have $H^1(\Delta, \CT)=0$, hence ${\bbbt}^2=0$ as well, and the complex 
\defB\ reduces to 
\eqn\defD{
0\ra Aut(\Delta) \ra H^0(\Delta, \CT)\ra {\bbbt}^1 \ra 0.}
The remaining term $H^0(\Delta,\CT)$ decomposes in representations of 
$T$ as follows 
\eqn\defE{
H^0(\Delta,\CT) \simeq \bigoplus_{n=0}^{d'_1}
\left(-\lambda_1+{n\over d'_1} \lambda_3\right)\oplus 
\bigoplus_{n=0}^{2d'_1}\left(
{n-d'_1\over d'_1}\lambda_3 \right).}
Note that for $n=d'_1$ in the second sum we obtain a term of weight zero
$(0)_{\IC}$ containing the image of the fixed part $(0)_\IR$ of 
$Aut(\Delta)$. It follows form the exact sequence \defD\ that 
we obtain a term of weight zero $(0)_\IR$ in the equivariant decomposition 
of ${\bbbt}^1$. 
The obstruction space $\CV_\Delta=H^0(\Delta, \CL)$ 
is also computed in appendix 
B, with the result 
\eqn\defF{
H^0(\Delta, \CL) = (0)_\IR \oplus \bigoplus_{n=1}^{3d'_1} \left({n\over d'_1} 
\lambda_3\right).}
Note that this is a real vector space containing a fixed direct summand 
$(0)_\IR$. 
The local contribution of such a fixed point should be of the form 
\eqn\locD{
\CC_{1,1'}(d'_1,d'_1h_1)= \int_{pt_{T}} e_T\left([\CV_\Delta]-[N_\Delta]\right)}
where $N_\Delta$ is the virtual normal bundle to $\Xi_\Delta$. 
A straightforward extrapolation of standard localization results would predict
that $N_\Delta$ is given by the moving part of ${\bbbt}^1$. However, we have to be 
more careful here. Since $\Xi_\Delta$ is an isolated fixed point, 
we will define 
$N_\Delta$ to be the tangent space ${\bbbt}^1$, including the fixed real 
deformation $(0)_\IR$ noticed above. 
The obstruction space contains an identical 
fixed real summand, and the two terms cancel off in the formula 
\locD,\ leaving a well-defined nonzero answer. 
This cancellation reflects the fact that the fixed infinitesimal deformation 
is obstructed. The situation encountered here is a bit unusual compared 
to standard localization computations for closed string maps, but we do not 
see other sensible solution for the moment. This unusual behavior is very
likely related to the fact that the cycles $\oL$ are not lagrangian 
middle dimensional cycles on $\CZ$. The solution proposed here should be 
regarded as an experimental result, which will be backed up 
by numerical computations in the next section. We leave a more conceptual 
treatment for future work\foot{Another point of view one could take is 
to consider open string maps to the singular threefold $\oY$ instead of 
$\CZ$. The problem is that the vertical discs intersect the singular divisor 
$D_\infty$ on $\oY$ and it is not clear how to write down the deformation 
complex of such a map. (Recall that $\oY$ contains nonreduced components). 
This approach has been effectively used for horizontal discs in the local 
case since they do not intersect the singular locus. According to our 
calculations in appendix B, for horizontal discs the two methods give 
the same answer. This is an encouraging sign, although much more 
work remains to be done.}.
Collecting the facts, this leaves us with the following formula for 
$d'_1$ multicovers of $D'_1$ 
\eqn\locE{
\CC_{1,1'}(d'_1, d'_1h_1)= 
{(-1)^{d'_1}\over d_1'^{d'_1+2}}{(3d'_1)!\over 
(d'_1!)^2} {\lambda_3^{d'_1+1}\over \prod_{n=0}^{d'_1}\left(-\l_1+
{n\over d'_1}\l_3\right)}.}
There is a subtle sign ambiguity in this expression reflecting the choice 
of a complex structure on moduli space of open string maps, as explained in 
appendix B. In the absence of a more rigorous construction, the only criterion 
for fixing the sign at the present stage is agreement with the closed string 
dual. This will be shown in the next section. 

$ii)$ If $\Sigma_0$ is not empty, we have to split $[{\bbbt}^1], [{\bbbt}^2]$ and $[\CV_\Xi]$ 
in open and closed string parts using the exact sequences 
\eqn\exseqA{
0\ra \CT \ra f_0^*T_\CZ \oplus \CT_\Delta \ra (f^*T_\CZ)_p\ra 0.}
\eqn\exseqB{
0\ra \CL \ra f_0^*\Lambda^4(T_\CZ)\oplus \CL_\Delta \ra 
(f^*\Lambda^4(T_\CZ))_p\ra 0.}
The associated long exact sequences read
\eqn\exseqC{\eqalign{ 
0 & \ra H^0(\Sigma, \CT) \ra H^0(\Sigma_0, f_0^* T_\CZ) \oplus 
H^0(\Delta, \CT_\Delta) \ra (T_\CZ)_{P'_1} \cr
& \ra H^1(\Sigma, \CT) \ra H^1(\Sigma, f_0^* T_\CZ)\oplus 
H^1(\Delta, \CT_\Delta)\ra 0.\cr}}
\eqn\exseqD{\eqalign{
0 & \ra H^0(\Sigma, \CL) \ra H^0(\Sigma_0,f_0^*\Lambda^4(T_\CZ))\oplus 
H^0( \Delta, \CL_\Delta) \ra \Lambda^4(T_\CZ)_{P'_1}\cr
&\ra H^1(\Sigma, \CL) \ra H^1(\Sigma_0,f_0^*\Lambda^4(T_\CZ))\oplus 
H^1(\Delta, \CL_\Delta) \ra 0.\cr}}
We claim that $H^1(\Sigma_0,f_0^*\Lambda^4(T_\CZ))=0$. This follows 
from a similar normalization sequence applied to the irreducible components
of $\Sigma_0$. The pull-back  
$f_0^*\Lambda^4(T_\CZ)$ has nonnegative degree on each irreducible component 
of $\Sigma_0$ since $\Lambda^4(T_\CZ) = \CO(-K_\CZ)$ is nef on $\CZ$. 
Since all components are rational, the claim follows easily from the 
associated long exact sequence. 
With more effort, we can also show that $H^1(\Delta, \CL_{\Delta})=0$
(according to appendix B), hence the last term of the exact sequence 
\exseqD\ is trivial. The morphism $H^0(\Sigma_0,f_0^*\Lambda^4(T_\CZ))\oplus 
H^0( \Delta, \CL_\Delta) \ra \Lambda^4(T_\CZ)_{P'_1}$ is surjective 
since $\Sigma_0$ has genus zero and $\Lambda^4(T_\CZ)$ is nef on $\CZ$. 
Therefore we conclude that $H^1(\Sigma,\CL)=0$ i.e. the obstruction bundle 
is convex, as promised before. The exact sequence \exseqD\
reduces to the first three terms. 

Moreover, note that $Aut(\Sigma)$ decomposes as $Aut(\Sigma_0,p)\oplus 
Aut(\Delta,p)$ in automorphisms of $\Sigma_0$, respectively 
$\Delta$ which preserve the point $p$. $Def(\Sigma)$ decomposes similarly as $Def(\Sigma_0, p)\oplus 
T_p\Sigma_0\otimes T_p\Delta$ in deformations of $\Sigma_0$ 
leaving $p$ fixed and deformations of the node. There are no 
deformations of $\Delta$. 
Using the exact sequences \defA,\ \exseqC\ and \exseqD,\ we find the following 
relations in the equivariant K-theory of $\Xi$ 
\eqn\locG{\eqalign{ 
[{\bbbt}^1]-[{\bbbt}^2] & = [H^0(\Sigma,\CT)]-[H^1(\Sigma,\CT)] + [Def(\Sigma)]
-[Aut(\Sigma)]\cr
& = [H^0(\Delta,\CT_\Delta)]-[Aut(\Delta)] + [\partial_{t'}]+ [H^0(\Sigma_0, f_0^*T_\CZ)]-[H^1(\Sigma_0,f_0^*T_\CZ)] 
\cr 
& \ \quad +[Def(\Sigma_0, p)]- [Aut(\Sigma_0,p)]-[(T_{\CZ})_{P'_1}] +[T_p\Sigma_0\otimes T_p\Delta]\cr}}
and
\eqn\locH{\eqalign{
[\CV_\Xi] = [H^0(\Delta, \CL)]+[H^0(\Sigma_0, f_0^*\Lambda^4(T_\CZ))]
-[\Lambda^4(T_\CZ)_{P'_1}].\cr}}
The combination $[H^0(\Delta,\CT_\Delta)]-[Aut(\Delta)]$ is precisely the 
K-theory class of the normal bundle $N_\Delta$ to the fixed point 
$(\Delta, f_\Delta)$ defined at point $(i)$ above. In order to identify the 
remaining terms in \locG,\ let us write down the tangent-obstruction 
complex for the stable map $(\Sigma_0,f_0,p)$ regarded as a point in 
$\om_{0,1}(\CZ, \beta')$
\eqn\defG{
0\ra Aut(\Sigma_0,p) \ra H^0(\Sigma_0, f_0^*T_\CZ) \ra {\bbbt}^1_0\ra 
Def(\Sigma_0,p)\ra H^1(\Sigma_0, f_0^*T_\CZ) \ra {\bbbt}^2_0\ra 0.} 
Using the K-theory relations derived from \defG,\ we can rewrite
\locG\ as follows
\eqn\locI{\eqalign{ 
[{\bbbt}^1]-[{\bbbt}^2] & = [N_\Delta] + [\partial_{t'}] + 
[{\bbbt}^1_0] - [{\bbbt}^2_0] -[(T_{\CZ})_{P'_1}] + 
[T_p\Sigma_0\otimes T_p\Delta].\cr}}
The combination $[{\bbbt}^1_0]-[{\bbbt}^2_0]-[(T_{\CZ})_{P'_1}]$ represents the image in 
K-theory of the deformation complex of $(\Sigma_0,f_0,p)$ regarded
as a point in $\om_{P'_1}$. This follows from the construction of 
$\om_{P'_1}$ as the subspace of $\om_{0,1}(\CZ,\beta')$ defined by 
$f_0(p)=P'_1$. Therefore the K-theory class of the virtual normal bundle 
$N_{\Xi_0}$ in $\om_{P'_1}$ is given by the moving parts 
\eqn\virnormA{
[N_{\Xi_0}]=[({\bbbt}^1_0)^m]-[({\bbbt}^2_0)^m]-[((T_{\CZ})_{P'_1})^m].}
We will define the class of 
the open string virtual normal bundle $N_{\Xi}$ to be 
\eqn\virnormB{
[N_{\Xi}]= [N_\Delta] +[({\bbbt}^1_0)^m]-[({\bbbt}^2_0)^m]-[((T_{\CZ})_{P'_1})^m] + 
[(\partial_{t'})^m] +[(T_p\Sigma_0\otimes T_p\Delta)^m].}
Up to the first term, which has been explained above, this is just the 
moving part of \locI,\ as expected. 
The fixed part of \locI,\ $[({\bbbt}^1_0)^f] -[({\bbbt}^2_0)^f] -[((T_{\CZ})_{P'_1})^f]$, 
enters the construction of the induced virtual cycle
$[\Xi_0]_T\in A_*^T(\Xi_0)$, $[\Xi_0]_T=\iota_{\Xi_0}^{!}[\om_{P'_1}]_T$
\GP.\ This is consistent with our earlier proposal $[\Xi] = \psi^{!} 
[\Xi_0]$ for the open string virtual cycle (see the paragraph below 
\commdiagA).\ 

The K-theory class of the obstruction bundle splits similarly in an 
open string part $[\CV_\Delta]$ which has been considered before, and a 
closed string part $[H^0(\Sigma_0, f_0^*\Lambda^4(T_{\CZ}))]$. The later 
is the K-theory class of the standard obstruction bundle over 
$\om_{0,1}(\CZ,\beta')$ restricted to $\Xi_0$. Recall that the 
fiber of $\CV^0$ over a point $(\Sigma_0,f_0,p)$ is given precisely by 
$H^0(\Sigma_0, f_0^*\Lambda^4(T_{\CZ}))$. 

We can now tie all loose ends together and write down a well defined 
local formula for the open string invariants $\CC_{1,1'}(d'_1, \beta)$
\eqn\locJ{
{\CC_{1,1'}(d'_1,d'_1h_1)\over (-\lambda_{t'}H)}
\int_{[\Xi_0]_T} {1\over e_T(N_{\Xi_0})} 
{e_T([\CV^0_{\Xi_0}-\Lambda^4(T_{\CZ})_{P'_1}])\over -\lambda_{t'}H -\psi_p}}
where $H$ is the generator of  $H^*(BT)$. The factor $(-\lambda_{t'}H)$ 
in the denominator represents the equivariant Euler class of 
$[\partial_{t'}]$, $\psi_p$ is the Mumford class associated to the point $p$, 
and $\Lambda^4(T_{\CZ})_{P'_1}$ should be regarded as an equivariant 
bundle over $\Xi_0$. Summing over all fixed loci $\Xi_0$ we obtain the 
following expression 
\eqn\locK{
\CC_{1,1'}(d'_1, \beta) = 
{\CC_{1,1'}(d'_1,d'_1h_1)\over (-\lambda_{t'}H)}
\int_{[\om_{P'_1}]_T}
{e_T([j_{P'_1}^*(\CV^0)-\Lambda^4(T_{\CZ})_{P'_1}])\over -\lambda_{t'}H -\psi_p}}
where $\Lambda^4(T_{\CZ})_{P'_1}$ should be regarded as an equivariant bundle 
over $\om_{P'_1}$. 

Although this formula is now well defined, it is also of little use 
for explicit computations. The major problem is that the fixed loci $\Xi_0$ 
can have a very complicated structure, since the $T$ action on $\CZ$ is 
not generic. Ideally, we would like to be able to compute the invariants 
\locK\ in terms of standard Kontsevich graphs for closed string maps 
to $\CZ$. In order to do so, we have to take one more step and
rewrite the formula \locK\ in terms of a generic torus action. 
Recall that $T$ is a subgroup of the torus $G=(S^1)^7$ which acts on 
$\CZ$ and $P'_1$ is a $G$-fixed point on $\CZ$. Then we have the following 
commutative diagram 
\eqn\commdiagB{
\xymatrix{ 
(\om_{P'_1})_T \ar[r]^S \ar[d] & (\om_{P'_1})_G \ar[d] \\
BT \ar[r]^s & BG.\\}}
Let $(\eta_1,\ldots,\eta_7)$ denote the generators of 
$H^*(BG)$ defined by the characters of $G$ as explained in 
chapter nine of \CK.\ Then we have 
\eqn\valA{{\cal K}_G \simeq \IQ\left(\eta_1,\ldots,\eta_7\right),\qquad 
{\cal K}_T \simeq \IQ\left(\l_1H, \l_3H\right).}
The map $s$ in the above diagram induces a pull-back map 
$s^*:{\cal K}_G\ra {\cal K}_T$ by localization, which factorizes as 
\eqn\valB{\xymatrix{
{\cal K}_G \ar[r]^{s^*} \ar[d]_{\pi} & {\cal K}_T \\
{\cal K}_G \otimes_{\CR_G} \left(\CR_G/{\cal J}\right) 
\ar[ur]^{{\overline s}}\\}}
where $\CJ\subset \CR_G$ is the ideal generated by $(\eta_2,\eta_4,
\eta_1+\eta_5,\eta_3+\eta_6,\eta_7)$, and ${\overline s}$ is defined by 
${\overline s}(\eta_1) = \lambda_1H$ and ${\overline s}(\eta_3)=\lambda_3H$. 
Using these relations, we can write the class $-\lambda_{t'}H -\psi_p=
-{\l_3\over d'_1}H-\psi_p$ 
in \locK\ as $s^*\left(-{\eta_3\over d'_1}-\psi_p\right)$. 
Moreover, the equivariant virtual class $[\om_{P'_1}]_T$ and the obstruction 
bundle $\CV^0$ are pulled back via $S$ from $(\om_{P'_1})_G$ (for the virtual 
cycles we have to use the Gysin map). Therefore, using the projection 
formula, we can rewrite \locK\ 
in the form 
\eqn\locL{
\CC_{1,1'}(d'_1, \beta) = {\CC_{1,1'}(d'_1,d'_1h_1)\over (-\lambda_{t'}H)}
s^*\left[\int_{[\om_{P'_1}]_G}
{e_G([j_{P'_1}^*(\CV^0)-\Lambda^4(T_\CZ)_{P'_1}])\over -{\eta_3\over d'_1} 
-\psi_p}
\right].}
We will explain how to perform explicit 
computations using this formula in the next section. Here let us note that we can rewrite 
\locL\ in a different form which is closer to the integrals written in \GZ\
\eqn\locM{
\CC_{1,1'}(d'_1, \beta) = {\CC_{1,1'}(d'_1,d'_1h_1)\over (-\lambda_{t'}H)}
s^*\left[{1\over e_G(\Lambda^4(T_\CZ)_{P'_1})}\int_{[\om_{0,1}(\CZ,\beta')]_G}
{e_G(\CV^0)ev_G^*(\phi_{P'_1})
\over -{\eta_3\over d'_1} -\psi_p}
\right]}
where $ev_{G}:\om_{0,1}(\CZ,\beta')_G\ra \CZ_G$ is the equivariant 
evaluation map, and $\phi_{P'_1}\in H^*_G(\CZ)$ is the equivariant Thom 
class of the fixed point $P'_1$. The equivalence of the two expressions  
follows from the commutative diagram \commdiagA\  by a series of formal
manipulations based on the  
functorial properties of Chow groups given for example in \YM\ 
chapter V, $\S 5-\S 8$.
Let 
$\alpha = {e_G(\CV^0)\over  -{\eta_3\over d'_1}-\psi_p}
\in A^*_G(\om_{0,1}(\CZ,\beta'))$ and note that the equivariant integral in 
\locL\ can be written as 
\eqn\locLA{
\int_{[\om_{P'_1}]_G}
{e_G([j_{P'_1}^*(\CV^0)-\Lambda^4(T_\CZ)_{P'_1}])\over 
-{\eta_3\over d'_1} -\psi_p}= 
{1\over e_G(\Lambda^4(T_\CZ)_{P'_1})} \pi_*j_{P'_1}^*(\alpha)}
where we use the notation of \commdiagA.\ Then, using the projection 
formula, we have 
\eqn\locMA{\eqalign{
 \pi_*j_{P'_1}^*(\alpha) = i_{P'_1}^{!}ev_{*}
\left(\alpha \cap [\om_{0,1}(\CZ,\beta')]\right).\cr}}
All these manipulations are carried out in equivariant setting, but we
suppressed the subscript $G$ for simplicity. 
If we denote by $\Pi: \CZ\ra pt$ the projection onto a point,
the equivariant integral in \locM\ can be written as 
\eqn\locMB{
\int_{[\om_{0,1}(\CZ,\beta')]_G}
{e_G(\CV^0)ev_G^*(\phi_{P'_1})
\over -{\eta_3\over d'_1} 
-\psi_p} = \Pi_*ev_*\left(\alpha\cup ev^*(\phi_{P'_1}) 
\cap [\om_{0,1}(\CZ,\beta')]\right).}
Now we can use localization on $\CZ_G$ to rewrite the right hand side 
of \locMB\ as 
\eqn\locMC{\eqalign{
&\Pi_*ev_*\left(\alpha\cup ev^*(\phi_{P'_1}) 
\cap [\om_{0,1}(\CZ,\beta')]\right)\cr
&\qquad\qquad\qquad\qquad\quad = \sum_{P_f} i_{P_f}^!\left(\phi_{P'_1}
\cap ev_*\left(\alpha\cup
\cap [\om_{0,1}(\CZ,\beta')]\right)\right)\cap {1\over e_G((T_{\CZ})_{P_f})}\cr
&\qquad\qquad\qquad\qquad\quad =\sum_{P_f} {i_{P_f}^*(\phi_{P'_1})\over e_G((T_\CZ)_{P_f})}\cap 
i_{P_f}^! ev_*\left(\alpha\cup
\cap [\om_{0,1}(\CZ,\beta')]\right).\cr}}
where the sum is over all fixed points $P_f$ of the $G$ action of $\CZ$. 
The equivariant Thom classes of the fixed points satisfy orthogonality 
conditions of the form 
\eqn\locMD{
i_{P_f}^*(\phi_{P'_1}) =\left\{\matrix{ e_G((T_\CZ)_{P_f})\qquad\hfill & \hbox{if}
\ P_f=P'_1\cr
0\qquad \hfill & \hbox{if}\ P_f\neq P'_1.}\right.}
Therefore \locMC\ reduces to 
\eqn\locME{
\Pi_*ev_*\left(\alpha\cup ev^*(\phi_{P'_1}) 
\cap [\om_{0,1}(\CZ,\beta')]\right)= i_{P_1'}^! ev_*\left(\alpha\cup
\cap [\om_{0,1}(\CZ,\beta')]\right)}
as claimed above. 

\bigskip\noindent
{\it Horizontal Discs} 
\medskip 

Let us now consider horizontal discs, i.e. $f_\Delta$ is an invariant 
$d'_3:1$ cover of $D_3'$. Most of the above considerations go through, 
but there is an important difference, namely the origin $Q_3$ 
of $D_3'$ is not a fixed point under the generic $G$ action. 
Instead, it is a degenerate $T$-fixed point lying on the fixed
curve $\overline{P_3P_3'}$. The contribution of the isolated fixed point 
$(\Delta, f_\Delta)$ can be computed along the same lines 
\eqn\locN{
\CC_{1,3'}(d_3', d_3'h_2)= {(-1)^{d'_3}\over {d'_3}^{3-d'_3}} 
{\prod_{n=1}^{d'_3-1}\left(\l_1-{n\over d'_3}\l_3\right)\over 
\l_3^{d'_3-1}}.} 
The overall sign is again ambiguous; the present choice should be 
regarded as part of the prescriptions of the duality map. 
The formula \locK\ goes through essentially unchanged 
\eqn\locO{
\CC_{1,3'}(d'_3, \beta) = 
{\CC_{1,3'}(d'_3,d'_3h_2)\over (-\lambda_{t}H)}
\int_{[\om_{Q_3}]_T}
{e_T([j_{Q_3}^*(\CV^0)-\Lambda^4(T_\CZ)_{Q_3}])\over -\lambda_{t}H -\psi_p}}
where $\lambda_t = {\lambda_3\over d'_3}$ is the weight of the 
induced torus action on the domain. 
Now we have to integrate against the virtual cycle of the moduli space
$\om_{Q_3}$ defined by the following diagram 
\eqn\commdiagC{
\xymatrix{ \om_{Q_3} \ar[r]^{j_{Q_3}} \ar[d]_\pi & \om_{0,1}(\CZ, \beta') 
\ar[d]^{ev}\\
Q_3\ar[r]^{i_{Q_3}} & \CZ\\}}
and $\beta' =\beta - d_3'h_2$. 

At this point, we encounter an extra complication, since $Q_3$ 
is not invariant under the generic torus action. Nevertheless, we would 
still like to rewrite \locN\ in terms of a generic torus action in order 
to perform the computations efficiently. 
A key observation is that there exists a $T$-equivariant automorphism 
$h:\CZ\ra \CZ$ mapping the $G$-fixed point $P'_3=\{Z_1=Z_4=U=W=0\}$
to $Q_3$. In terms of homogeneous coordinates, $h$ is 
given by 
\eqn\torautA{
h:[Z_1,Z_2,Z_3,Z_4,U,V]\ra [Z_1,Z_2,Z_3,Z_4,U,V,W-\mu V_2V_3].}
It is straightforward to check that \torautA\ is compatible 
with the $(\IC^*)^3$ action \toricA\ and the $T$-action on $\CZ$. 
Let $\om_{P_3'}$ be the subspace of the moduli space $\om_{0,1}(\CZ,\beta')$ 
defined by a commutative diagram of the form \commdiagC,\
with $Q_3$ replaced by $P_3'$. Since $P'_3$ is a fixed point, there 
is an induced $T$-action on $\om_{P'_3}$. 
Then composition by $h:\CZ\ra \CZ$ induces a $T$-equivariant 
automorphism $\Psi_h:\om_{0,1}(\CZ, \beta') \ra \om_{0,1}(\CZ, \beta')$ 
and we obtain a commutative diagram of the form 
\eqn\commdiagD{
\xymatrix{
& \om_{Q_3} \ar[rr]^{j_{Q_3}}\ar'[d][dd] & & \om_{0,1}(\CZ,\beta') \ar[dd] \\
\om_{P'_3} \ar[ur]^{\psi_h}\ar[rr]^{~j_{P'_3}}\ar[dd] & & 
\om_{0,1}(\CZ,\beta') \ar[ur]^{\Psi_h}\ar[dd]  \\
& Q_3 \ar'[r][rr]^{i_{Q_3}} & & \CZ. \\
P'_3 \ar[rr]^{i_{P'_3}}\ar[ur] & & \CZ \ar[ur]^h}}
The obstruction bundle $\CV_0$ over $\om_{0,1}(\CZ,\beta')$ and
the virtual cycle $[\om_{0,1}(\CZ,\beta')]$ are invariant under $h$
by construction.
That is we have $\Psi_h^* (\CV_0) = \CV_0$ and $h^![\om_{0,1}(\CZ,\beta')]
=[\om_{0,1}(\CZ,\beta')]$. 
The virtual cycles $[\om_{P'_3}]$ and $[\om_{Q_3}]$ are determined by Gysin
maps $[\om_{P'_3}] = i_{P'_3}^! [\om_{0,1}(\CZ,\beta')]$, 
$[\om_{Q_3}] = i_{Q_3}^! [\om_{0,1}(\CZ,\beta')]$. 
Therefore we have $\psi_h^![\om_{Q_3}]=[\om_{P'_3}]$ and 
$\psi_h^*j_{Q_3}^*(\CV_0)=j_{P'_3}^*(\CV_0)$.
Moreover, 
$\Lambda^4(T_\CZ)_{Q_3}\simeq \Lambda^4(T_\CZ)_{P'_3}$ as $T$-vector spaces. 
We conclude that the equivariant integral \locO\ is equal to
\eqn\locP{
\CC_{1,3'}(d'_3, \beta) = 
{\CC_{1,3'}(d'_3,d'_3h_2)\over (-\lambda_{t}H)}
\int_{[\om_{P_3'}]_T}
{e_T([j_{P'_3}^*(\CV^0)-\Lambda^4(T_\CZ)_{P_3'}])\over -\lambda_{t}H -\psi_p}.}
Since $P_3'$ is fixed under the generic torus action, we can further 
rewrite this expression as 
\eqn\locQ{
\CC_{1,3'}(d'_3, \beta) = 
{\CC_{1,3'}(d'_3,d'_3h_2)\over (-\lambda_{t}H)} s^*\left[
\int_{[\om_{P_3'}]_G} 
{e_T([j_{P'_3}^*(\CV^0)-\Lambda^4(T_\CZ)_{P_3'}])\over 
{\eta_3\over d'_3} -\psi_p}\right],}
or, equivalently, as 
\eqn\locR{
\CC_{1,3'}(d'_1, \beta) = {\CC_{1,3'}(d'_3,d'_3h_2)\over (-\lambda_{t}H)}
s^*\left[{1\over e_G(\Lambda^4(T_\CZ)_{P_3'})}\int_{[\om_{0,1}(\CZ,\beta')]_G}
{e_G(\CV^0)ev^*(\phi_{P'_3})
\over {\eta_3\over d'_3} -\psi_p}\right].}
This formula will be used for explicit computations in the next section. 
\bigskip\noindent
{\it Multiple boundary components} 
\medskip 

We are left with open string maps with several boundary components. 
Let $f:\Sigma\ra \CZ$, $\Sigma=\Sigma_0 \cup \Delta_1\cup\ldots \cup \Delta_h$
be such a map of type $(\rho,d_a)$. Recall that  
$\rho:\{1,\ldots,h\} \ra \{1',\ldots,4',1'',\ldots,4''\}$ specifies the 
image of each disc $\Delta_a$, $a=1,\ldots,h$ of the domain, and 
the $d_a$ are the corresponding degrees. 

If $\Sigma_0$ is nonempty, we proceed by analogy with the previous 
two cases. A similar sequence of arguments shows that the sum over 
all open string local contributions can be written as a moduli space 
integral of the form 
\eqn\locS{\eqalign{
\CC_{h,\rho}(d_a,\beta) = &
{1\over |\CP|} 
\prod_{a=1}^h{\CC_{1,\rho(a)}(d_a, d_a[D_{\rho(a)}]) \over -(\lambda_{t_a}H)}\cr
& \times s^*\left[{1\over \prod_{a=1}^h e_G(\Lambda^4(T_\CZ)_{P(a)})}
\int_{[\om_{0,h}(\CZ,\beta')]_G}{e_G(\CV^0)
\prod_{a=1}^h{ev_a^*(\phi_{P(a)})
\over -\kappa_a -\psi_{p_a}}}\right].\cr}}
Let us explain the notation. The coefficient 
$\CC_{1,\rho(a)}(d_a, d_a[D_{\rho(a)}])$ represents the multicover 
contribution of the $a$-th disc of the domain which is mapped 
$d_a:1$ to the disc $D_{\rho(a)}$ in the target space. $P$ is a function 
from the index set $\{1,\ldots,h\}$ to the set of fixed points of $\CZ$ under 
the $G$ action. If $D_{\rho(a)}$ is a vertical disc, $P(a)$ is the 
origin of $D_{\rho(a)}$; if $D_{\rho(a)}$ is horizontal, then we 
have $P(a)= P_3'$ if $\rho(a) = 3', 4'$, and $P(a)=P''_4$ 
if $\rho(a)=3'', 4''$. 
$\om_{0,h}(\CZ,\beta')$ is the moduli space of stable closed string maps 
to $\CZ$ with $h$ marked points $\{p_a\}$. The homology class $\beta'\in 
H_2(\CZ, \IZ)$ is given by $\beta' = \beta -\sum_{a=1}^h [D_{\rho(a)}]$. 
$ev:\om_{0,h}(\CZ,\beta') \ra \CZ^h$ is the evaluation map, 
and $ev_a: \om_{0,h}(\CZ,\beta')\ra \CZ$ denotes its $a$-th component.
$\kappa_a$, $a=1,\ldots,h$ is an equivariant class in ${\cal K}_G$ 
such that $s^*(\kappa_a) = \l_{t_a}H$, where $\l_{t_a}$ is the 
weight of the induced torus action on the disc $\Delta_a$.
One can easily check that $\kappa_a= {\eta_3\over d_a}$ if $\rho(a)=1',3''$, 
$\kappa_a={\eta_1\over d_a}$ if $\rho(a)=1'',4'$, 
$\kappa_a=-{\eta_3\over d_a}$ if $\rho(a)=2'',3'$ and $\kappa_a=
-{\eta_1\over d_a}$  if $\rho(a)= 2',4''$.
Finally, ${1\over |\CP|}$ is a symmetry factor which takes into account the 
automorphism group of a fixed locus. 
The automorphism group of an invariant open 
string map as above is of the form $\prod_{a=1}^h \left(\IZ/d_a\IZ\right) 
\times \CP$. The first $h$ factors represent deck transformations of 
the Galois cover $f_{\Delta_a} :\Delta_a \ra D_{\rho(a)}$. Their effect would 
be a prefactor $\prod_{a=1}^h {1\over d_a}$ in $\locS$ which is 
absorbed in the coefficients $\CC_{1,\rho(a)}(d_a, d_a[D_{\rho(a)}])$. 
$\CP$ is a subgroup of the permutation group $\CS_h$ which permutes 
the marked points $p_a$ leaving the map $d_a$ unchanged. More precisely,
if we have $n$ discs of the domain mapping to the same disc in $\CZ$, 
$\CP$ contains a factor $\CS_{n}$. Therefore we obtain a factor $1\over n!$ 
in \locS\ for each such group of $n$ discs. There are eight discs in the 
target space, so we can have up to eight factors of this form. 

If $\Sigma_0$ is empty, the domain of $f$ can be either a single disc
or a two discs with common origin forming a nodal cylinder. The first case 
has been treated above. A map of the form $f:\Delta_1\ra \CZ$
contributes $\CC_{1,\rho(1)}(d_1, d_1[D_{\rho(1)}])$ to the 
instanton expansion. 
The second case follows easily from the first 
using normalization exact sequences. Given a map $f:\Delta_1\cup_p 
\Delta_2\ra \CZ$ of type $(\rho, d_1, d_2)$ as above, its contribution 
can be easily shown to be of the form 
\eqn\locSB{\eqalign{
&\CC_{1,2,\rho(1),\rho(2)}(d_1,d_2,d_1[D_{\rho(1)}]+d_2[D_{\rho(2)}])\cr
& \qquad =\CC_{1,\rho(1)}(d_1,d_1[D_{\rho(1)}]) 
\CC_{1,\rho(2)}(d_2, d_2[D_{\rho(2)}]) 
{e_T([(T_\CZ)_{P(1)}-\Lambda^4(T_\CZ)_{P(1)}])\over 
(-\lambda_{t_1}H)(-\lambda_{t_2}H)(-\lambda_{t_1}-\lambda_{t_2})H}.\cr}}
For a uniform treatment, we can extend the formula $\locS$ to encompass 
all possible cases. By convention, if $\Sigma_0$ is empty, $\locS$ 
should be interpreted as explained in the current paragraph. 

This concludes our discussion of open string instantons
on compact threefolds. Since this section is rather long and complicated, 
let us summarize the main points. The open string instanton corrections 
take the form \instcorrB,\ where the enumerative invariants 
$\CC_{h,\rho}(d_a,\beta)$ are given by \locS,\ which is our main formula.
Although this expression has been derived starting from 
heuristic considerations, it is a well defined equivariant integral over a
moduli space (or stack) of stable maps. In the next section we will find 
very convincing evidence for the approach proposed here by direct computations 
and comparison with the closed string dual. 

\newsec{Explicit Computations and Large $N$ Duality}

At this stage, we have all the ingredients needed for computing  
topological open string amplitudes and testing large $N$ duality. 
Some preliminary remarks on large $N$ duality for compact 
threefolds have been included in section two. Let us start 
with a more precise account of the duality predictions  
for the extremal transition under consideration. 

\subsec{Large $N$ duality predictions} 

As explained in section four, we have an extremal transition 
between smooth projective Calabi-Yau threefolds $Y, {\wY}$ which 
can be represented as hypersurfaces in toric varieties 
$\CZ, \wcz$. $\wcz$ is the blow-up of $\CZ$ along the section 
$U=V=0$, and $\wY$ is the strict transform of the singular hypersurface 
$Y_0$. $\wcz$ has the following toric presentation 
\eqn\toricB{ 
\matrix{ & \wZ_1 & \wZ_2 & \wZ_3 & \wZ_4 & \wU & \wV & \wW & \wT \cr
\IC^* & 1 & 1 & 0 & 0 & -1 & -1 & 0 & 0 \cr
\IC^* & 0 & 0 & 1 & 1 & -1 & -1 & 0 & 0 \cr
\IC^* & 0 & 0 & 0 & 0 & 1 & 1 & 1 & 0 \cr
\IC^* & 0 & 0 & 0 & 0 & 1 & 1 & 0 & -1. \cr}}
One can easily check that $h^{1,1}(\wZ) = h^{1,1}(Z)+1$, therefore 
$r=1$ in the notation of section two. 
The toric contraction $\pi:\wcz\ra \CZ$ is given in terms of homogeneous 
coordinates by 
\eqn\tormapA{
Z_i=\wZ_i,\ i=1,\ldots,4,\quad U=\wU \wT, \quad V=\wV\wT,\quad W=\wW} 
with exceptional divisor $(\wT)$. Note that $(\wT)$ is isomorphic 
to $\IP^1\times \IF_0$ and the map \tormapA\ contracts the $\IP^1$ 
fibers. 

Geometrically, ${\widetilde \CZ}$ is a fibration over $\IF_0$ with
$\IF_1$ fibers. This follows from the fact that each $\IP^2$ fiber of 
$\CZ\ra \IF_0$ undergoes an embedded one point blow-up. 
The Mori cone of $\wcz$ is generated by 
(see equation (A.1)) 
\eqn\MorigenB{\vbox{\halign{ $#$ \hfill &\qquad  $#$ \hfill \cr
\wh_1:\quad \wZ_1=\wZ_3=\wT=0, & \wh_2:\quad \wZ_1=\wU=\wT=0,\cr
\wh_3:\quad \wZ_3=\wU=\wT=0, & \wh_4:\quad \wZ_1=\wZ_3=\wU=0.\cr}}}
Note that $\wh_1, \wh_4$ are vertical classes and $\wh_2,\wh_3$ are 
horizontal classes on $\wcz$. Moreover, $\wh_1$ is the fiber class 
of $(\wT)$, and $\wh_1\cdot (\wT)=-1$. Therefore in the notation of section 
two, we have a single exceptional curve class $[C]=\wh_1$, and 
one exceptional divisor $D=-(\wT)$ so that $C\cdot D = 1$.  

The K\"ahler cones of $\CZ,\wcz$ are generated by the toric 
divisors (see formulae (A.1), (A.3) in appendix A) 
\eqn\kahlergenA{\vbox{\halign{ $#$ \hfill &\qquad  $#$ \hfill &\qquad  $#$ \hfill &\qquad  $#$ \hfill\cr
J_1= (W), & J_2=(Z_1), & J_3=(Z_3) & \cr
\wJ_1= (\wU), & \wJ_2=(\wZ_1), & \wJ_3=(\wZ_3), & \wJ_4=(\wW).\cr}}}
Using \tormapA\ and standard linear relations among toric divisors, 
we find the following relations 
\eqn\kahlergenB{\eqalign{
\wJ_1= \pi^*(J_1-J_2-J_3)+(\wT),\quad \wJ_2=\pi^*(J_2), \quad 
\wJ_3=\pi^*(J_3),\quad \wJ_4 = \pi^*(J_1).}}
Therefore, the geometric part of the 
duality map \dualitymapA\ becomes in this case 
\eqn\dualitymapB{
\wt_1 = -i\l,\quad \wt_2 -\wt_1= t_2,\quad \wt_3-\wt_1=t_3,\quad 
\wt_4 +\wt_1 = t_1.} 
In order to test the 
duality predictions, \dualitymapB\ must be supplemented 
with extra relations between the open string K\"ahler parameters 
$\tau_i', \tau_i''$, $i=1,\ldots,4$ and the closed string parameters 
$\wt_\gamma$, $\gamma=1,\ldots,4$. These will be determined later. 
We have to compare the genus 
zero Gromov-Witten expansion of $\wY$ 
\eqn\GWC{
\CF_{\wY;cl}^{(0)}(g_s, \wt_\gamma) = g_s^{-2} 
\sum_{\wbeta\in H_2(\wcz, \IZ)} \wC_{\wbeta} e^{-\langle \wJ,\wbeta\rangle}}
to the genus zero open string expansion of $(Y,L)$, or, more precisely, 
to a generating functional attached to the triple $(\CZ,\oY,\oL)$. 

Since the discussion in section two was very schematic, let us 
summarize the main points of the construction. We start with a 
generating functional of the form 
\eqn\genfctA{\eqalign{
& F_{inst}(g_s, t_\alpha, \tau_i',\tau_i'') \cr 
& =\sum_{h=1}^\infty \sum_{(d_a,\beta)}
\sum_{\rho} C_{h,\rho}(d_a,\beta) 
e^{-\langle J, \beta-\sum_{a=1}^hd_a[D_{\rho(a)}]
\rangle } e^{-\sum_{a=1}^hd_a \tau_{\rho(a)}} \prod_{a=1}^h 
\Tr V_{\rho(a)}^{d_a}\cr}}
where the notation has been explained in detail in section six, 
and will not be reviewed here. The coefficients 
$C_{h,\rho}(d_a,\beta)=i_{pt}^*\CC_{h,\rho}(d_a,\beta)$ 
are defined by our main formula \locS,\ which is reproduced below 
for convenience 
\eqn\genfctB{\eqalign{
\CC_{h,\rho}(d_a,\beta) = &
{1\over |\CP|} 
\prod_{a=1}^h{\CC_{1,\rho(a)}(d_a, d_a[D_{\rho(a)}]) \over -\lambda_{t_a}H}\cr
& \times s^*\left[{1\over \prod_{a=1}^h e_G(\Lambda^4(T_\CZ)_{P(a)})}
\int_{[\om_{0,h}(\CZ,\beta')]_G}{e_G(\CV^0)
\prod_{a=1}^h{ev_a^*(\phi_{P(a)})
\over -\kappa_a -\psi_{p_a}}}\right].\cr}}
The formal series \genfctA\ is called the open string instanton sum
and should be interpreted as a series of corrections to Chern-Simons 
theory\foot{Note that it suffices to included only genus zero corrections 
in this formula, in agreement with the footnote at page 24.}.
The final open string  generating functional is the Chern-Simons 
free energy with all these corrections taken into account.
Therefore the final expression for the genus zero free energy 
will be of the form 
\eqn\genfctC{\eqalign{
& \CF_{(Y,L);op}^{(0)}(g_s,t_\alpha,\l,\tau_i',\tau''_i) 
\cr
&=\CF_{Y;cl}^{(0)}(g_s,t_\alpha)+\CF_{CS,1}^{(0)}+\CF_{CS,2}^{(0)}+
\sum_{d'_i,d''_i,\beta'} F_{d'_i,d''_i,\beta}(g_s,\l) 
e^{-\langle J,\beta'\rangle}
e^{-\sum_{i=1}^4 (d'_i\tau'_i+d''_i\tau''_i)}\cr}}
where 
\eqn\GWD{
\CF_{Y;cl}^{(0)} = g_s^{-2}\sum_{\beta\in H_2(\CZ,\IZ)} C_{\beta}
e^{-\langle J,\beta\rangle}}
is the genus zero Gromov-Witten expansion of $Y$.  
The next two terms in \genfctC\ represent the genus zero contributions 
of the uncorrected Chern-Simons theories supported on $\oL_1,\oL_2$
and the last sum encodes the effect of open string instantons. 
Note that we sum over all relative homology classes $\beta = \beta' + 
\sum_{i=1}^4 (d_i'[D'_i] + d''_i[D''_i])$, with $d'_i,d''_i\geq 0$. 
In the remaining part of this section we will show by explicit 
computations that \GWC\ and \genfctC\ are in exact agreement. This is very 
strong evidence for the large $N$ duality, as well as the open string techniques developed here. 

Let us start with some remarks on the instanton coefficients 
\genfctB.\ This formula factorizes in open and closed string contributions 
reflecting the structure of a generic invariant open string map. 
The open string contributions are represented by the prefactor
$\prod_{a=1}^h{\CC_{1,\rho(a)}(d_a, d_a[D_{\rho(a)}]) \over -(\lambda_{t_a}H)}$
while the closed string contribution is the equivariant integral in 
square brackets, which takes values in 
${\cal K}_G \simeq \IQ\left(\eta_1,\ldots,\eta_7\right)$. 
This integral can be evaluated by localization 
with respect to the $G$-action on the moduli space 
$\om_{0,h}(\CZ, \beta')$. The fixed loci of this action can be 
classified using the graph method developed in 
\refs{\MK} and the evaluation of 
local contributions is standard material. The open string factors 
can be represented graphically by adding extra legs to the closed 
string graphs as in \GZ.\ 
There is however a subtlety in this 
approach, as the closed string graphs represent fixed loci under
the induced $G$-action on the moduli space of marked stable maps. 
On the other hand, $G$ does not act on the moduli space of open string 
maps, as explained in detail in section six. Hence a closed string
graph with extra legs should not be interpreted naively as a 
graphical representation of an open string fixed locus. Instead
one should think of such a graph as encompassing two sets of data
corresponding to the factorization of \genfctB\ into open and 
closed string contributions. 
The closed string data is encoded in a conventional closed 
string graph, while the open string data is encoded 
in the extra legs. We will shortly discuss concrete examples. 

The homomorphism $s^*:{\cal K}_G \ra {\cal K}_T$ has been described in 
detail in section six, below equation \valB.\ We have $s^*(\eta_{1,3}) 
=\lambda_{1,3}H$, $s^*(\eta_{2,4,7})=s^*(\eta_1+\eta_5)=s^*(\eta_3+\eta_6)=0$. 
These formulae must be extended to ${\cal K}_G$ by localization. 
Since we also have to take the nonequivariant limit of \genfctB,\ 
the final answer will be a homogeneous rational function of 
$(\lambda_1, \lambda_3)$ of degree zero. It is very convenient to express 
the answer as a rational function of the ratio $z= \l_3/\l_1$. 
Following the local examples discussed in section five, $z$ will be related 
to framing in Chern-Simons theory, provided that the later can be 
thought of as a formal variable. We will show that this algorithm gives 
rise to sensible results, although we will find some subtleties 
along the way. 

Each sphere ${\oL_1, \oL_2}$ contains two knots $\Gamma'_1, \Gamma''_1$ and 
respectively $\Gamma'_2, \Gamma''_2$ which form Hopf links with linking number 
$+1$. As explained in the paragraph below \instcorrB,\ these knots are 
endowed with the orientation induced by the canonical orientation 
of the vertical discs with respect to the inner normal convention. 
This means that the multicover contributions of the vertical discs 
will be weighted by holonomy factors of the form $\Tr (V_1')^d$, 
$\Tr (V_1'')^d$ and respectively $\Tr (V_2')^d$, $\Tr(V_2'')^d$, 
where $d>0$ is the degree. The 
contributions of the horizontal discs will be weighted by holonomy factors 
of the form $\Tr (V_1')^{-d}= \Tr (\overline{V'_1})^d$ etc.\foot{Here 
we denote by convention $\Tr_R\overline{V}=\Tr_{\overline R}V$
for any $U(N)$ group element $V$.}
The framing of $\Gamma_1',\ldots,\Gamma_2''$ can be related to the
torus weights following the same considerations as in the local case. 
We find the following relations 
\eqn\framP{
p'_1={1\over z},\quad p_1''=z,\quad p_2'=z, \quad p_2''={1\over z}.} 

After these preliminary remarks, let us turn to concrete computations. 
For a systematic approach we will distinguish several cases, depending 
on the class $\beta\in H_2(\CZ, \IZ)$. 

\subsec{Vertical Classes} 

We start with instanton corrections associated to vertical homology classes. 
This means that the discs $\Delta_a$ are mapped to the vertical discs 
$D_1', D_1'', D_2', D_2''$, and the class $\beta'$ associated to the 
closed curve $\Sigma_0$ is a multiple of $h_1$. Let us denote by 
$n$ the total degree of such a map, i.e. $\beta = nh_1$, $n\geq 0$. 
The class $\beta'$ introduced in section six will be of the form 
$\beta'= n'h_1$, with $0\leq n'<n$. For a fixed $n$ we have to sum over 
all values of $n'$. 
Except for the individual disc factors in \locS\ the computation reduces 
to the evaluation of the equivariant integral by localization.
The fixed loci in the moduli space of stable maps with marked points 
consist of points of the form $(\Sigma_0, f_0, p_a)$ 
where the marked points $p_a$ are mapped to $\{P_1', P_1'',P_2', P_2''\}$ 
and $f_0(\Sigma_0)$ is a $G$-invariant vertical curve on $\CZ$. 
The structure of the $G$-fixed locus on $\CZ$ and the 
$G$-invariant curves have been described in detail in section 6.1. 
There we found a 'skeleton' consisting of $24$ invariant curves 
which is reproduced below for convenience. 

\ifig\monster{The toric skeleton of $\CZ$ and the configuration 
of vertical discs attached to the fiber $F_1$. The color coding has been 
explained in section six.
}{\epsfxsize5.0in\epsfbox{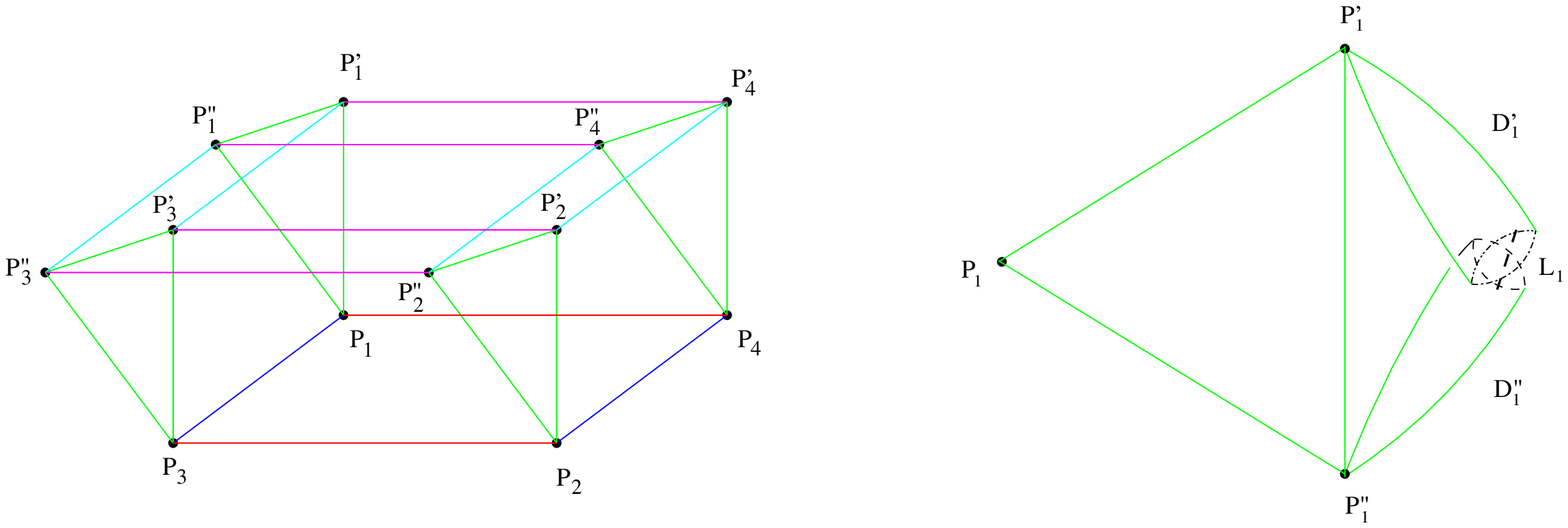}}

The vertical invariant curves form four connected (reducible) 
components lying in four distinct $\IP^2$ fibers which were denoted 
by $F_1, \ldots, F_4$. These are the 
green triangles in the above figure. Since the marked points must be 
mapped to either $\{P'_1,P''_1\}$ or $\{P_2',P_2''\}$, it follows that 
$\Sigma_0$ can be mapped either to $F_1$ or to $F_2$, which are 
disjoint fibers. Therefore the fixed loci naturally divide into two classes
depending on their image. Moreover, there is a $\IZ/2$ symmetry which 
exchanges the fixed loci in different classes. For each fixed locus mapping 
to $F_1$ there is an identical fixed locus mapping to $F_2$. A straightforward 
local computation shows that their contributions are also identical, 
therefore it suffices to consider only one class, say $G$-invariant maps 
to $F_1$. The local geometry near $F_1$ is sketched in fig. 8.
The fixed loci are represented by closed string graphs composed of 
continuous line segments. 
Additionally, we have extra legs corresponding to the open string 
factors, as explained below \valA.\  
In particular we can have pure open string graphs when the domain 
of the map $f$ consists only of disc components. 

The image of each irreducible component of the domain is specified 
by the inclination angle of the corresponding line segment and by 
the color code. Uncontracted components are represented by 
green lines with marked endpoints. 
Contracted components are represented by black line segments
with no marking at the endpoints. The degree 
of the map onto its image 
is specified by a number $d$ attached to each segment unless 
$d=0,1$ in which case the segment is left unlabeled. 
For example the graph $({\rm f}3)$ in fig. 9 represents an invariant 
map $f:\IP^1\ra 
\CZ$ which is mapped $1:1$ to ${\overline {P_1'P_1''}}$. The marked 
point  $p_1$ is mapped to $P'_1$. This corresponds to a certain 
local contribution to the equivariant integral 
\eqn\equivintT{
s^*\left[{1\over e_G(\Lambda^4(T_\CZ)_{P(1)})}
\int_{[\om_{0,h}(\CZ,h_1)]_G}{e_G(\CV^0)
{ev_1^*(\phi_{P(1)})
\over -{\eta_3} -\psi_{p_1}}}\right].}
In addition we have an open string prefactor of the form 
$C_{1,1'}(1,[D'_1])$ corresponding to the dashed line segment. 
According to the discussion below \valA,\ this should not be 
interpreted as graphical representation corresponding to a
fixed invariant map. The correct point of view is to regard 
this graphs as a pair $(\hbox{\tt closed string graph, open string graph})$ 
representing a local contribution to the instanton coefficient 
$C_{1,1'}(1,2h_1)$ defined in \genfctB.\ 
Similarly, $({\rm f}5)$ consists of a closed string graph which represents 
a $f:\IP^1 \ra 
\CZ$ which is $1:1$ onto ${\overline {P_1P_1'}}$. The marked point 
$p_1$ is mapped to $P'_1$. The extra dashed leg represents an open 
string factor given by a degree one cover of $D'_1$.  
The graph $({\rm f}15)$ has a similar interpretation, 
except that the domain of the closed string map has 
three irreducible components. 
One component is contracted (the black line), and the two other components are 
are mapped to ${\overline {P'_1P''_1}}$ with degree 
$1$. The open string data consists again of a degree one cover of 
$D'_1$. 
 
We list below all contributions of vertical maps to
$F_1$ of total degree up to $3$. Each local contribution carries 
a subscript which corresponds to a fixed locus represented in fig. 9, fig. 10 and fig. 11. 

\ifig\disksandcurves{Stable maps: degree one and two fiber class.}
{\epsfxsize4.0in\epsfbox{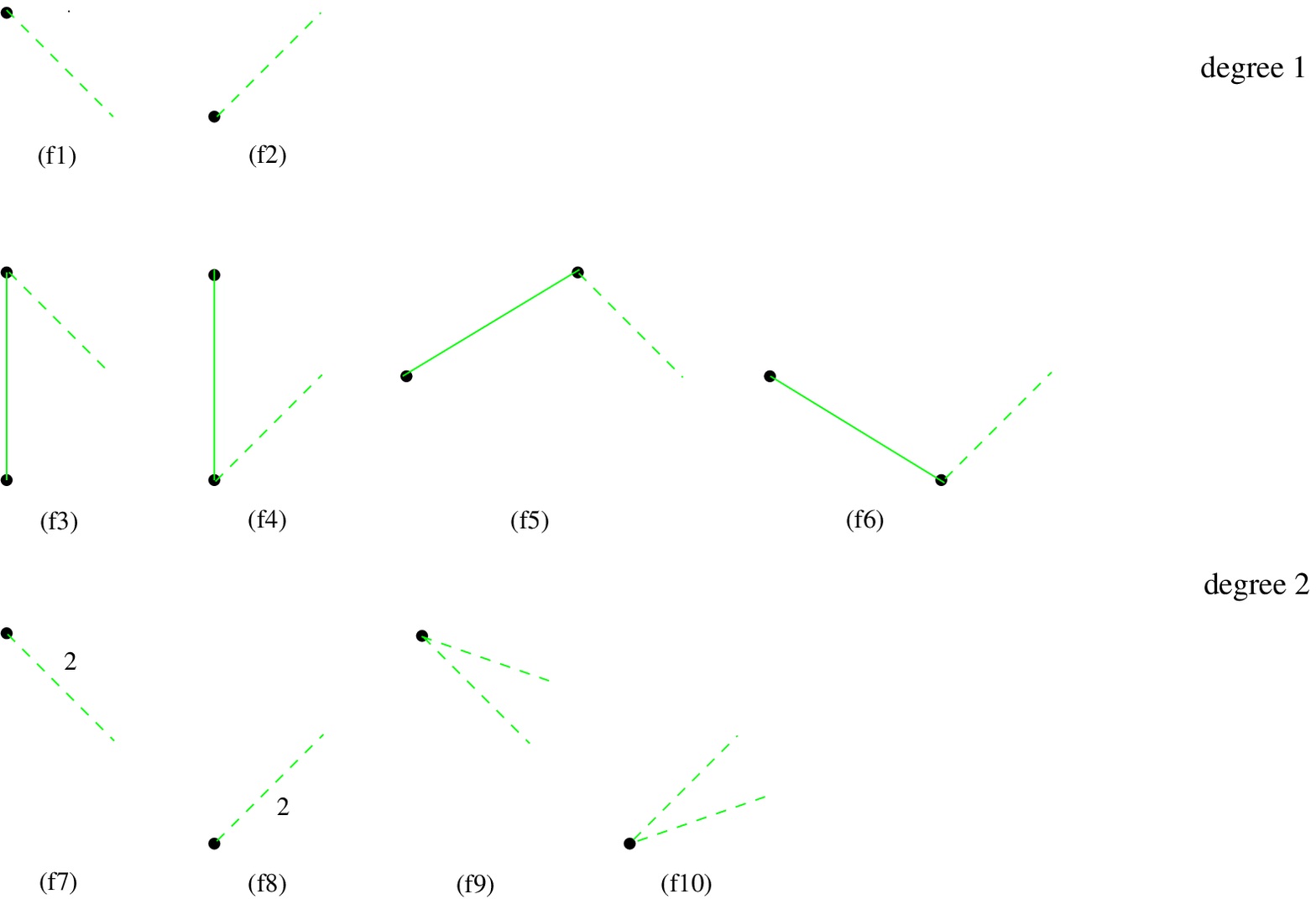}}

\noindent Degree one:
\eqn\fiberdegone{
C_{({\rm f}1)}={6z^2\over z-1},\qquad  C_{({\rm f}2)}=-{6\over z(z-1)}.}

\noindent Degree two:
\eqn\fiberdegtwo{\vbox{\halign{ $#$ \hfill &\qquad  $#$ \hfill &\qquad  $#$ \hfill \cr
C_{({\rm f}3)}={18z(z+2)(2z+1)\over (z-1)(2z-1)}, & C_{({\rm f}4)}={18(z+2)(2z+1)\over z(z-1)(z-2)}, & C_{({\rm f}5)}=C_{({\rm f}6)}=0, \cr
C_{({\rm f}7)}=-{45z^3\over 2(z-1)(z-2)}, & C_{({\rm f}8)}=-{45\over 2z(z-1)(2z-1)}, &  C_{({\rm f}9)}={3z^2\over z-1},\cr
C_{({\rm f}10)}=-{3\over z(z-1)}. & \cr }}}
\noindent Degree three:
$$\vbox{\halign{ $#$ \hfill &\qquad  $#$ \hfill \cr
C_{({\rm f}11)}={27z(z+2)^2(2z+1)^2\over (z-1)^3(2z-1)}, & C_{({\rm f}12)}={27(z+2)^2(2z+1)^2\over z(z-1)^3(z-2)}, \cr
C_{({\rm f}13)}=C_{({\rm f}14)}=0, & C_{({\rm f}15)}={27(z+2)^2(2z+1)^2\over (z-1)^3},\cr
C_{({\rm f}16)}=-{27(z+2)^2(2z+1)^2\over z(z-1)^3}, & C_{({\rm f}17)}=C_{({\rm f}18)}=0,\cr  
C_{({\rm f}19)}=C_{({\rm f}20)}=0, & C_{({\rm f}21)}=-{27z(z+2)(z+5)(2z+1)(5z+1)\over (z-1)^3(3z-1)},\cr
C_{({\rm f}22)}=-{27(z+2)(z+5)(2z+1)(5z+1)\over z(z-1)^3(z-3)}, & C_{({\rm f}23)}=C_{({\rm f}24)}=C_{({\rm f}25)}=0\cr
C_{({\rm f}26)}=C_{({\rm f}27)}=C_{({\rm f}28)}=0 & C_{({\rm f}29)}={18z(z+2)(2z+1)\over (z-1)^2},\cr
C_{({\rm f}30)}={18(z+2)(2z+1)\over z(z-1)^2}, & C_{({\rm f}31)}=C_{({\rm f}32)}=0,\cr
C_{({\rm f}33)}={36z(z+2)(2z+1)\over (z-1)^2(z-2)(2z-1)}, & C_{({\rm f}34)}=-{270z^2(z+2)(2z+1)\over (z-1)(z-2)(3z-2)},\cr
}}$$

\eqn\fiberdegthree{\vbox{\halign{ $#$ \hfill &\qquad  $#$ \hfill \cr
C_{({\rm f}35)}= {270(z+2)(2z+1)\over z(z-1)(2z-1)(2z-3)}, & C_{({\rm f}36)}={4z^2\over z-1},~~C_{({\rm f}37)}=-{4\over z(z-1)},\cr
C_{({\rm f}38)}=-{60z^3\over (z-1)(z-2)}, & C_{({\rm f}39)}=-{60\over z(z-1)(2z-1)},\cr
C_{({\rm f}40)}={1120z^4\over 3(z-1)(z-3)(2z-3)}, & C_{({\rm f}41)}=-{1120\over 3z(z-1)(3z-1)(3z-2)}.\cr
}}}

\ifig\disksandcurvesA{Stable maps: degree three fiber class - I.} 
{\epsfxsize5.0in\epsfbox{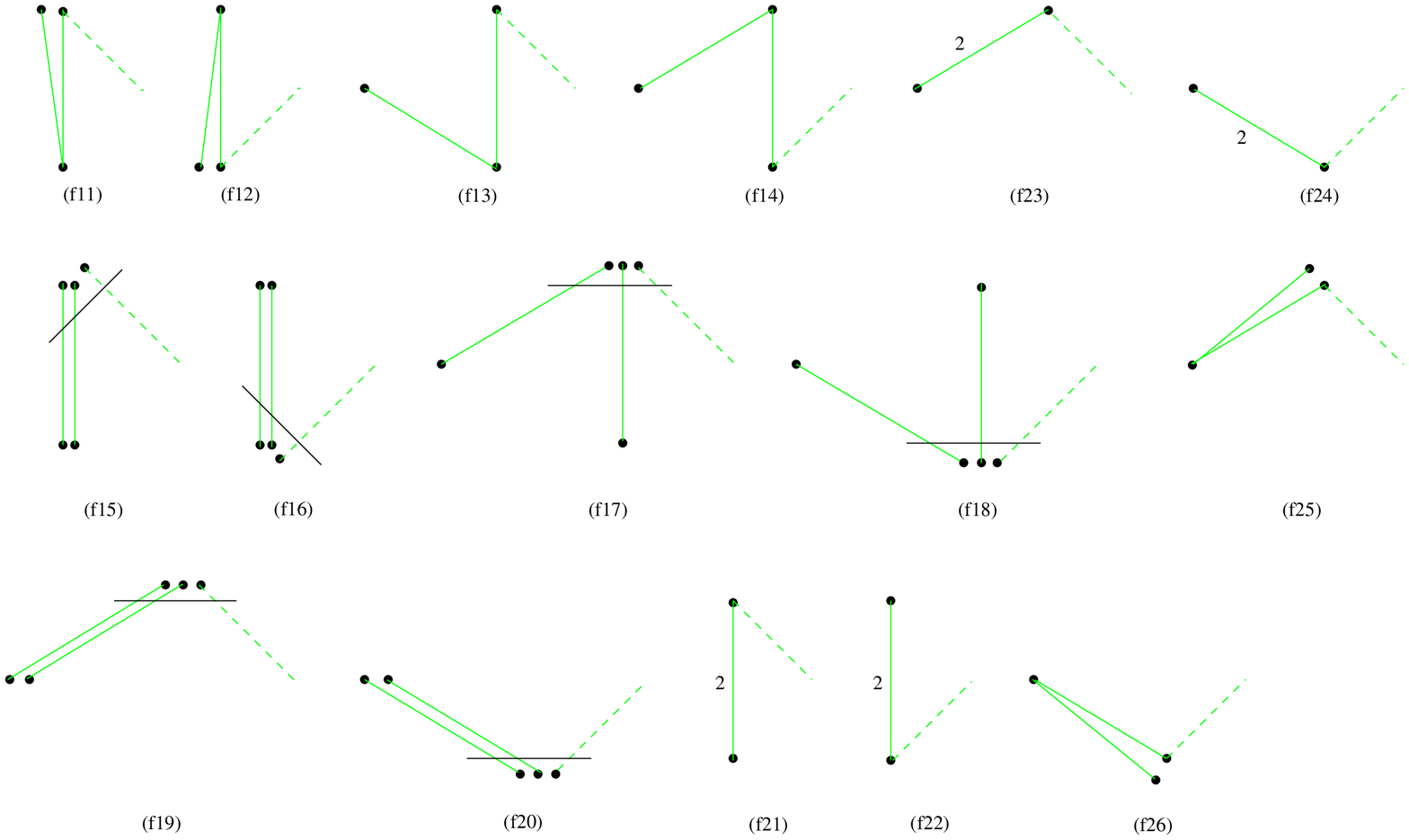}}

\noindent
We have identical corrections for vertical maps to the fiber $F_2$, 
except that the holonomy variables are different. 
Collecting all localization results, we can write the 
vertical instanton corrections in the form 
\eqn\instcorrfiberclass{\eqalign{
& F_{inst}^{(f)}=-ia_{(t'_1)}q'_1-ia_{(t_1,t'_1)}q_1q'_1-ia_{(2t'_1)}q'^2_1
 -ia_{(2t_1,t'_1)}q_1^2q'_1-ia_{(t_1,2t'_1)}q_1q'^2_1-ia_{(3t'_1)}q'^3_1,
}}
where 
$$\eqalign{
& a_{(t'_1)}={1\over g_s}\left[C_{({\rm f}1)}(\Tr V'_1+\Tr V''_2)+C_{({\rm f}2)}(\Tr V''_1+\Tr V'_2)\right],\cr
& a_{(t_1,t'_1)}={1\over g_s}\left[(C_{({\rm f}3)}+C_{({\rm f}5)})(\Tr V'_1+\Tr V''_2)+(C_{({\rm f}4)}+C_{({\rm f}6)})
(\Tr V''_1+\Tr V'_2)\right],\cr
& a_{(2t'_1)}={1\over g_s}\left[{C_{({\rm f}7)}}(\Tr V'^2_1+\Tr V''^2_2)+{C_{({\rm f}8)}}(\Tr V''^2_1+\Tr V'^2_2)\right]
-iC_{({\rm f}9)}\left[(\Tr V'_1)^2+(\Tr V''_2)^2\right]\cr
& \qquad\quad~~ -iC_{({\rm f}10)}\left[(\Tr V''_1)^2+(\Tr V'_2)^2\right],\cr
}$$
\eqn\instcorrfibfcns{\eqalign{
& a_{(2t_1,t'_1)}= {1\over g_s}\left[\mathop{\sum_{k=11}^{27}}_{k~{\rm odd}}C_{({\rm f}k)}(\Tr V'_1+\Tr V''_2)+
\mathop{\sum_{k=2}^{28}}_{k~{\rm even}}
C_{({\rm f}k)}(\Tr V''_1+\Tr V'_2)\right],\cr
& a_{(t_1,2t'_1)}=-i(C_{({\rm f}29)}+C_{({\rm f}31)})\left[(\Tr V'_1)^2+(\Tr V''_2)^2\right]
-i(C_{({\rm f}30)}+C_{({\rm f}32)})\left[(\Tr V''_1)^2+(\Tr V'_2)^2\right]\cr
&\quad-iC_{({\rm f}33)}(\Tr V'_1\Tr V''_1+\Tr V''_2\Tr V'_2)+{1\over g_s}\big[C_{({\rm f}34)}(\Tr V'^2_1+\Tr V''^2_2)
+C_{({\rm f}35)}(\Tr V''^2_1+\Tr V'^2_2)\big],\cr
& a_{(3t'_1)}= -g_s\bigg[C_{({\rm f}36)}\left[(\Tr V'_1)^3+(\Tr V''_2)^3\right]+C_{({\rm f}37)}\left[(\Tr V''_1)^3+(\Tr V'_2)^3
\right]\bigg]
-iC_{({\rm f}38)}(\Tr V'_1\Tr V'^2_1\cr
&\qquad\quad~~ +\Tr V''_2\Tr V''^2_2)-iC_{({\rm f}39)}(\Tr V''_1\Tr V''^2_1+\Tr V'_2\Tr V'^2_2)
+{1\over g_s}\big[C_{({\rm f}40)}(\Tr V'^3_1+\Tr V''^3_2)\cr
&\qquad\quad~~ +C_{({\rm f}41)}(\Tr V''^3_1+\Tr V'^3_2)\big].
}}

\ifig\disksandcurvesB{Stable maps: degree three fiber class - II.}
{\epsfxsize5.0in\epsfbox{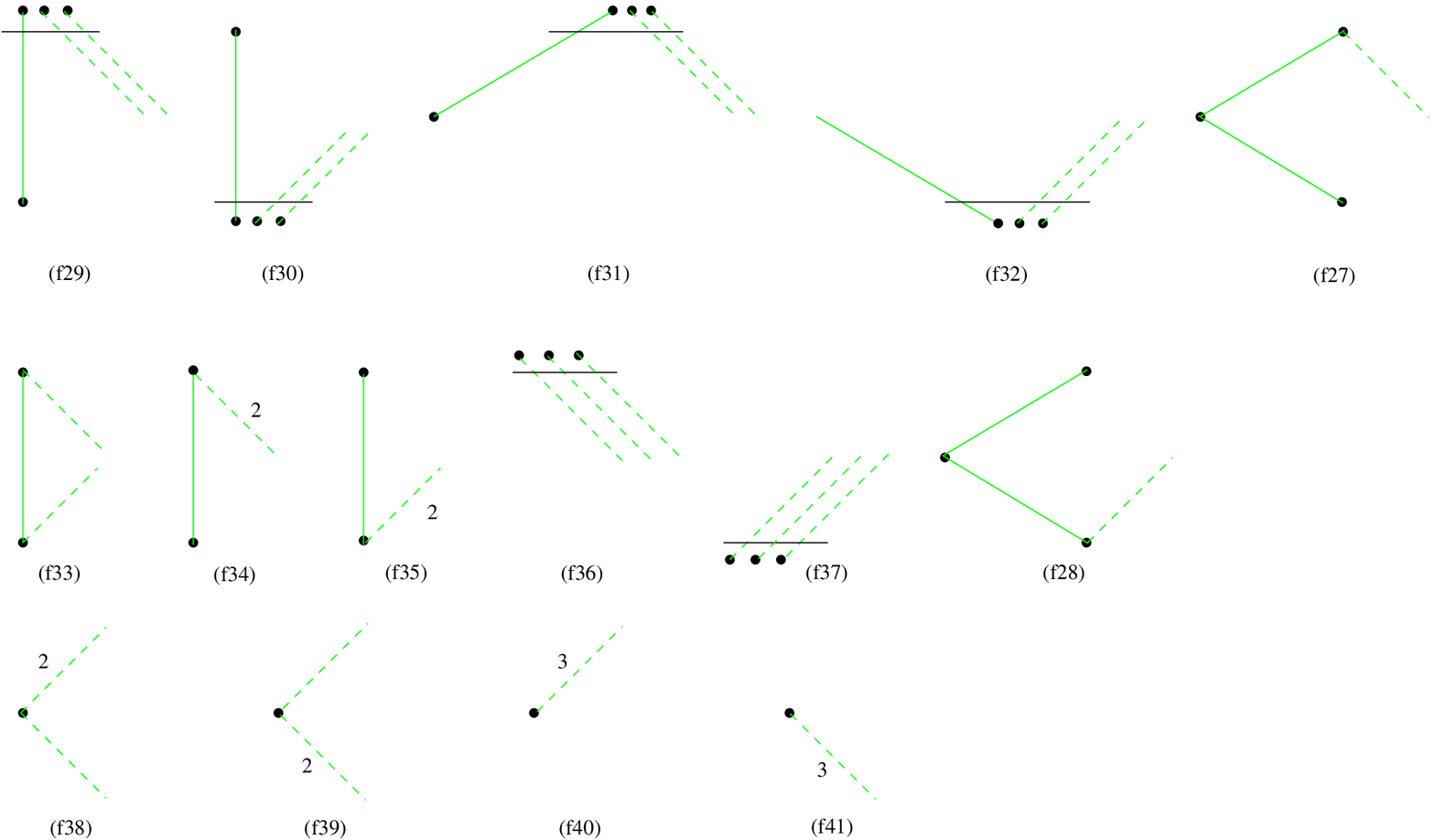}}

\noindent Some points in this formula deserve explanation. 
As we mentioned earlier,
for each disc $\Delta_a$ of the domain we have to write down an instanton 
factor of the form  $e^{-d_a\tau_{\rho(a)}}$, where $\tau_{\rho(a)}$ are 
flat open string K\"ahler parameters, which are generally different from 
the flat closed string moduli. For the vertical classes under consideration, 
we should have four
such parameters $\tau_1', \tau_1'',\tau_2',\tau_2''$ corresponding to 
$D_1', D_1'',D_2', D_2''$. 
These parameters depend linearly on the framing of the corresponding 
knots, hence a priori they have different values although all vertical 
discs are in the same homology class. 
However, once 
we absorb the trivial framing dependence by redefining the holonomy 
variables, as explained below equation \framdepB,\ we can take them to be 
equal. We let $q_1'= e^{-\tau_1'} = e^{-\tau_1''}=e^{-\tau_2'}=e^{-\tau_2''}$,
and $q_1=e^{-t_1}$ 
be the closed string instanton factor for the vertical class $h_1$. 

The next step is to compute the Chern-Simons free energy including the 
instanton corrections 
\instcorrfiberclass.\ All link and knot invariants should be 
expanded in powers of $g_s$ as explained in section five. The resulting 
expansion will automatically be a power series of the framings 
$p_1', p_1'', p'_2,p''_2$, which take the values \framP.\ 
Let us first record the final result including corrections up to 
second order in $q_1, q_1'$
\eqn\degtworesA{\eqalign{
g_s^2\ln \left<e^{ F_{inst}^{(f2)}}\right>= &
-12q_1'(y-y^{-1}){z^2+z+1\over z} \cr
& +{3\over 2} (q_1')^2y^2{14z^4+51z^3+173z^2+51z+14\over z(z-2)(2z-1)}\cr
& +3(q_1')^2{-8z^4-24z^3-86z^2-24z-8\over z(z-2)(2z-1)} \cr
& +{3\over 2}(q'_1)^2y^{-2}{2z^4-3z^3-z^2-3z+2\over z(z-2)(2z-1)}\cr
& +3q'_1q_1(y-y^{-1}){-24z^4-36z^3+12z^2-36z-24\over z(z-2)(2z-1)}\cr}}
where $y=e^{i\lambda/2}$ denotes the exponentiated 't Hooft coupling constant 
of the two Chern-Simons theories supported on $\oL_1, \oL_2$. (The two 
coupling constants must be equal as a result of the zero charge condition, 
as explained above \openB).\ 

Having reached this point, we have to face a new puzzle: unlike the local 
examples studied in section five, the final answer is a function of $z$. 
If this answer has to be taken at face value, what is the correct 
interpretation? To shed some light on this question, let us 
perform a change of variables of the form 
\eqn\dualitymapC{
q_1={\tq_1}{\tilde q_4},\qquad y=({\tilde q_1})^{1/2},
\qquad q'_1=({\tilde q_1})^{1/2}{\tilde q_4}.}
Note that the first two relations follow from the duality map 
\dualitymapB.\ The third relation involving the open string instanton 
factor $q'_1$ takes into account an extra shift in the K\"ahler 
parameters due to open string quantum corrections \DFGi.\ 
As a function of the new parameters, 
\degtworesA\ becomes 
\eqn\degtworesB{\eqalign{
g_s^2\ln \left<e^{ F_{inst}^{(f2)}}\right>= & 12{\tilde q_4}(1-{\tilde q_1})
{z^2+z+1\over z} -
{3\over 2}{\tilde q_1}^2{\tilde q_4}^2{17z^2+53z+17\over z} \cr
& +6{\tilde q_1}{\tilde q_4}^2{4z^2+13z+4\over z} +
{3\over 2}{\tilde q_4}^2{2z^4-3z^3-z^2-3z+2\over z(z-2)(2z-1)}.\cr}}
In order to understand the meaning of this expression in the context of 
large $N$ duality, it is very helpful to take a closer look at the closed 
string expansions on both sides, i.e. the series 
$\CF_{Y;cl}^{(0)}(g_s,t_\alpha)$ and $\CF_{\wY;cl}^{(0)}(g_s,\wt_\gamma)$. 

In both cases, the coefficients $C_\beta,\wC_{\wbeta}$
can be computed by localization 
using the convex obstruction bundle approach reviewed in section three.
We have studied in detail the $G$ action on $\CZ$.  
There is a similar ${\widetilde G}=(S^1)^8$ action on ${\widetilde \CZ}$ 
given by 
\eqn\toractB{
\left(e^{i\wphi_1},e^{i\wphi_2},\ldots,e^{i\wphi_8}\right) 
\cdot \left(\wZ_1, \wZ_2, \ldots, \wT\right) = \left(e^{i\wphi_1}\wZ_1,
e^{i\wphi_2}\wZ_2, \ldots, e^{i\wphi_8}\wT\right).}
whose fixed locus consists of all isolated points on $\wcz$ which can 
be represented as a quadruple intersection of toric divisors. The 
${\widetilde G}$-invariant curves on $\wcz$ form a skeleton which can be 
obtained from the skeleton of 
$\CZ$ by replacing the fibers 
$F_1,\ldots,F_4$ with $\IF_1$ surfaces 
${\widetilde F}_1,\ldots,{\widetilde F}_4$. 
For later use, note that there is a lift of the action of $T$ to 
$\wcz$ given by 
\eqn\liftactA{
\left(e^{i\wphi_1},e^{i\wphi_2},\ldots,e^{i\wphi_8}\right) =
\left(e^{i\l_1\phi}, 1, e^{i\l_3\phi}, 1, e^{-i\l_1\phi}, 
e^{-i\l_3\phi}, 1,1\right).}

Now let us write down the vertical Gromov-Witten expansions for $Y,\wY$. 
We have $\beta = nh_1$, $n\geq 0$, on $Y$ and $\wbeta = n_1\wh_1+ n_4\wh_4$,
$n_1, n_4\geq 0$ on $\wY$. Therefore we obtain 
\eqn\vertGWA{
\CF_{Y;cl}^{(0),f}(g_s, t_1) = \sum_{n\geq 0} C_n q_1^n,\qquad 
\CF_{\wY;cl}^{(0),f} = \sum_{n_1, n_4\geq 0} {\widetilde C}_{n_1n_4} 
{\tilde q_1}^{n_1} {\tilde q_4}^{n_4}}
where the coefficients $C_n, {\widetilde C}_{n_1n_4}$
have integral representations of the form 
\eqn\vertGWB{
C_n = \int_{[\om_{0,0}(\CZ, \beta)]} e(\CV),\qquad 
{\widetilde C}_{n_1n_4}= 
\int_{[\om_{0,0}({\widetilde \CZ}, {\wbeta})]}e({\widetilde \CV}).}
$\CV$, ${\widetilde \CV}$ are the obstruction bundles. 

Given the structure of the toric skeleton in both cases the fixed loci 
in the moduli spaces $\om_{0,0}(\CZ, \beta)$ and $\om_{0,0}({\widetilde \CZ}, {\wbeta})$ fall naturally in four classes,
depending on their image in $\CZ,\wcz$. 
A component of the fixed locus in $\om_{0,0}(\CZ, \beta)$ will be called 
of type $i$, $i=1,\ldots,4$ if its image is embedded in the invariant 
fiber $F_i\subset \CZ$. Similarly, a component of the fixed locus in 
$\om_{0,0}({\widetilde \CZ}, {\wbeta})$ will be called of type $i$, 
$i=1,\ldots, 4$ if its image is embedded in the invariant 
fiber $\wF_i\subset \wcz$.
The coefficients $C_n, \wC_{n_1n_4}$ receive contributions 
from all such fixed loci, that is we have 
\eqn\vertGWC{\eqalign{
& C_n = i^*_{pt} \sum_{i=1}^4 \sum_{\Xi_i} \int_{[\Xi_i]_G} {e_G(\CV_{\Xi_i})
\over e_G(N_{\Xi_i})},\qquad 
\wC_{n_1n_4} = i^*_{pt}   \sum_{i=1}^4 \sum_{\wXi_i} \int_{[\wXi_i]_\wG} 
{e_\wG(\wcv_{\wXi_i})
\over e_\wG(N_{\wXi_i})}.\cr}}
The local contributions in \vertGWC\ are homogeneous rational functions 
of degree zero of the torus weights. In order to understand the meaning of 
the open string expansion \degtworesB,\ we will specialize these 
local contributions to ${\cal K}_T$ as explained below equation \valB,\
obtaining rational functions of $z$.

Naively, large $N$ duality predicts a relation of the form 
\eqn\largeNA{
\sum_{n_1, n_4\geq 0} {\widetilde C}_{n_1n_4} 
{\tilde q_1}^{n_1} {\tilde q_4}^{n_4} = \sum_{n\geq 0} C_n q_1^n
+ \CF_{CS,1}^{(0)}+\CF_{CS,2}^{(0)}+ \ln \left<e^{ F_{inst}^{(f2)}}\right>}
where the coefficients $ C_n, {\widetilde C}_{n_1n_4}$ are given by 
\vertGWC.\
Clearly, such a relation cannot be true since $C_n, {\widetilde C}_{n_1n_4}$
are rational numbers, while the coefficients in the open string 
expansion are functions of $z$. 
However, let us recall an interesting geometric fact. 
The closed component $\Sigma_0$ of the open string maps 
which contribute to \degtworesB\ is mapped either to $F_1$ or to $F_2$. 
There are no such maps to $F_3, F_4$ since all these fibers are supported 
away from the vertical discs. This suggests that in \largeNA\ one should 
take a similar truncation of the coefficients 
$C_n, {\widetilde C}_{n_1n_4}$ by only summing over invariant 
maps to $F_1,F_2$ and respectively $\wF_1, \wF_2$. 
More precisely, the truncated coefficients are given by 
\eqn\vertGWD{
\eqalign{
& C^{tr}_n = i^*_{pt} s^*\sum_{i=1}^2 \sum_{\Xi_i} 
\int_{[\Xi_i]_G} {e_G(\CV_{\Xi_i})
\over e_G(N_{\Xi_i})},\qquad 
\wC^{tr}_{n_1n_4} = i^*_{pt}  {\widetilde s}^* 
\sum_{i=1}^2 \sum_{\wXi_i} \int_{[\wXi_i]_\wG} 
{e_\wG(\wcv_{\wXi_i})
\over e_\wG(N_{\wXi_i})}\cr}}
where $s^*:{\cal K}_G\ra {\cal K}_T$, $ {\widetilde s}^*:{\cal K}_{\wG} 
\ra {\cal K}_T$ are specialization morphisms. 
Clearly these are no longer rational numbers since we do not sum over all 
fixed loci. After specialization to ${\cal K}_T$ we will obtain
rational functions of $z$, as discussed before. 
One can regard $C^{tr}_n, \wC^{tr}_{n_1n_4}$ as equivariant refinements of 
standard Gromov-Witten invariants which probe a finer structure of the 
moduli space of maps. 
Then, we propose the following modified large $N$ duality conjecture 
\eqn\largeNB{
\sum_{n_1, n_4\geq 0} {\widetilde C}^{tr}_{n_1n_4} 
{\tilde q_1}^{n_1} {\tilde q_4}^{n_4} = \sum_{n\geq 0} C^{tr}_n q_1^n
+ \CF_{CS,1}^{(0)}+\CF_{CS,2}^{(0)}+\ln \left<e^{ F_{inst}^{(f2)}}\right>.}
Ideally, one would like to have a conceptual proof of this conjecture, 
which seems to be very difficult. Here we will restrict ourselves to a
numerical test. The truncated 
closed string expansions can be evaluated by summing over Kontsevich 
graphs \MK.\ This is a standard computation, so we will omit the details. 
Up to second degree terms, one finds the following expressions  
\eqn\vertGWD{\eqalign{
&\sum_{n\geq 0} C^{tr}_n q_1^n = 18{(z+2)(2z+1)\over z} q_1 
+ {81\over 4}{2z^2+5z+2\over z}q_1^2 + \CO(q_1^3)\cr
&\sum_{n_1, n_4\geq 0} {\widetilde C}^{tr}_{n_1n_4}
{\tilde q_1}^{n_1} {\tilde q_4}^{n_4}=
2{\tilde q_1} + {1\over 4}{\tilde q_1^2}+ 12{z^2+z+1\over z}{\tilde q}_4 +
{3\over 2}{2z^4-3z^3-z^2-3z+2\over z(z-2)(2z-1)}{\tilde q_4}^2\cr
&\qquad\qquad\qquad \qquad\qquad \ 
+6{4z^2+13z+4\over z} {\tilde q_1}{\tilde q}_4+
{3\over 4}{20z^2+29z+20\over z}{\tilde q_1}^2{\tilde q_4}^2 +
\CO({\tilde q_1}^3,\ldots, {\tilde q_4}^3).}}
Using the duality relation $q_1={\tilde q}_1{\tilde q}_4$ and 
\degtworesB,\ it is
straightforward to check that \largeNB\ is satisfied up to terms of degree 
two.  
This is positive evidence for the modified duality conjecture.
Obviously, one would like to test this conjecture at higher order 
in the expansion. The main problem is that the closed string
computations become very tedious since we have to sum over 
large numbers of graphs. Although a more thorough investigation 
is possible, 
it would be preferable to develop a more 
conceptual approach.  We leave this aspect for future work. 

Before we can continue our analysis, we should try to understand 
the meaning of the modified duality conjecture. The truncation 
\vertGWD\ seems to be necessary because the open string expansion 
cannot capture all the closed string information. For pure geometric 
reasons, on the open string side, we cannot take into account 
the effect of the fixed loci mapping to $\wF_3, \wF_4$. 
Clearly, this phenomenon 
is very specific to the present example. In other examples 
one may find that various truncations of Gromov-Witten 
invariants are needed. For example we expect that no truncation is
necessary for the compact $dP_5$ model described in section four. 
This is so because in that case we have four vanishing cycles 
rather than two, and the invariant open string maps can take values 
in the fibers $F_3, F_4$ as well. 
The general assertion one could make in this context is that 
the open string expansion computes the sum over those closed string 
graphs which are geometrically accessible. The notion of geometrically 
accessible graphs depends on the peculiarities of the model. 
We expect that such a notion and a refined duality conjecture 
can be formulated for all transitions in which the nodal points 
are fixed points of the generic torus action. This is a very interesting 
subject for future work. 

Returning to our model, there is an important observation one could 
make. While the unrefined duality conjecture \largeNA\ cannot be true 
for arbitrary values of $z$, quite remarkably, it holds true for 
the special value $z=1$! First note that if we specialize $z=1$ in 
\vertGWD\ we obtain
\eqn\vertGWE{\eqalign{
& \sum_{n\geq 0} C^{tr}_n q_1^n = 162 q_1 
+ {729\over 4}q_1^2 + \ldots \cr
& \sum_{n_1, n_4\geq 0} {\widetilde C}^{tr}_{n_1n_4}
{\tilde q_1}^{n_1} {\tilde q_4}^{n_4}=
2{\tilde q_1} + {1\over 4}{\tilde q_1^2}+ 36{\tilde q}_4 +
{9\over 2}{\tilde q_4}^2+
126{\tilde q_1}{\tilde q}_4+
{207\over 4}{\tilde q_1}^2{\tilde q_4}^2 +\ldots.}}
Although written in terms of truncated coefficients, 
these are the full genus zero Gromov-Witten 
expansions of the two models (see appendix A). The explanation is that
for $z=1$, the contributions of the fixed loci of type $3,4$ cancel each 
other, leaving only the contributions of fixed loci of type $1,2$. 
This phenomenon is not very uncommon in Gromov-Witten theory. Sometimes  
one can exploit the symmetry properties of the target space 
in order to simplify the local contributions of the fixed loci
by making a special choice of weights. See for example \FP.\  

Since the refined duality conjecture was shown to be true up to 
order two for any $z$, 
it follows that the unrefined conjecture also holds up to 
order two if we set $z=1$. 
Exploiting this feature, we can perform higher order tests of the duality 
more efficiently. Below we record the open string expansion
including terms up to order three in K\"ahler classes for $z=1$ 
\eqn\resultdegthreefiber{\eqalign{
&g_s^2\ln \left<e^{ F_{inst}^{(f)}}\right>=-36q'_1(y-y^{-1})+324q_1q'_1(y-y^{-1})
+450q'^2_1+{9\over 2}q'^2_1y^{-2}-{909\over 2}q'^2_1y^2\cr
&\quad +8748q_1^2q'_1(y-y^{-1})-6804q_1q'^2_1+5346q_1q'^2_1y^{-2}+8262q_1q'^2_1y^2 
+{4\over 3}q'^3_1y^{-3}-5344q'^3_1y^{-1}\cr
&\quad +22508q'^3_1y-{51496\over 3}q'^3_1y^3+\cdots.
}}
Using the duality map \dualitymapC\ we can rewrite this expression 
in the following form 
\eqn\degthreevert{\eqalign{
g_s^2\ln \left<e^{ F_{inst}^{(f)}}\right>=
&36\tq_4-36\tq_1\tq_4+{9\over 2}\tq_4^2+
126\tq_1\tq_4^2-{261\over 2}\tq_1^2\tq_4^2
+{4\over 3}\tq_4^3+2\tq_1\tq_4^3+152\tq_1^2\tq_4^3\cr
&-{466\over 3}\tq_1^3\tq_4^3+\cdots .
}}
This formula is to be compared to the closed string genus zero 
expansion worked out in appendix A. Before running this test,  
it is more convenient to compute the open string expansions for 
mixed and horizontal classes as well. 

\subsec{Mixed and Horizontal Classes} 

We have to perform analogous computations for instanton corrections
in homology classes of the form $\beta = n_1h_1+n_2h_2+n_3h_3$
with $n_1,n_2,n_3\geq 0$. 
The technology is very similar, except that now one has to 
consider horizontal discs as well. The equivariant integrals 
\genfctB\ can be computed as above using the graph method. 
We record below the instanton expansion and enumerate the 
relevant graphs, which should be interpreted as explained below 
\valA\ and \equivintT.\ 

\ifig\mixeddisksandcurvesoneandtwo{Stable maps: degree two mixed and degree one and two horizontal classes. 
The color coding is as before.}{\epsfxsize3.7in\epsfbox{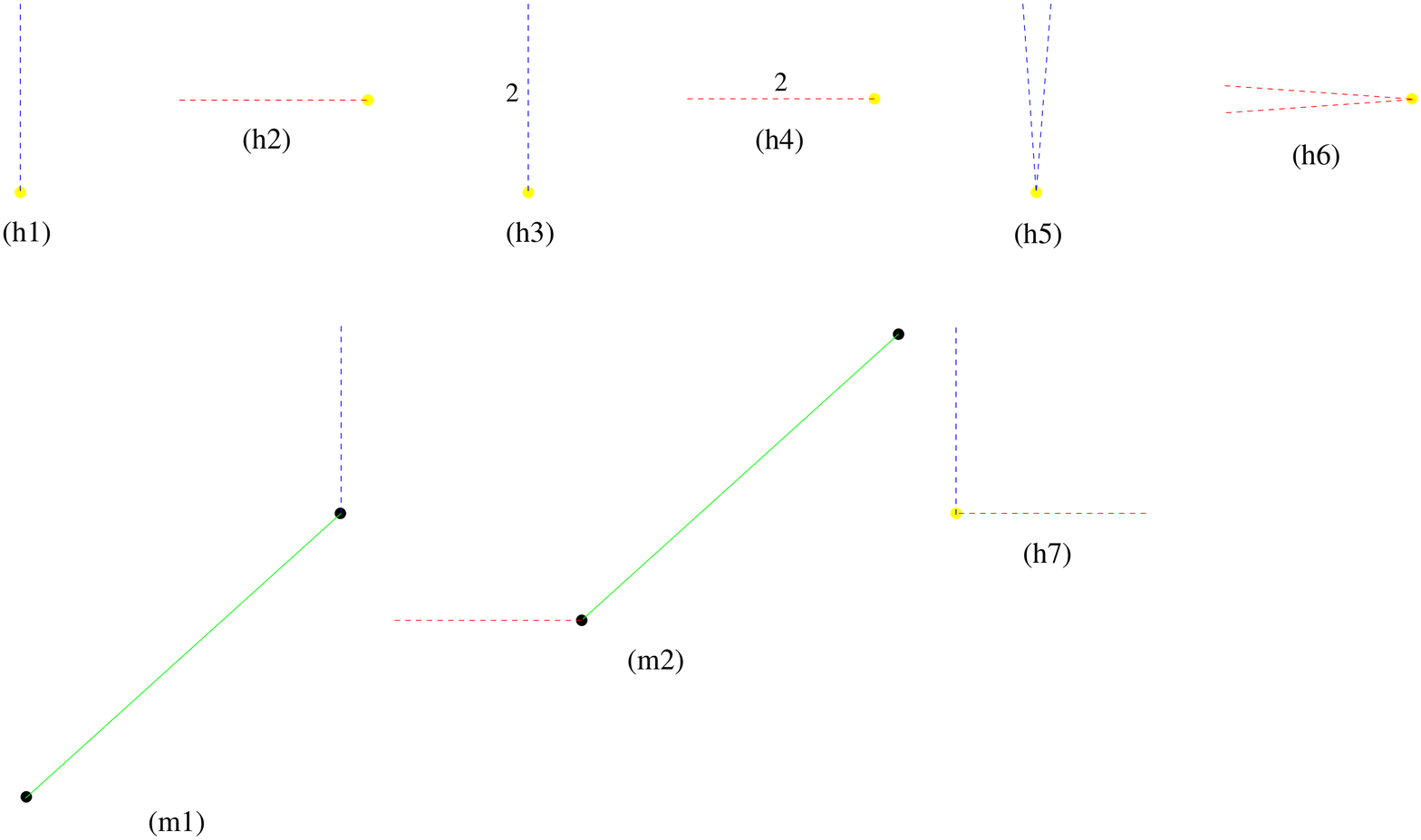}}

The closed string data is encoded in continuous line segments 
representing irreducible components of the domain. 
Each such component is either contracted or mapped to an
invariant curve in $\CZ$ as indicated by color, inclination angle 
and degree. The conventions for the degree of the map are the same as 
before, that is for $d=0,1$ the segment is left unmarked, 
while for $d\geq 2$ the segment is marked. 
The extra legs drawn with dashed colored lines represent the open 
string contributions to the instanton coefficients \genfctB.\ 
The colors correspond to different homology classes, as explained 
below fig. 6.
For example, the graphs in the first line of fig. 12 represent pure
horizontal open string contributions. The vertical blue lines 
correspond to multicover contributions of either $D'_3$ or $D''_3$, 
and the horizontal red lines correspond to multicover contributions 
of $D'_4$ or $D''_4$. Due to the symmetry properties of $\CZ$, 
the moduli space integrals are identical for $D'_3, D''_3$ and 
respectively $D'_4, D''_4$. However, the holonomy variables 
are different. This is reflected in the coefficients 
$a_{(t_2')}, a_{(t_3')}$ in equation (7.34). 
$({\rm h}7)$ is again a pure open string contribution representing a 
$1:1$ cover of either $D'_3\cup_{Q_3}D''_3$ or $D'_4\cup_{Q_4} D''_4$. 
This gives rise to $a_{(t_2',t_3')}$. $({\rm m}1)$ 
represents a contribution of the form \genfctB\ 
with $h=1$, $a=1$, $\rho(1)=3'$ (or $\rho(1)=4''$), $d'_3=1$ 
(or $d'_4=1$) and $\beta=h_1+h_2$. Therefore we have one degree 
one disc factor corresponding to a $1:1$ cover of either $D'_3$ or 
$D'_4$ and an equivariant integral on $\om_{0,1}(\CZ,h_1)$ which 
localizes on a vertical curve in $F_3$ ($F_4$) passing through 
$P'_3$ ($P''_4$). There are two such curves ${\overline {P'_3P_3}}$ and 
${\overline {P'_3P_3''}}$ in $F_3$, respectively 
${\overline {P''_4P_4}}$, ${\overline {P''_4P_4'}}$ in $F_4$.
However it can be easily checked that whenever a component 
maps to either ${\overline {P'_3P_3}}$ or ${\overline {P''_4P_4}}$
the contribution of the corresponding graph vanishes identically 
upon specialization to ${\cal K}_T$. This follows form the fact that these 
curves are fixed under $T$, hence the tangent toric weight vanishes. 
Such graphs will not be included in the figures. 
Overall, we are left with the contribution of the vertical 
curves ${\overline {P'_3P_3''}}$ and ${\overline {P''_4P_4'}}$; they determine 
the term $a_{(t_1,t'_2)}$. 
The last graph $({\rm m}2)$ is very similar and yields $a_{(t_1, t'_3)}$. 
The same rules apply to the third degree graphs represented in fig. 13 
and fig. 14. The final expression for all relevant instanton corrections 
is\foot{Note that in this formula we have to include pure vertical 
corrections up to degree two as well. Such terms multiply the 
existing mixed and horizontal 
corrections in the Chern-Simons expansion giving rise to new mixed 
terms in the final answer.}

\ifig\mixedbetaA{Stable maps: degree three mixed classes - I. 
The color coding is as before.}
{\epsfxsize5.3in\epsfbox{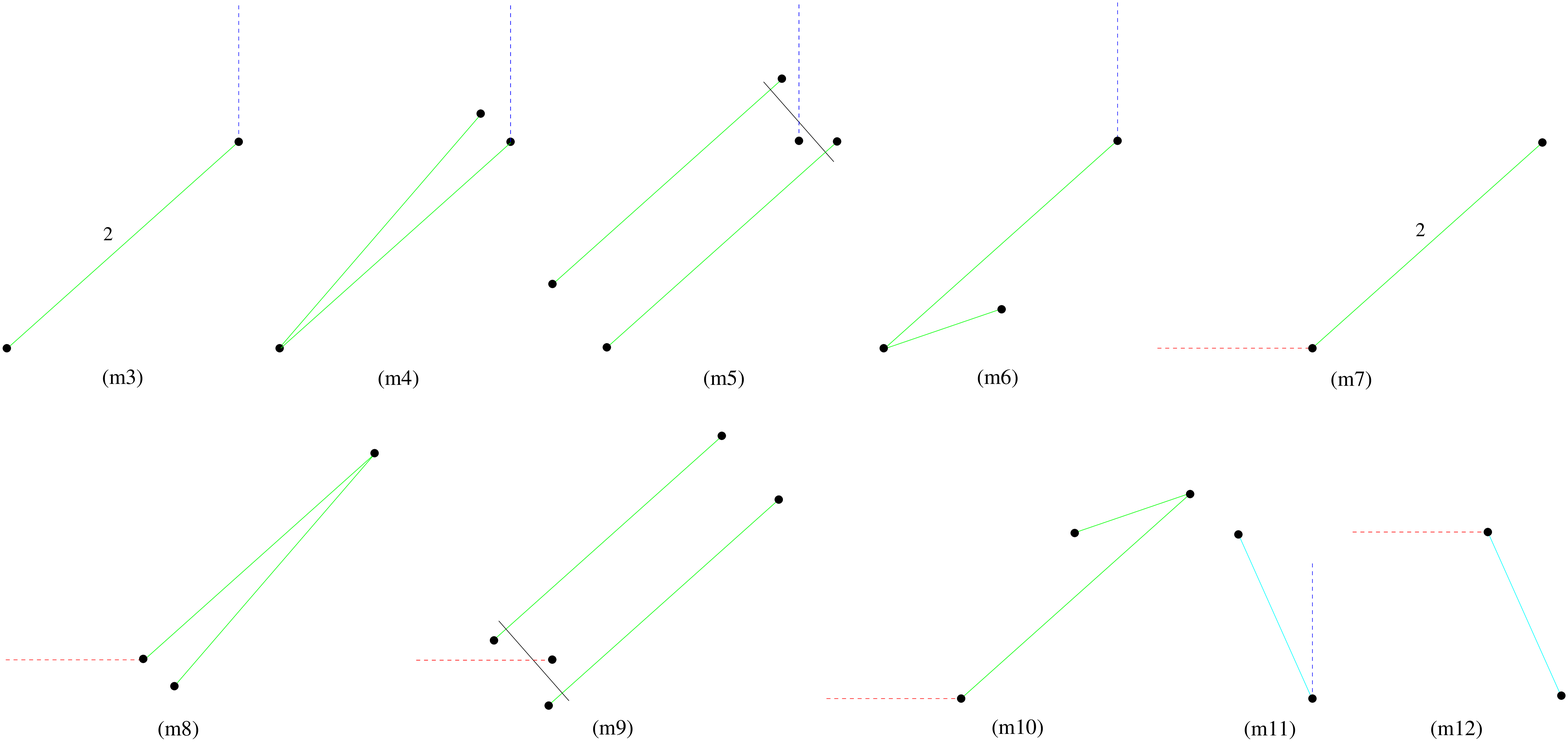}}

\eqn\instcorrmixed{\eqalign{
F_{inst}^{(m,h)}=&-ia_{(t'_1)}q'_1-ia_{(t'_2)}q'_2-ia_{(t'_3)}q'_3-ia_{(t_1,t'_1)}q_1q'_1-ia_{(t_1,t'_2)}q_1q'_2
-ia_{(t_1,t'_3)}q_1q'_3\cr 
& -ia_{(2t'_1)}q'^2_1-ia_{(2t'_2)}q'^2_2-ia_{(2t'_3)}q'^2_3-ia_{(t'_2,t'_3)}q'_2q'_3
-ia_{(2t_1,t'_2)}q_1^2q'_2-ia_{(2t_1,t'_3)}q_1^2q'_3\cr
&-ia_{(t_1,t_2,t'_1)}q_1q_2q'_1 -ia_{(t_1,t_2,t'_2)}q_1q_2q'_2-ia_{(t_1,t_2,t'_3)}q_1q_2q'_3
-ia_{(t_1,t_3,t'_1)}q_1q_3q'_1\cr
& -ia_{(t_1,t_3,t'_2)}q_1q_3q'_2-ia_{(t_1,t_3,t'_3)}q_1q_3q'_3-ia_{(t_1,2t'_2)}q_1q'^2_2-ia_{(t_1,2t'_3)}q_1q'^2_3
}}
\noindent where 
\eqn\mixedcoeffsoneandtwo{\eqalign{
& a_{(t'_2)}=-{1\over g_s}C_{({\rm h}1)}(\Tr {\overline V'_1}+\Tr {\overline V''_2}),\qquad~~~~
a_{(t'_3)}=-{1\over g_s}C_{({\rm h}2)}(\Tr {\overline V''_1}+\Tr {\overline V'_2}),\cr
& a_{(t_1,t'_2)}=-{1\over g_s}C_{({\rm m}1)}(\Tr {\overline V'_1}+\Tr {\overline V''_2}),\qquad~ 
a_{(t_1,t'_3)}=-{1\over g_s}C_{({\rm m}2)}(\Tr {\overline V''_1}+\Tr {\overline V'_2}),\cr
& a_{(2t'_2)}=-{1\over g_s}C_{({\rm h}3)}(\Tr {\overline V'^2_1}+\Tr {{\overline V''^2_2}})
-iC_{({\rm h}5)}\left[(\Tr {\overline V'_1})^2+(\Tr {\overline V''_2})^2\right],\cr
& a_{(2t'_3)}=-{1\over g_s}C_{({\rm h}4)}(\Tr {\overline V''^2_1}+\Tr {{\overline V'^2_2}})
-iC_{({\rm h}6)}\left[(\Tr {\overline V''_1})^2+(\Tr {\overline V'_2})^2\right],\cr
& a_{(t'_2,t'_3)}=-iC_{({\rm h}7)}(\Tr {\overline V'_1}\Tr {\overline V'_2}
+\Tr {{\overline V''_1}}\Tr {\overline V''_2}),\cr
& a_{(2t_1,t'_2)}=-{1\over g_s}\left(\sum_{k=3}^6C_{({\rm m}k)}\right)(\Tr {\overline V'_1}+\Tr {\overline V''_2}),\cr
& a_{(2t_1,t'_3)}=-{1\over g_s}\left(\sum_{k=7}^{10}C_{({\rm m}k)}\right)(\Tr {\overline V''_1}+\Tr {\overline V'_2}),\cr
& a_{(t_1,t_2,t'_1)}={1\over g_s}\big[(C_{({\rm m}19)}+C_{({\rm m}20)})(\Tr V'_1+\Tr V''_2)+
(C_{({\rm m}21)}+C_{({\rm m}22)})(\Tr V''_1+\Tr V'_2)\big],\cr
& a_{(t_1,t_2,t'_2)}=-{1\over g_s}C_{({\rm m}11)}(\Tr {\overline V'_1}+\Tr {\overline V''_2}),\quad a_{(t_1,t_2,t'_3)}=
-{1\over g_s}C_{({\rm m}12)}(\Tr {\overline V''_1}+\Tr {\overline V'_2}),\cr
& a_{(t_1,t_3,t'_1)}={1\over g_s}\big[(C_{({\rm m}23)}+C_{({\rm m}24)})(\Tr V'_1+\Tr V''_2)+
(C_{({\rm m}25)}+C_{({\rm m}26)})(\Tr V''_1+\Tr V'_2)\big],\cr
& a_{(t_1,t_3,t'_2)}=-{1\over g_s}C_{({\rm m}13)}(\Tr {\overline V'_1}+\Tr {\overline V''_2}),\quad a_{(t_1,t_3,t'_3)}=
-{1\over g_s}C_{({\rm m}14)}(\Tr {\overline V''_1}+\Tr {\overline V'_2}),\cr
& a_{(t_1,2t'_2)}=-{1\over g_s}C_{({\rm m}15)}(\Tr {\overline V'^2_1}+\Tr {{\overline V''^2_2}})
-iC_{({\rm m}16)}\left[(\Tr {\overline V'_1})^2+(\Tr {\overline V''_2})^2\right],\cr
& a_{(t_1,2t'_3)}=-{1\over g_s}C_{({\rm m}17)}(\Tr {\overline V''^2_1}+\Tr {{\overline V'^2_2}})
-iC_{({\rm m}18)}\left[(\Tr {\overline V''_1})^2+(\Tr {\overline V'_2})^2\right].
}}
The expressions for the coefficients $C$ obtained by 
localization are listed below.
\noindent Degree one:
\eqn\horizdegone{
C_{({\rm h}1)}=C_{({\rm h}2)}=-1.
}

\noindent Degree two:
\eqn\horizdegtwo{\vbox{\halign{ $#$ \hfill &\qquad  $#$ \hfill \cr
C_{({\rm h}3)}=-{z-2\over 4z}, & C_{({\rm h}4)}={2z-1\over 4},\cr
C_{({\rm h}5)}={z-1\over 4z^2}, & C_{({\rm h}6)}=-{z(z-1)\over 4},\cr 
C_{({\rm h}7)}=-1, & C_{({\rm m}1)}=C_{({\rm m}2)}=-{6(z-1)^2\over z}.\cr
}}}

\noindent Degree three:
\eqn\horizdegthree{\vbox{\halign{ $#$ \hfill &\qquad  $#$ \hfill \cr
C_{({\rm m}3)}={45(z-1)^2\over z(z+1)}, & C_{({\rm m}4)}=C_{({\rm m}6)}=-{6(z-1)^2\over z},\cr
C_{({\rm m}5)}=-{36(z-1)^2\over z^2}, & C_{({\rm m}7)}={45(z-1)^2\over (z+1)},\cr
C_{({\rm m}9)}=-36(z-1)^2, & C_{({\rm m}8)}=C_{({\rm m}10)}=-{6(z-1)^2\over z},\cr
C_{({\rm m}11)}=C_{({\rm m}13)}=3, & C_{({\rm m}12)}=-{6\over z(z-1)},\cr
C_{({\rm m}14)}={6z^2\over z-1}, & C_{({\rm m}15)}={6(z-1)^2\over z^2},\cr
C_{({\rm m}16)}=-{6(z-1)^2\over z^3}, & C_{({\rm m}17)}= 6(z-1)^2,\cr
C_{({\rm m}18)}=-6z(z-1)^2, & C_{({\rm m}19)}={6z^2\over z-1},\cr
C_{({\rm m}20)}={12z^2\over z-1}, & C_{({\rm m}21)}=-{9(2z-3)(z-3)\over z-1},\cr
C_{({\rm m}22)}=C_{({\rm m}24)}=0, & C_{({\rm m}23)}={9(3z-1)(3z-2)\over z(z-1)},\cr
C_{({\rm m}25)}=-{6\over z(z-1)}, & C_{({\rm m}26)}=-{12\over z(z-1)}.\cr
}}}

\ifig\mixedbetaB{Stable maps: degree three mixed classes - II. The color coding is as before.}
{\epsfxsize5.0in\epsfbox{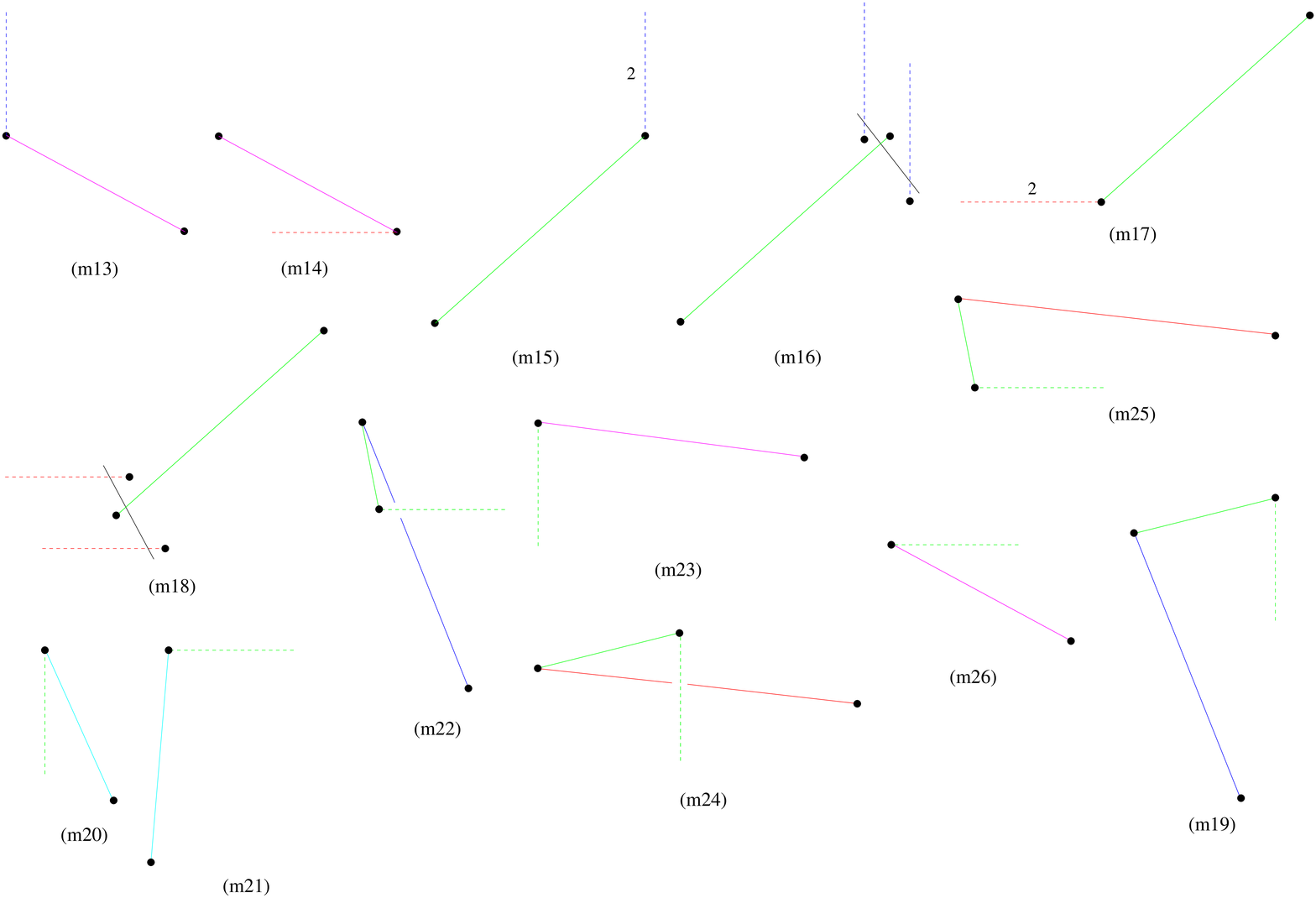}}

\noindent For the final result, we have to compute the Chern-Simons free energy 
with all these corrections included. Following the rules given so far, 
this is a straightforward, although quite tedious computation. 
In order to simplify this process we have considered only pure horizontal 
corrections up to order two instead of three. The only new aspect 
in this calculation is that one has to
evaluate Chern-Simons expectation values of 
the form $\langle \Tr_{R_1}(V_1') \Tr_{R_2}({\overline V}'_1)\rangle_{p'_1}$ 
and 
$\langle \Tr_{R_1}(V'_1) \Tr_{R_2}({\overline V}''_1)\rangle_{p'_1,p''_1}$, 
where the knots $\Gamma_1', \Gamma_1''$ form a Hopf link with linking 
number $+1$. Such expectation values are typically avoided in 
local geometric transitions by performing an analytic continuation 
\refs{\MV,\OV}. In our situation, this is not possible, hence we 
have to perform the computations directly. This entails an exercise 
in representation theory which is discussed in appendix C. 
The final result is\foot{The subscript $|(m,h)$ means that 
we only keep mixed and horizontal terms in the final answer. 
The pure vertical terms have been already taken into account in 
\degthreevert.}
$$\eqalign{
&g_s^2\ln \left<e^{ F_{inst}^{(m,h)}}\right>_{|(m,h)}=
-2q'_2(y-y^{-1})-2q'_3(y-y^{-1})-{1\over 4}q'^2_2(y^2-y^{-2})-{1\over 4}q'^2_3(y^2-y^{-2})\cr
&\quad -2q'_2q'_3(2y^2-3+y^{-2})+36q'_1q'_2(1-y^{-2})+36q'_1q'_3(1-y^{-2})
-324q_1q'_1q'_2(1-y^{-2})\cr
&\quad -324q_1q'_1q'_3(1-y^{-2})+90q'^2_1q'_2
(5y-4y^{-1}-y^{-3})+90q'^2_1q'_3(5y-4y^{-1}-y^{-3})\cr
}$$
\eqn\degthreemhA{\eqalign{
&\quad -216q_1q_2q'_1(y-y^{-1})-216q_1q_3q'_1(y-y^{-1})+36q'_1q'_2q'_3
(3y-5y^{-1}+2y^{-3})\cr
&\quad +18q_1q'_2q_3(y-y^{-1})+18q_1q_2q'_3(y-y^{-1}).
}}
Using \dualitymapC\ supplemented with the extra relations 
\eqn\dualitymapD{
q_2={\tq_2\over \tq_1},\quad q_3={\tq_3\over \tq_1},\quad q'_2=(\tq_1)^{1/2}\tq_2,
\quad q'_3=(\tq_1)^{1/2}\tq_3,}
we find the following expression in terms of closed string variables 
\eqn\degthreemhB{\eqalign{
&g_s^2\ln \left<e^{ F_{inst}^{(m,h)}}\right>_{|(m,h)}=
2(\tq_2+\tq_3)-2(\tq_1\tq_2+\tq_1\tq_3+\tq_2\tq_3)
+{1\over 4}(\tq^2_2+\tq^2_3)+6\tq_1\tq_2\tq_3-
{1\over 4}(\tq_1^2\tq_2^2\cr
&\qquad\qquad\quad +\tq_1^2\tq_3^2)-4\tq^2_1\tq_2\tq_3-36(\tq_2+\tq_3)\tq_4+
126(\tq_2+\tq_3)\tq^2_4
+36(\tq_1\tq_2+\tq_1\tq_3+\tq_2\tq_3)\tq_4\cr
&\qquad\qquad\quad -252\tq_1(\tq_2+\tq_3)
\tq^2_4-144\tq_1\tq_2\tq_3\tq_4+126\tq^2_1(\tq_2+\tq_3)\tq^2_4+
108\tq^2_1\tq_2\tq_3\tq_4.
}}
The contribution of the pure Chern-Simons sector reads
\eqn\degthreeCS{\eqalign{
&{\cal F}^{(0)}_{CS,1}+{\cal F}^{(0)}_{CS,2}=y^2_1+y^2_2+{1\over 8}
(y_1^4+y_2^4)+{1\over 27}(y_1^6+y_2^6)+\cdots,}}
which in closed string variables becomes
\eqn\resultdegthreetwochernsimons{\eqalign{
&{\cal F}^{(0)}_{CS,1}+{\cal F}^{(0)}_{CS,2}=2\tq_1+{1\over 4}\tq^2_1
+{2\over 27}\tq^3_1+\cdots.}}
The full open closed expansion of the topological free energy is obtained 
by adding \degthreevert,\ \degthreemhB,\ \degthreeCS,\ and the genus 
zero closed string contribution (A.15) expressed in terms of closed 
string variables ${\tilde q_1},\ldots, {\tilde q}_4$. 
This gives the following expression 
\eqn\finalresult{\eqalign{
& {\cal F}_{Y;cl}^{(0);inst}+{\cal F}^{(0)}_{CS,1}+{\cal F}^{(0)}_{CS,2}+g_s^2\ln \left<e^{ F_{inst}^{(f)}}\right> 
+g_s^2\ln \left<e^{ F_{inst}^{(m,h)}}\right>_{|(m,h)}\cr
&=2(\tq_1+\tq_2+\tq_3)+36\tq_4
-2(\tq_1\tq_2+\tq_1\tq_3+\tq_2\tq_3)+126(\tq_1+\tq_2+\tq_3)\tq_4+{9\over 2}\tq^2_4+{1\over 4}(\tq^2_1\cr
&+\tq^2_2+\tq^2_3)-{1\over 4}(\tq^2_1\tq^2_2+\tq^2_1\tq^2_3)
+{4\over 3}\tq^3_4+126(\tq_1+\tq_2+\tq_3)\tq^2_4+36(\tq_1\tq_2+\tq_1\tq_3\cr
&+\tq_2\tq_3)\tq_4+6\tq_1\tq_2\tq_3
+{2\over 27}\tq^3_1+2\tq_1\tq^3_4+{207\over 4}\tq^2_1\tq^2_4
+2178(\tq_1\tq_2+\tq_1\tq_3)\tq_4^2-144\tq_1\tq_2\tq_3\tq_4\cr
&-4\tq^2_1\tq_2\tq_3+152\tq^2_1\tq^3_4+126(\tq_1^2\tq_2+\tq_1^2\tq_3)\tq_4^2+108\tq_1^2\tq_2\tq_3\tq_4+{20\over 3}\tq^3_1\tq^3_4.}}
This result should be compared to the closed string expansion (A.14), 
keeping carefully track of the degrees of the terms in the two 
formulae. It can be easily seen that we obtain an exact agreement, 
except for the following terms in the closed string formula (A.14) 
\eqn\extraterms{\eqalign{
& {2\over 27}(\tq_2^3+\tq_3^3) + 2(\tq_2+\tq_3)\tq_4^3 + 
{207\over 4}(\tq_2^2+\tq_3^2)\tq_4^2 + 2178\tq_2\tq_3\tq_4^2 
-4\tq_1\tq_2\tq_3(\tq_2+\tq_3)+
152(\tq_2^2\cr &+\tq_3^2)\tq_4^3
+126(\tq_1\tq_2^2+\tq_1\tq_3^2+\tq_2^2\tq_3+\tq_2\tq_3^2)\tq_4^2
+ 108(\tq_1\tq_2^2\tq_3+\tq_1\tq_2\tq_3^2)\tq_4 + 
{20\over 3}(\tq_2^3+\tq_3^3)\tq_4^3.\cr}} 
These are either 
pure horizontal terms of degree three (the first two terms) 
or mixed and vertical terms of degree at least four. 
Such terms cannot be obtained in the open string string expansion
unless we include higher degree corrections in the open string instanton 
series. (For example, we have checked that the first two terms in 
\extraterms\ are recovered if we include degree three pure horizontal 
corrections in \instcorrmixed).\ Therefore we can conclude that there 
is perfect agreement, once the degrees are matched consistently. 
This is a very convincing evidence for the duality conjecture, and 
of the open string techniques developed in this paper.

\appendix{A}{Genus Zero Closed String Partition Functions}

In this appendix, we will compute the genus zero closed string partition functions 
of $\wy$ and $Y$ using mirror symmetry. First, 
we determine the K\"ahler and Mori cones of the 
ambient toric varieties and then we write down and 
solve the GKZ systems. From this, we will determine the 
Gromov-Witten invariants of the Calabi-Yau hypersurfaces.

\subsec{The K\"ahler and Mori Cones of $Z$ and $\wz$} 

Following the method of piecewise linear functions of \OP\ we obtain
the following basis for the Mori cone of $\IP_{\nabla_{Y}}$ 
\eqn\linrelone{
\matrix{ & \zeta_1 & \zeta_2 & \zeta_3 & \zeta_4 & \zeta_5 & \zeta_6 & \zeta_7 \cr
l^{(1)}= & 0 & 0 & 1 & 0 & 0 & 1 & 1 \cr
l^{(2)}= & 0 & 1 & -1 & 0 & 1 & -1 & 0 \cr
l^{(3)}= & 1 & 0 & -1 & 1 & 0 & -1 & 0 \cr
}}
where $\zeta_i$, $i=1,\dots, 7$ are the divisor classes corresponding
respectively to the $i$-th vertex of $\nabla_{Y}$. Therefore the 
K\"ahler cone of $\IP_{\nabla_{Y}}$ is generated by the divisor 
classes $\zeta_1$, $\zeta_2$ and $\zeta_7$. The only nonvanishing intersection 
numbers in the Calabi-Yau hypersurface $Y$ are:
\eqn\intnumbsone{
\zeta_7^3=18,\qquad \zeta_7^2\zeta_1=\zeta_7^2\zeta_2=6,\qquad\zeta_1\zeta_2\zeta_7=3.
}
Similarly, the basis of the Mori cone of 
$\IP_{\nabla_{\wy}}$ is given by
\eqn\linreltwo{
\matrix{ & \tzeta_1 & \tzeta_2 & \tzeta_3 & \tzeta_4 & \tzeta_5 & \tzeta_6 & \tzeta_7 & \tzeta_8 \cr
{{\tilde l}}^{(1)}= & 0 & 0 & 1 & 0 & 0 & 1 & 0 & -1 & \cr
{{\tilde l}}^{(2)}= & 0 & 1 & 0 & 0 & 1 & 0 & 0 & -1 & \cr
{{\tilde l}}^{(3)}= & 1 & 0 & 0 & 1 & 0 & 0 & 0 & -1 & \cr
{{\tilde l}}^{(4)}= & 0 & 0 & 0 & 0 & 0 & 0 & 1 & 1 & \cr
}}
where $\tzeta_i$, $i=1,\dots, 8$ are the divisor classes corresponding
respectively to the $i$-th vertex of $\nabla_{\wy}$. Then the K\"ahler 
cone of $\IP_{\nabla_{\wy}}$ is generated by the divisor classes
$\tzeta_4$, $\tzeta_5$, $\tzeta_6$ and $\tzeta_7$. The triple intersections can be
easily computed using {\tt SCHUBERT} \KSAS . The nonvanishing intersection 
numbers in the Calabi-Yau hypersurface $\wy$ are:
\eqn\intnumbstwo{\vbox{\halign{ $#$ \hfill &\qquad  $#$ \hfill &\qquad  $#$ \hfill &\qquad  $#$ \hfill \cr
{\tzeta_7}^3=18, & \tzeta_4{\tzeta_7}^2=6, & \tzeta_5{\tzeta_7}^2=6, & \tzeta_6{\tzeta_7}^2=6,\cr
\tzeta_4\tzeta_5\tzeta_7=3, & \tzeta_4\tzeta_6\tzeta_7=3, & \tzeta_5\tzeta_6\tzeta_7=3, & \tzeta_4\tzeta_5\tzeta_6=2.\cr
}}}
We also have $c_2(\wy)\tzeta_7=72$ and $c_2(\wy)\tzeta_4=c_2(\wy)\tzeta_5=c_2(\wy)\tzeta_6=24$.

\subsec{The GKZ Operators and the Prepotential}

The GKZ operators
associated with the Mori cone generators \linreltwo\ read
\eqn\gkzops{\eqalign{
&{\widetilde{\cal D}}_i={\wTheta_i}^2-\tz_i(\wTheta_4-\wTheta_1-\wTheta_2-\wTheta_3)(2\wTheta_4+\wTheta_1+\wTheta_2+\wTheta_3+1),
~~~i=1,\dots, 3,\cr
&{\widetilde{\cal D}}_4=\wTheta_4(\wTheta_4-\wTheta_1-\wTheta_2-\wTheta_3)-\tz_4\prod_{i=1}^2(2\wTheta_4+\wTheta_1+\wTheta_2
+\wTheta_3+i),
}} 
where $\tz_i$, $i=1,\dots, 4$ are the large complex structure coordinates
and $\wTheta_i=\tz_i{\partial\over\partial \tz_i}$, $i=1,\dots, 4$. Note that
the GKZ operators are sufficient to determine a complete set of period
integrals for the mirror of $\wy$ since $\nabla_{\wy}$ does not have
any points interior to codimension one faces and admits a (unique)
maximal triangulation with all the cones of unit volume
\refs{\HKTY,\HLY}. Also note that, as a consistency check, it is
possible to rederive the intersection numbers \intnumbstwo\
starting from the principal part of the GKZ operators \gkzops .

We denote by $\varpi_0$ the solution of \gkzops\ that is analytic at
$\tz_i=0$, $i=1,\dots, 4$. This solution is given by the series 
\refs{\CdFKM,\HKTY}
\eqn\periodzero{\eqalign{
\varpi_0=&\sum_{n_1,n_2,n_3,n_4}f_0\cr
=&\sum_{n_1,n_2,n_3,n_4}{\tz^{n_1}_1\tz^{n_2}_2\tz^{n_3}_3\tz^{n_4}_4
\Gamma(1+n_1+n_2+n_3+2n_4)\over \Gamma(1+n_1)^2\Gamma(1+n_2)^2\Gamma(1+n_3)^2
\Gamma(1+n_4)\Gamma(1-n_1-n_2-n_3+n_4)}.
}}
There are four solutions of \gkzops\ that are asymptotically like $\ln
\tz_i$, $i=1,\dots, 4$
\eqn\degonepers{
\varpi_i=\varpi_0\ln \tz_i+f_i,
}
where
\eqn\thefs{\eqalign{
&f_i=\sum_{n_1,n_2,n_3,n_4}\left[f_0(S_{n_1+n_2+n_3+2n_4}-2S_{n_i}+S_{-n_1-n_2-n_3+n_4})+h\right],~~~i=1,\dots, 3,\cr
&f_4=\sum_{n_1,n_2,n_3,n_4}\left[f_0(2S_{n_1+n_2+n_3+2n_4}-S_{n_4}-S_{-n_1-n_2-n_3+n_4})-h\right].
}}
In the above we have introduced the notation $S_n=\sum_{k=1}^n{1\over
k}$ and we have defined
\eqn\theh{
h={(-\tz_1)^{n_1}(-\tz_2)^{n_2}(-\tz_3)^{n_3}(-\tz_4)^{n_4}
\Gamma(1+n_1+n_2+n_3+2n_4)\Gamma(n_1+n_2+n_3-n_4)\over\Gamma(1+n_1)^2
\Gamma(1+n_2)^2\Gamma(1+n_3)^2\Gamma(1+n_4)}.}
Following \refs{\CdFKM,\HKTY} we can now write down the mirror map
\eqn\mirrormap{
\ttt_i=-{\varpi_i\over\varpi_0}=-\ln \tz_i -{f_i\over\varpi_0},~~~i=1,\dots, 4.}

In order to compute the genus zero Gromov-Witten invariants of $\wy$,
we need to find the second order solutions of the system of
differential equations \gkzops . These are easily obtained to
be given by
\eqn\secordgkz{\eqalign{
&\varpi_{ij}=\varpi_0\ln \tz_i\ln \tz_j+\ln \tz_if_j+\ln \tz_jf_i+f_{ij},~~~i,j=1,\dots,
3,~~~i\neq j,\cr
&\varpi_{i4}=\varpi_0\ln \tz_i\ln \tz_4+\ln \tz_if_4+\ln \tz_4f_i+f_{i4},~~~i=1,\dots,
3,\cr
&\varpi_{44}=\varpi_0(\ln \tz_4)^2+2\ln \tz_4f_4+f_{44},\cr
}} 
where
\eqn\thesecondfs{\eqalign{
&f_{ij}=\sum_{n_1,n_2,n_3,n_4}\big[f_0\big(S_{n_1+n_2+n_3+2n_4}^2
-S_{(n_1+n_2+n_3+2n_4)^2}-2S_{-n_1-n_2-n_3+n_4}(S_{n_i}+S_{n_j})\cr
&~~~~+2S_{n_1+n_2+n_3+2n_4}(S_{-n_1-n_2-n_3+n_4}-S_{n_i}-S_{n_j})+S_{-n_1-n_2-n_3+n_4}^2+S_{(-n_1-n_2-n_3+n_4)^2}\cr
&~~~~+4S_{n_i}S_{n_j}+2S_{(n_i)^2}\delta_{ij}\big)+2h\big(S_{n_1+n_2+n_3-n_4-1}+S_{n_1+n_2+n_3+2n_4}-S_{n_i}-S_{n_j}\big),~~~i,j=1,\dots, 3,\cr
&~~~~i\neq j,\cr
&f_{i4}=\sum_{n_1,n_2,n_3,n_4}\big[f_0\big(2S_{n_1+n_2+n_3+2n_4}^2
-2S_{(n_1+n_2+n_3+2n_4)^2}+S_{-n_1-n_2-n_3+n_4}(2S_{n_i}-S_{n_4})\cr
&~~~~+S_{n_1+n_2+n_3+2n_4}(S_{-n_1-n_2-n_3+n_4}-4S_{n_i}-S_{n_4})-S_{-n_1-n_2-n_3+n_4}^2-S_{(-n_1-n_2-n_3+n_4)^2}\cr
&~~~~+2S_{n_i}S_{n_4}\big)+h\big(-2S_{n_1+n_2+n_3-n_4-1}+S_{n_1+n_2+n_3+2n_4}+2S_{n_i}-S_{n_4}\big)\big],~~~i=1,\dots, 3,\cr
&f_{44}=\sum_{n_1,n_2,n_3,n_4}\big[f_0\big(4S_{n_1+n_2+n_3+2n_4}^2
-4S_{(n_1+n_2+n_3+2n_4)^2}+2S_{n_4}S_{-n_1-n_2-n_3+n_4}+S_{n_4}^2\cr
&~~~~+S_{(n_4)^2}-4S_{n_1+n_2+n_3+2n_4}(S_{n_4}+S_{-n_1-n_2-n_3+n_4})+S_{-n_1-n_2-n_3+n_4}^2
+S_{(-n_1-n_2-n_3+n_4)^2}\big)\cr
&~~~~+2h(S_{n_1+n_2+n_3-n_4-1}-2S_{n_1+n_2+n_3+2n_4}+S_{n_4})\big].
}}
In \thesecondfs\ we have introduced the notation 
$S_{(n)^2}=\sum_{k=1}^n{1\over k^2}$. 

Taking into account the triple intersections \intnumbstwo,\ a
consequence of mirror symmetry is that the following equations hold true 
\refs{\CdFKM,\HKTY,\HLY}:
\eqn\derivsprepot{\eqalign{
&\partial_{\ttt_i}{\cal
F}_{\wy;cl}^{(0)}=-{1\over\varpi_0}\left[2\varpi_{jk}+3\sum_{j=1}^3
\varpi_{j4}+3\varpi_{44}\right],~~~i=1\dots 3,~~~j,k\neq i,4,\cr
&\partial_{\ttt_4}{\cal F}_{\wy;cl}^{(0)}=-{3\over\varpi_0}
\left[\mathop{\sum_{i,j,k=1}^3}_{i<j<k}\varpi_{ij}+2\sum_{i=1}^3\varpi_{i4}+3\varpi_{44}\right].}}

The system of equations \derivsprepot\ completely determines the 
prepotential. Up to degree six in the instanton 
expansion\foot{We only include terms that correspond to instanton corrections of degree less or equal than three on $Y$.}, 
the prepotential for $\wy$ reads 

\vfill\eject 
\noindent
\eqn\theprepotential{\eqalign{
{\cal
F}_{\wy;cl}^{(0)}&=-3\ttt^3_4-3(\ttt_1+\ttt_2+\ttt_3)\ttt^2_4-3(\ttt_1\ttt_2+\ttt_1\ttt_3+\ttt_2\ttt_3)\ttt_4
-2\ttt_1\ttt_2\ttt_3+\ttt_1
+\ttt_2+\ttt_3+3\ttt_4\cr
&+2(\tq_1+\tq_2+\tq_3)+36\tq_4
-2(\tq_1\tq_2+\tq_1\tq_3+\tq_2\tq_3)+126(\tq_1+\tq_2+\tq_3)\tq_4+{9\over 2}\tq^2_4\cr
&+{1\over
4}(\tq^2_1+\tq^2_2+\tq^2_3)+{4\over
3}\tq^3_4+126(\tq_1+\tq_2+\tq_3)\tq^2_4+36(\tq_1\tq_2+\tq_1\tq_3+\tq_2\tq_3)\tq_4+6\tq_1\tq_2\tq_3\cr
&+{2\over
27}(\tq^3_1+\tq^3_2+\tq^3_3)+2(\tq_1+\tq_2+\tq_3)\tq^3_4
-{1\over 4}(\tq_1^2\tq_2^2+\tq_1^2\tq_3^2+\tq_2^2\tq^2_3)
+{207\over 4}(\tq^2_1+\tq^2_2+\tq^2_3)\tq^2_4
\cr
&+2178(\tq_1\tq_2+\tq_1\tq_3+\tq_2\tq_3)\tq_4^2-144\tq_1\tq_2\tq_3\tq_4-4\tq_1\tq_2\tq_3(\tq_1+\tq_2+\tq_3)\cr
&+152(\tq^2_1+\tq^2_2+\tq^2_3)\tq^3_4+126
(\tq_1^2\tq_2+\tq_1\tq_2^2+\tq_1^2\tq_3+\tq_1\tq_3^2+\tq_2^2\tq_3+\tq_2\tq_3^2)\tq_4^2\cr
&+108(\tq_1^2\tq_2\tq_3+\tq_1\tq_2^2\tq_3+\tq_1\tq_2\tq_3^2)\tq_4+{20\over 3}(\tq^3_1+\tq^3_2+\tq^3_3)\tq^3_4\cdots ,
}}
where $\tq_i=e^{-\ttt_i}$, $i=1,\ldots, 4$.

Proceeding in a similar manner, we obtain the prepotential for $Y$ up to degree three in the instanton 
expansion 
\eqn\theprepotentialbefore{\eqalign{
{\cal
F}_{Y;cl}^{(0)}&=-3t_1^3-3t_1^2t_2-3t_1^2t_3-3t_1t_2t_3+3t_1+t_2+t_3+162q_1+{729\over 4}q_1^2
+162(q_1q_2+q_1q_3)\cr
&+162q_1^3+2430(q_1^2q_2+q_1^2q_3)+\cdots ,
}}
where $q_i=e^{-t_i}$, $i=1,\ldots, 3$.

\appendix{B}{Disc Multicovers}

Here we summarize the computation of disc multicover 
contributions using the convex obstruction bundle  
approach. As explained in section six, it suffices to consider 
only the discs $D'_1, D'_3$ since all other cases are similar. 

\subsec{Multicovers of $D_1'$}

Recall that the disc $D'_1$ is obtained by intersecting the curve 
$C'_{13}$ with the 3-cycle $\oL_1$, according to the local analysis 
of section 6.2. The defining equations of $D'_1$ in the coordinate 
patches $\CU_1=\{Z_2\neq 0, Z_4\neq 0, W\neq 0\}$, 
$\CU_1' = \{Z_2\neq 0, Z_4\neq 0, V \neq 0\}$ are 
\eqn\discsA{
{\vbox{\halign{ $#$ \hfill &\qquad  $#$ \hfill &\qquad   $#$ \hfill &
\qquad  $#$ \hfill &\qquad   $#$ \hfill \cr
{\cal U}_1: & x_1=u_1=0, & y_1v_1=\mu, & |y_1|\leq \mu^{1/2} & 
|v_1|\geq {\mu}^{1/2}\cr
{\cal U}'_1: &  x'_1=u'_1=0, & y'_1=\mu v'_1, &|y'_1| \leq \mu^{1/2} , 
& |v'_1|\leq \mu^{-1/2}.\cr
}}}}

We have to compute the cohomology groups $H^0(\Delta, \CT)$, $H^1(\Delta, \CT)$ 
and the obstruction groups $H^0(\Delta, \CL)$, $H^1(\Delta, \CL)$ defined in section 6.3. 
The computation 
is performed in {\v C}ech cohomology, as in \refs{\DFGii,\LS}. 
We will work with the following two set cover of the domain $\Delta$  
\eqn\openA{
\Upsilon'_1=\{0\leq |t'|<\mu^{-1/2d'_1}\},\qquad 
\Upsilon_1=\{\mu^{1/2d'_1}\leq |t|<(\mu+\epsilon^2)^{1/2d'_1}\}.}
A Galois cover of 
degree $d'_1$ of $D'_1$ is given in local coordinates by
\eqn\localcoordA{\vbox{\halign{ $#$ \hfill &\qquad  $#$ \hfill 
&\qquad   $#$ \hfill &\qquad  $#$ \hfill &\qquad   $#$ \hfill \cr
{\cal U}'_1: & x'_1(t')=0, & y'_1(t')=\mu t'^{d'_1}, & u'_1(t')=0, 
& v'_1(t')=t'^{d'_1}\cr
{\cal U}_1: &  x_1(t)=0, & y_1(t)=\mu/ t^{d'_1}, & u_1(t)=0, 
& v_1(t)=t^{d'_1}.\cr
}}}
Proceeding as in \DFGii\ we construct the {\v C}ech complex
\eqn\cechBnew{ 0\ra \CT_{\Delta}(\Upsilon'_1)\oplus 
\CT_{\Delta}(\Upsilon_1) {\buildrel \kappa
\over \ra} \CT_{\Delta}(\Upsilon'_{1}\cap\Upsilon_1) \ra 0}
where the generic sections in $\CT_{\Delta}(\Upsilon'_1)$, $\CT_{\Delta}(\Upsilon_1)$ 
can be written as 
\eqn\locsectCnew{\eqalign{
& s'_1 = \left(\sum_{n=0}^{\infty}\alpha'_n {t'}^n\right)
\partial_{x'_1}+ \left(\mu\sum_{n=0}^\infty \beta'_n {t'}^n\right)
\partial_{y'_1} + \left(\sum_{n=0}^\infty \gamma'_n {t'}^n\right)
\partial_{u'_1} + \left(\sum_{n=0}^\infty \delta'_n {t'}^n\right)
\partial_{v'_1}\cr & s_1 = \left(\sum_{n\in \IZ}\alpha_n
t^n\right) \partial_{x_1}+ \left(\mu\sum_{n\in \IZ}\beta_n
t^n\right) \partial_{y_1} + \left(\sum_{n\in \IZ}\gamma_n
t^n\right) \partial_{u_1} + \left(\sum_{n\in \IZ}\delta_n
t^n\right) \partial_{v_1}.\cr}}
Note that we sum over $n\geq 0$ for sections in 
$\CT_{\Delta}(\Upsilon'_1)$. In order to have a 
uniform notation, we can extend these sums to $n \in \IZ$, 
with the convention that
$\alpha'_n, \ldots, \delta'_n$ are zero for $n<0$.
The boundary conditions at $|t|=\mu^{1/2d'_1}$ yield the following 
relations between the coefficients
\eqn\boundcondA{
\alpha_n=\mu^{-{n\over d'_1}}{\bar\gamma}_{-n},\qquad 
\beta_n=\mu^{-{d'_1+n\over d'_1}}{\bar\delta}_{-n}.
}
Now we have to compute the kernel and cokernel of $\kappa$. 
Using the linear transformations
\eqn\lintransfnew{
\partial_{x_1}=\partial_{x'_1},~~~\partial_{y_1}=\partial_{y'_1},
~~~\partial_{u_1}=v'_1\partial_{u'_1},
~~~\partial_{v_1}=-u'_1v'_1\partial_{u'_1}-{v'_1}^2\partial_{v'_1}
}
and taking into account the local equations \discsA,\
we find that $\kappa$ is given by
\eqn\mapkappanew{\eqalign{
\kappa(s'_1,s_1)&= \left(\sum_{n\in \IZ} (\alpha'_n-
\alpha_{-n}){t'}^n \right)\partial_{x'_1}+ \left(\mu\sum_{n\in \IZ}
(\beta'_n- \beta_{-n}){t'}^n\right)\partial_{y'_1}\cr
&+ \left(\sum_{n\in \IZ}
(\gamma'_n- \gamma_{-n+d'_1}){t'}^n\right)\partial_{u'_1}+
\left(\sum_{n\in \IZ} (\delta'_n
+\delta_{-n+2d'_1}){t'}^n\right)\partial_{v'_1}.
}}
Therefore, in order to find the kernel of $\kappa$ 
we have to solve the following system of 
equations
\eqn\kernelnew{
\alpha'_n-\alpha_{-n}=0,~~~\beta'_n- \beta_{-n}=0,~~~\gamma'_n- 
\gamma_{-n+d'_1}=0,~~~\delta'_n
+\delta_{-n+2d'_1}=0.
}
Since $\alpha'_n, \ldots, \delta'_n=0$ for $n<0$, we obtain
\eqn\constraintsA{\eqalign{
&\alpha_n=0,~n>0,\quad \beta_n=0,~n>0,\quad\gamma_n=0,~n>d'_1,
\quad \delta_n=0,~n>2d'_1.
}}
Moreover, combining \boundcondA\ and \constraintsA,\
we find the following relations 
\eqn\constraintsB{
\alpha_n=0,~n<-d'_1,\quad \beta_n=0,~n<-2d'_1,\quad \gamma_n=0,~n<0,
\quad \delta_n=0,~n<0.
}
Therefore $\hbox{Ker}(\kappa)=H^0(\Delta, \CT)$ is the 
$2(3d'_1+2)$-dimensional real vector space spanned by sections of the form
\eqn\kernelkappaA{
s_1=\left(\sum_{n=-d'_1}^0 {\alpha}_{n}t^n\right)
\partial_{x_1}+
\left(\mu\sum_{n=-2d'_1}^0 {\beta}_{n}t^n\right) 
\partial_{y_1}+\left(\sum_{n=0}^{d'_1}\gamma_nt^n\right)
\partial_{u_1}+\left(\sum_{n=0}^{2d'_1}
\delta_nt^n\right)\partial_{v_1}
}
with the coefficients $\alpha_n,\ldots,\delta_n$ subject to the boundary conditions \boundcondA .
Next, we compute the cokernel of $\kappa$. This is generated by 
local sections with coefficients 
$a_n,\ldots,d_n$ for which the following equations have no solutions
\eqn\cokernelnew{
\alpha'_n-\alpha_{-n}=a_n,~~~\beta'_n- \beta_{-n}=b_n,~~~\gamma'_n- 
\gamma_{-n+d'_1}=c_n,~~~\delta'_n
+\delta_{-n+2d'_1}=d_n.
}
Taking into account \boundcondA\ and the constraints 
$\alpha'_n,\ldots,\gamma'_n=0$, it is straightforward to check that 
these equations admit nontrivial solutions for any values of $a_n,\ldots,d_n$. 
Therefore, $\hbox{Coker}(\kappa)=H^1(\Delta,{\cal T}_{\Delta})$ is trivial. 

In conclusion, as shown in section 6.3, the space of infinitesimal 
deformations of the multicover \localcoordA\ is isomorphic to 
$H^0(\Delta, \CT)$. Although this is a priori only a real vector 
space, it should carry a complex structure reflecting the 
choice of an orientation on the moduli space of open string maps. 
This is a very subtle issue since we do not have a rigorous construction
of the moduli space and the virtual fundamental class.
Here we will simply choose the complex structure defined by the 
isomorphism $H^0(\Delta, \CT)\ra \IC^{3d'_1+1}$, 
$s_1\ra (\alpha_m, \beta_n),\ {m=1,\ldots,d'_1,\ n=1,\ldots, 2d'_1}$. 
With this choice, 
$H^0(\Delta,{\cal T}_{\Delta})$ is $S^1$-isomorphic to
\eqn\zerocohA{
\bigoplus_{n=0}^{d'_1}\left(-\lambda_1+{n\over d'_1}
\lambda_3\right)\oplus
\bigoplus_{n=0}^{2d'_1}\left(-\lambda_3+{n\over d'_1}\lambda_3\right).}
One could in principle make other choices by choosing different
isomoprhisms to $\IC^{3d'_1+2}$, for example 
$s_1\ra ({\gamma}_{m},{\delta}_{n}),\
{m=1,\ldots,d'_1,\ n=1,\ldots, 2d'_1}$. 
This would change the representations in \zerocohA\ by complex conjugation, 
and it would reflect in a different sign of the multicover contribution to the instanton sum. 

Now we turn to the computation of $H^0(\Delta,{\cal L}_{\Delta})$. 
We construct the {\v C}ech complex
\eqn\cechBnew{ 0\ra \CL_{\Delta}(\Upsilon'_1)\oplus \CL_{\Delta}(\Upsilon_1) 
{\buildrel \kappa
\over \ra} \CL_{\Delta}(\Upsilon'_{1}\cap\Upsilon_1) \ra 0.}
The generic sections of $\CL$ over $\Upsilon'_1$ and $\Upsilon_1$ have the 
form
\eqn\locsectDnew{\eqalign{
&{\tilde s'_1}= \left(\sum_{n=0}^{\infty}\epsilon'_n {t'}^n\right)
\Lambda'^{max},\cr
&{\tilde s_1}= \left(\sum_{n\in \IZ}\epsilon_n {t}^n\right)\Lambda^{max},
}}
where $\Lambda'^{max}=\partial_{x'_1}\wedge\partial_{y'_1}
\wedge\partial_{u'_1}\wedge\partial_{v'_1}$ 
and $\Lambda^{max}=\partial_{x_1}\wedge\partial_{y_1}
\wedge\partial_{u_1}\wedge\partial_{v_1}$. 
Using \lintransfnew\ we find that $\kappa$ is given by 
\eqn\mapkappaBnew{
\kappa({\tilde s'_1},{\tilde s_1})=\left(\sum_{n\in\IZ}
(\epsilon'_n-\epsilon_{-n+3d'_1})t'^n\right)\Lambda'^{max},
}
where it is understood that $\epsilon'_n=0$ for $n<0$. 
Then, after using the boundary condition 
$\epsilon_n={\bar\epsilon}_{-n}$ we find that the kernel of 
$\kappa$ is generated by sections of the form
\eqn\kernelkappaB{
{\tilde s}_1=\left(\epsilon_0+\sum_{n=1}^{3d'_1}
(\epsilon_nt^n+{\bar \epsilon}_nt^{-n})\right)
\Lambda^{max}.}
This is a $6d'_1+1$ real vector space which splits naturally 
in a direct sum of the form $\IR\oplus \IR^{6d'_1}$. The first summand 
is generated by sections of the form $\epsilon_0\Lambda^{max}$, 
and the second summand is defined by sections ${\tilde s}_1$ with 
$\epsilon_0=0$. One can define a complex structure on the subspace 
$\epsilon_0=0$ by an isomorphism $\IR^{6d'_1}\ra \IC^{3d'_1}$, 
${\tilde s}_1\ra (\epsilon_n),\ {n=1,\ldots,3d'_1}$. 
Then $H^0(\Delta,\CL_{\Delta})$ is $S^1$-isomorphic to
\eqn\zerocohB{
\left( 0\right)_{\IR}\oplus\bigoplus_{n=1}^{3d'_1}
\left({n\lambda_3\over d'_1}\right).}
Analogous considerations show that the cokernel of $\kappa$ is empty in 
this case, therefore  $H^1(\Delta, \CL_{\Delta})=0$. 

\subsec{Multicovers of $D'_3$}

Since $D'_1, D'_3$ are complementary discs in $C'_{13}$, $D'_3$ 
is covered by the coordinate patches 
$\CU_1=\{Z_2\neq 0, Z_4\neq 0, W\neq 0\}$, 
$\CU_3=\{Z_2\neq 0,Z_3\neq 0, W\neq 0\}$. 
The local coordinates in the second patch are  
\eqn\coordlocalnewA{
{\cal U}_3:\qquad x_3={Z_1\over Z_2},\quad y_3={Z_4\over Z_3},
\quad u_3={UZ_2Z_3\over W},\quad 
v_3={VZ_2Z_3\over W}}
and $D'_3$ is given by 
\eqn\discsB{
{\vbox{\halign{ $#$ \hfill &\qquad  $#$ \hfill &\qquad   $#$ \hfill &
\qquad  $#$ \hfill &\qquad   $#$ \hfill \cr
{\cal U}_1: & x_1=u_1=0, & y_1v_1=\mu, & |y_1|\geq \mu^{1/2} & 
|v_1|\leq {\mu}^{1/2}\cr
{\cal U}_3: &  x_3=u_3=0, & v_3=\mu,  &|y_3| \leq \mu^{-1/2}. \cr
}}}}
We choose an open cover of the domain of the form 
\eqn\openB{
\Upsilon_1=\{\mu^{1/2d'_3}\leq |t|<(\mu+\epsilon^2)^{1/2d'_3}\},\qquad 
\Upsilon_3=\{0\leq |t'|<\mu^{-1/2d'_3}\}.}
A Galois cover of degree $d'_3$ of $D'_3$ is given in local coordinates by
\eqn\localcoordB{\vbox{\halign{ $#$ \hfill &\qquad  
$#$ \hfill &\qquad   $#$ \hfill &\qquad  $#$ \hfill &\qquad   $#$ \hfill \cr
\CU_1:& x_1(t)=0, & y_1(t)=t^{d'_3}, & u_1(t)=0, & v_1=\mu t^{-d'_3}\cr
\CU_3: & x_3(t')=0, & y_3(t')=t'^{d'_3}, & u_3(t')=0, & v_3(t')=\mu.\cr
}}}
The {\v C}ech complex for $\CT$ is  
\eqn\cechBnew{ 0\ra \CT_{\Delta}(\Upsilon_1)\oplus \CT_{\Delta}(\Upsilon_3) 
{\buildrel \kappa
\over \ra} \CT_{\Delta}(\Upsilon_{1}\cap\Upsilon_3) \ra 0}
with generic sections in $\CT_{\Delta}(\Upsilon_1)$, 
$\CT_{\Delta}(\Upsilon_3)$ given by  
\eqn\locsectCnew{\eqalign{
& s_1 = \left(\sum_{n\in\IZ}\alpha_n {t}^n\right)
\partial_{x_1}+ \left(\sum_{n\in\IZ} \beta_n {t}^n\right)
\partial_{y_1} + \left(\sum_{n\in\IZ} \gamma_n {t}^n\right)
\partial_{u_1} + \left(\mu\sum_{n\in\IZ} \delta_n {t}^n\right)
\partial_{v_1}\cr & s_3 = \left(\sum_{n=0}^{\infty}\alpha'_n
t'^n\right) \partial_{x_3}+ \left(\sum_{n=0}^{\infty}\beta'_n
t'^n\right) \partial_{y_3} + \left(\sum_{n=0}^{\infty}\gamma'_n
t'^n\right) \partial_{u_3} + \left(\mu\sum_{n=0}^{\infty}\delta'_n
t'^n\right) \partial_{v_3}.\cr}}
Note that we sum over $n\geq 0$ for sections in $\CT_{\Delta}(\Upsilon_3)$. 
In order to have a 
uniform notation, we can extend these sums to $n \in \IZ$, with the 
convention that
$\alpha'_n, \ldots, \delta'_n$ are zero for $n<0$.
Using the linear transformations
\eqn\lintransfii{
\partial_{x_1}=\partial_{x_3},~~~\partial_{y_1}=-y_3^2\partial_{y_3}+
y_3v_3\partial_{v_3},~~~\partial_{u_1}=y_3^{-1}\partial_{u_3},~~~
\partial_{v_1}=y_3^{-1}\partial_{v_3}}
and proceeding as in the previous case, 
we obtain the following form for the {\v C}ech differential
\eqn\mapkappaCnew{\eqalign{
\kappa(s_1,s_3)&= \left(\sum_{n\in \IZ} (\alpha'_n-
\alpha_{-n}){t'}^n \right)\partial_{x_3}+ \left(\sum_{n\in \IZ}
(\beta'_n+ \beta_{-n+2d'_3}){t'}^n\right)\partial_{y_3}\cr
&+ \left(\sum_{n\in \IZ}
(\gamma'_n- \gamma_{-n-d'_3}){t'}^n\right)\partial_{u_3}+
\left(\mu\sum_{n\in \IZ}(\delta'_n -\beta_{-n+d'_3}-
\delta_{-n-d'_3}){t'}^n\right)\partial_{v_3}.
}}
Again, in order to determine $H^0(\Delta,\CT_{\Delta})$, 
we have to compute the kernel of $\kappa$, which is determined by the 
following system of equations 
\eqn\kernelnewii{
\alpha'_n=\alpha_{-n},~~~\beta'_n=-\beta_{-n+2d'_3},
~~~\gamma'_n=\gamma_{-n-d'_3},~~~\delta'_n=\beta_{-n+d'_3}+\delta_{-n-d'_3}.
}
But $\alpha'_n,\ldots,\delta'_n=0$ for $n<0$, and we obtain that
\eqn\constraintsAi{
\alpha_n=0,~n>0,\quad \beta_n=0,~n>2d'_3,\quad \gamma_n=0,~n>-d'_3,
\quad \beta_{-n+d'_3}+\delta_{-n-d'_3}=0,~n<0.
}
We will also need to employ boundary conditions at $|t|=\mu^{1/2d'_3}$
which yield the following relations 
\eqn\boundcondAi{
\alpha_n=\mu^{-{n\over d'_3}}{\bar\gamma}_{-n},\qquad 
\beta_n=\mu^{{d'_3-n\over d'_3}}{\bar\delta}_{-n}.
}
Using \constraintsAi\ and \boundcondAi,\ we obtain 
\eqn\constraintsAii{
\alpha_n=0,~n<d'_3,\quad\beta_n=0,~n<0,
\quad\gamma_n=0,~n<0,\quad\delta_n=0,~n<-2d'_3~{\rm or}~n>0
}
\eqn\constraintsAiii{
\beta_{d'_3+1}=\bar\beta_{d'_3-1},\quad\delta_{-d'_3+1}=\bar\delta_{-d'_3-1}.
}
Taking into account all constraints, we find that the kernel of 
$\kappa$ is generated by sections of the form
\eqn\kernelkappaBi{
s_1=\left[\sum_{n=0}^{d'_3-1}(\beta_nt^n-\bar\beta_nt^{n+d'_3+1})+
\beta_{d'_3}t^{d'_3}\right]\partial_{y_1}+\left[\mu\sum_{n=-d'_3+1}^{0}
(\delta_nt^n-\bar\delta_nt^{-n-d'_3-1})+\delta_{-d'_3}t^{-d'_3}\right]
\partial_{v_1},
}
with the coefficients $\beta_n,\delta_n$ subject to the boundary conditions 
\boundcondAi .
Choosing $(\delta_n),\ n=-d_3',\ldots,0$ to be 
holomorphic coordinates, the $S^1$-equivariant decomposition of 
$H^0(\Delta,{\cal T}_{\Delta})$ reads 
\eqn\zerocohAi{
\bigoplus_{n=0}^{d'_3}\left(-\lambda_3+{n\over d'_3}\lambda_3\right).}

Next, proceeding as in the previous case, we find that the cokernel of 
$\kappa$ is generated by sections of the form
\eqn\cokernelkappaBi{
s_1=\left(\sum_{n=1}^{d'_3-1} \alpha_nt^n\right)\partial_{x_1}+
\left(\sum_{n=-d'_3+1}^{-1}
\gamma_nt^n\right)\partial_{u_1},
}
with the coefficients $\alpha_n,\beta_n$ subject to the boundary 
conditions \boundcondAi.\ If we choose $\gamma_n,\ n=-d'_3+1,\ldots,-1$,
to be holomorphic coordinates, $H^1(\Delta,{\cal T}_{\Delta})$ 
is $S^1$-isomorphic to
\eqn\onecohAi{
\bigoplus_{n=1}^{d'_3-1}\left(\lambda_1-{n\over d'_3}\lambda_3\right).
}
Finally, we are left with $H^0(\Delta,{\cal L}_{\Delta})$. 
In this case the {\v C}ech complex reads 
\eqn\cechBnewi{ 0\ra \CL_{\Delta}(\Upsilon_1)\oplus 
\CL_{\Delta}(\Upsilon_3) {\buildrel 
{\kappa}
\over \ra} \CL_{\Delta}(\Upsilon_{1}\cap\Upsilon_3) \ra 0.}
The generic sections of $\CL$ over $\Upsilon_1$ and $\Upsilon_3$ have the form
\eqn\locsectDnewi{\eqalign{
&{\tilde s_1}= \left(\sum_{n\in \IZ}\epsilon_n {t}^n\right)
\Lambda_1^{max},\cr
&{\tilde s_3}= \left(\sum_{n=0}^{\infty}\epsilon'_n {t'}^n\right)
\Lambda_3^{max},\cr
}}
where $\Lambda_1^{max}=\partial_{x_1}\wedge\partial_{y_1}
\wedge\partial_{u_1}\wedge\partial_{v_1}$ 
and $\Lambda_3^{max}=\partial_{x_3}\wedge\partial_{y_3}
\wedge\partial_{u_3}\wedge\partial_{v_3}$. 
Using \lintransfii\ we find that $\kappa$ is given by 
\eqn\mapkappaBnewi{
\kappa({\tilde s_1},{\tilde s_3})=\left(\sum_{n\in\IZ}
(\epsilon'_n-\epsilon_{-n})t'^n\right)\Lambda_3^{max},
}
where it is understood that $\epsilon'_n=0$ for $n<0$. Then, 
also using the boundary condition 
$\epsilon_n={\bar\epsilon}_{-n}$, it follows that the kernel 
of $\kappa$ is generated by sections of the form 
\eqn\kernelkappaB{
{\tilde s}_1=\epsilon_0\Lambda_1^{max}.
}
Therefore, $H^0(\Delta,\CL_{\Delta})$ is $S^1$-isomorphic to
$\left( 0\right)_{\IR}$. A similar analysis shows that 
$H^1(\Delta,\CL_{\Delta})$ is trivial. 

To conclude this section, note that horizontal discs such as 
$D'_3$ are embedded in the smooth component $\oY_1$ of $\oY$, 
and have no common points with the singular divisor. 
This is already clear from the local analysis in section five. 
Then one can compute the same multicover contributions by taking 
open string maps to $\oY_1$ as in \DFGii\ instead of $\CZ$. 
If the formalism proposed here is robust, this point of view 
should give the same results as above. Since this computation 
has been performed in detail in \DFGii,\ let us just record the 
results. With suitable choices of complex structures, we have 
\eqn\loccoverA{\eqalign{
& H^0(\Delta, \CT)_{\oY} \simeq (0)_\IR\oplus \bigoplus_{n=0}^{d'_3-1}
\left(-\lambda_3+{n\over d'_3}\lambda_3\right)\cr
& H^1(\Delta, \CT)_{\oY} \simeq \bigoplus_{n=1}^{d'_3-1} 
\left(\lambda_1-{n\over d'_3}\lambda_3\right).\cr 
}}
Note that in this case there is no obstruction bundle since 
$D'_3$ is a rigid disc on $\oY_1$ supported away from the 
canonical class. Therefore the expected dimension of the moduli space 
is zero \KL.\ The formulae \loccoverA\ are very similar to 
\zerocohAi,\ \onecohAi\ except for a small discrepancy 
in the fixed parts of the deformation spaces
$H^0(\Delta, \CT)$, $H^0(\Delta, \CT)_{\oY}$. 
For maps to $\CZ$, one finds that the fixed part is $(0)_\IC$ 
corresponding to the term $n=d'_3$ in \zerocohAi,\ while 
for maps to $\oY$ the fixed part is $(0)_\IR$. This discrepancy 
is accounted for by the fixed part of the obstruction space, 
as explained in section 6.3. The final result is that the two 
contributions agree, as expected.

\appendix{C}{Chern-Simons Expectation Values}

The Chern-Simons expansion performed in section seven involves 
products of holonomy variables in $U(N)$ representations 
of the form $R_i, {\overline R_j}$, where $R_i, R_j$ are defined 
by Young tableaux.  
Such expressions can usually be avoided  
in the context of large $N$ duality literature \refs{\LMi,\LMii, \MV, \OV, \RS} by analytic continuation. 
This is no longer true in compact examples, hence we have to  
evaluate such knot and link invariants by direct methods. 
We only need to consider two cases. 

$i)$ Consider unknot invariants of the form 
\eqn\unknotA{
\left<\Tr_{R_1}V\Tr_{{\overline R}_2}V\right>_p}
where $p$ denotes the framing. 
Proceeding by analogy with \refs{\LMi,\LMii, \MV, \OV, \RS}, we write  
\eqn\repdecompA{
\left<\Tr_{R_1}V\Tr_{{\overline R}_2}V\right>_p = \sum_{\rho} 
\left<\Tr_{\rho}V\right>_p}
where 
\eqn\repdecompB{R_1\otimes {\overline  R}_2 = \oplus_\rho  \rho}
is the irreducible 
decomposition of the product representation. 
Each term in the right hand side of \repdecompA\
can then be evaluated according to the 
rules explained in \MV\ 
\eqn\unknotB{
\left<\Tr_{\rho}V\right>_p = x^{pC\rho} \left<\Tr_{\rho}V\right>_0}
where $x=e^{i\pi\over k+N}$ and $C_\rho$ is the second Casimir of $\rho$. 
Therefore the computation reduces to some standard representation theory 
and expectation values of the unknot in the canonical framing. 
Given a $SU(N)$ representation $R$ defined by a Young tableau $\Pi$, 
$\overline R$ is isomorphic to the $SU(N)$ representation defined 
by the complementary Young tableau $\overline \Pi$. Two Young tableaux 
$\Pi, {\overline \Pi}$ are called complementary if by adjoining the 
$(N-k)$-th row of ${\overline \Pi}$ to the $k$-th row of $\Pi$ 
we obtain a square tableau with $N\times N$ boxes. Then, for 
two $SU(N)$ representations $R_1, {\overline R}_2$, one can compute 
the irreducible decomposition \repdecompB\ using the standard rules 
for Young tableaux. The irreducible decomposition for $U(N)$ 
representations can be obtained by tensoring the $SU(N)$ decomposition 
by appropriate $U(1)$ representations. 
Using these rules we find the following relations 
\eqn\ctwo{\eqalign{
&\tableau{1}\otimes{\overline {\tableau{1}}}=\bbbone\oplus A_1,\cr
&\tableau{2}\otimes{\overline {\tableau{1}}}=\tableau{1}\oplus A_2,\cr
&\tableau{1 1}\otimes{\overline {\tableau{1}}}=\tableau{1}\oplus A_3,\cr
}}
where $A_1, A_2, A_3$ are irreducible representations of $U(N)$ 
given by 
\eqn\ctwoB{
A_1=\matrix{\tableau{2 1}\hfill\cr\vdots\hfil\cr \tableau{1}\hfill\cr}
\otimes (-\sqrt{N}),\quad
A_2=\matrix{\tableau{3 1}\hfill\cr\vdots\hfil\cr \tableau{1}\hfill\cr}
\otimes (-\sqrt{N}),\quad
A_3=\matrix{\tableau{2 2 1}\hfill\cr\vdots \hfil\cr \tableau{1}\hfill\cr}
\otimes (-\sqrt{N}).}
The number of boxes in the first column of each tableau in \ctwoB\ 
is $N-1$ and $(w)$ denotes a one dimensional $U(1)$ representation 
of charge $w$. A routine computation shows that the 
dimensions and the quadratic Casimir operators of 
$A_1, A_2, A_3$ are 
\eqn\cthr{\vbox{\halign{ $#$ \hfill &\qquad  $#$ \hfill &\qquad $#$ \hfill\cr
{\rm dim}(A_1)=N^2-1, & {\rm dim}(A_2)={N(N-1)(N+2)\over 2}, & {\rm dim}(A_3)={N(N+1)(N-2)\over 2},\cr
C_2(A_1)=2N, & C_2(A_2)=3N+2, & C_2(A_3)=3N-2.\cr
}}}
Furthermore, we have the following relations 
\eqn\cfou{\vbox{\halign{ $#$ \hfill &\qquad $#$ \hfill\cr
\tableau{1}\otimes A_1=\tableau{1}\oplus A_2\oplus A_3, & 
{\overline {\tableau{1}}}\otimes {\overline A}_1={\overline {\tableau{1}}}\oplus{\overline A}_2\oplus{\overline A}_3.\cr
}}}

The unknot expectation values in canonical framing can be computed using 
the rules explained in \refs{\MV,\OV}. The $U(1)$ factor decouples, 
hence
$U(N)$ expectation values are the same as $SU(N)$ expectation values. 
The later are of the form $\left<\Tr_{R}V\right> = \Tr_R U_0$, 
where $U_0$ is a fixed group element. Using this property, one can 
easily show that 
\eqn\cfiv{\eqalign{
& \left<\Tr_{A_1}V\right>_0=(\left<\Tr_{\tableau{1}}V\right>_0)^2-1,\cr
& \left<\Tr_{A_2}V\right>_0=\left<\Tr_{\tableau{1}}V\right>_0
(\left<\Tr_{\tableau{2}}V\right>_0-1),\cr
& \left<\Tr_{A_3}V\right>_0=\left<\Tr_{\tableau{1}}V\right>_0
\left(\left<\Tr_{\tableau{1 1}}V\right>_0-1\right).\cr
}}
Using \unknotB\ and \cfiv\ and using the normalization explained in 
section five, we find 
\eqn\unknotC{\eqalign{
&\left<\Tr_\ybox V\Tr_{\overline \ybox} {V}\right>=y^{-2}+\left<\Tr_\ybox 
V\right>_0^2-1,\cr
&\left<\Tr_{\tableau{2}}V\Tr_{\overline \ybox} {V}\right>=
\left<\Tr_{\overline {\tableau{2}}}V\Tr_\ybox {V}\right>=
y^{-2}\left<\Tr_\ybox V\right>_0 +x^{2p}\left<\Tr_\ybox V\right>_0
(\left<\Tr_{\tableau{2}}V\right>_0-1),\cr
& \left<\Tr_{\tableau{1 1}}V\Tr_{\overline \ybox} {V}\right>=
\left<\Tr_{\overline{\tableau{1 1}}} V\Tr_\ybox {V}\right>=
y^{-2}
\left<\Tr_\ybox V\right>_0
+x^{-2p}\left<\Tr_\ybox V\right>_0\left
(\left<\Tr_{\tableau{1 1}}V\right>_0-1\right).\cr}}

$ii)$ We also encounter expectation values of the form 
\eqn\linkA{
\left< \Tr_{R_1} V \Tr_{R_2} U \right>^{+1}_{p_1, p_2}} 
for a Hopf link with linking number $+1$ and framings $(p_1, p_2)$. 
This computation can be reduced to a Hopf link with $l=-1$ and canonical 
framing on both components as follows 
\eqn\linkB{
\left< \Tr_{R_1} V \Tr_{R_2} U \right>^{+1}_{p_1, p_2}=
x^{p_1C_{R_1}+p_2C_{R_2}} 
\left< \Tr_{R_1} V \Tr_{R_2} U \right>^{-1}_{0,0}(x\rightarrow x^{-1}).}
The correctly normalized link invariant in the context of geometric transitions
\refs{\LMi-\LMii} reads 
\eqn\csix{
\left<\Tr_{\tableau{1}}V\Tr_{\overline {\tableau{1}}}U\right>^{-1}_{0,0}=x^{C_2({\tableau{1}})+C_2({\overline {\tableau{1}}})}
\sum_{\rho\in{\tableau{1}}\otimes{\overline {\tableau{1}}}}x^{-C_2({\rho})}{\rm dim}_q\rho.}
Dropping again the trivial framing factors, we finally obtain 
\eqn\csev{\eqalign{
&\left<\Tr V\Tr_{\overline \ybox} {U}\right>=
\left<Tr_{\overline \ybox} {V}\Tr_\ybox U \right>
=y^{-2}+\left<\Tr_\ybox V\right>_0^2-1,\cr
&\left<\Tr_{\tableau{2}}V\Tr_{\overline \ybox}{V}\right>=
\left<\Tr_{\overline {\tableau{2}}}V\Tr_{\ybox}{V}\right>=
y^{-2}x^{2(p-1)}\left<\Tr V\right>_0
+x^{2p}\left<\Tr V\right>_0(\left<\Tr_{\tableau{2}}V\right>_0-1),\cr
&\left<\Tr_{\tableau{1 1}}V\Tr_{\overline \ybox}{V}\right>=
\left<\Tr_{\overline {\tableau{1 1}}}V\Tr_{\ybox}{V}\right>=
y^{-2}x^{-2(p-1)}\left<\Tr V\right>_0
+x^{-2p}\left<\Tr V\right>_0
\left(\left<\Tr_{\tableau{1 1}}V\right>_0-1\right),\cr
&\left<\Tr V\Tr_{\overline\ybox} {V}\Tr U\right>=
\left<\Tr V\Tr_{\overline\ybox} {V}\Tr_{\overline\ybox} {U}\right>=
2y^{-2}\left<\Tr V\right>_0
+x^2\left<\Tr V\right>_0(\left<\Tr_{\tableau{2}}V\right>_0-1)\cr
&\qquad\qquad\qquad\qquad \qquad\qquad\qquad\qquad \qquad
+x^{-2}\left(\left<\Tr_{\tableau{1 1}}V\right>_0-1\right).
}}
\vfill\eject


\listrefs
\end